\documentclass[journal]{IEEEtran}
\usepackage{mathrsfs}
\usepackage{epsfig}
\usepackage{graphicx}
\usepackage{color}
\usepackage{placeins}
\usepackage{float}
\usepackage{latexsym}
\usepackage{boxedminipage}
\usepackage{cite}
\usepackage{tabularx,colortbl}
\usepackage{algorithm}
\usepackage{algorithmic}
\usepackage{amsthm}
\usepackage{enumerate}
\usepackage{subfigure}
\usepackage{amssymb}
\usepackage{bm}
\usepackage{array}
\usepackage{amsfonts}
\usepackage{amsmath}
\usepackage{multirow}
\usepackage{color}
\usepackage{bm}
\usepackage{multicol}
\usepackage{booktabs}
\usepackage[pagewise,switch]{lineno}
\newtheorem{definition}{Definition}

\newtheorem{proposition}{Proposition}

\makeatletter

\newenvironment{myproof}[1][\proofname]{\par
  \pushQED{\qed}%
  \normalfont \topsep6\p@\@plus6\p@\relax
  \trivlist
  \item[\hskip\labelsep
        \itshape
    #1\@addpunct{:}]\ignorespaces
}{%
  \popQED\endtrivlist\@endpefalse
}
\makeatother
\begin{document}
\title{Graph Learning for Spatiotemporal Signals with Long- and Short-Term Characterization}
\author{Yueliang~Liu,~\IEEEmembership{Student Member,~IEEE,}
        Wenbin~Guo,~\IEEEmembership{Member,~IEEE,}
        Kangyong~You,
        Lei~Zhao,
        Tao~Peng,
        Wenbo~Wang,~\IEEEmembership{Senior~Member,~IEEE}

\thanks{
This work was supported by
the National Natural Science Foundation of China (61271181,61571054),
the Science and Technology on Information Transmission and Dissemination in Communication Networks Laboratory Foundation and
BUPT Excellent Ph.D. Students Foundation (Grant CX2018101) (\it Corresponding author: Wenbin Guo.)}

\thanks{Y. Liu and W. Guo are with the School of Information and Communication Engineering, Beijing University of Posts and Telecommunications, Beijing 100876, China, and also with the Science and Technology on 	Information Transmission and Dissemination in Communication Networks Laboratory, Shijiazhuang 050000, China (e-mail: \{liuyueliang, gwb\}@bupt.edu.cn).}
}

\maketitle

\begin{abstract}

Mining natural associations from high-dimensional spatiotemporal signals plays an important role in various fields including biology, climatology, and financial analysis.
However, most
existing works have mainly studied time-independent signals without considering the correlations of spatiotemporal signals that achieve high learning accuracy.
This paper aims to learn graphs 
that better reflect underlying data relations by
leveraging
the long- and short-term characteristics of spatiotemporal signals.
First, a spatiotemporal signal model is presented that considers both spatial and temporal relations. In particular, 
we integrate
a low-rank representation and a Gaussian Markov process 
to describe the temporal correlations.
Then, the graph learning problem is formulated as a
joint low-rank component estimation and graph Laplacian inference.
Accordingly, we propose a
low rank and spatiotemporal smoothness-based graph learning method (GL-LRSS), which
introduces a spatiotemporal smoothness prior into time-vertex signal analysis.
By jointly exploiting the low rank of long-time observations and the
smoothness of short-time observations, the overall learning performance can be effectively improved.
Experiments on both synthetic and real-world datasets
demonstrate substantial improvements in the learning accuracy of the
proposed method over the state-of-the-art low-rank component estimation and
graph learning methods.

\end{abstract}


\IEEEpeerreviewmaketitle

\section{Introduction}
\IEEEPARstart{A}{pplications} \hspace{-0.2cm}
in a variety of fields, \hspace{-0.1cm}from finance and sociology to transportation and sensor networks,
rely on statistics, modeling, and processing of spatiotemporal signals.
These signals often represent 
long time series
measured over a certain spatial range.
Examples include biomedical imaging data \cite{ref1}, video sequences \cite{ref2}, social interactions among
individuals \cite{ref3}, and environmental sensing \cite{ref4}.
The usually complex spatiotemporal correlations and interactions can hinder the analysis of spatiotemporal signals.

Graphs can be useful for data analysis due to their ability to provide flexible descriptions in irregular domains.
In recent years, graph signal processing (GSP)\cite{ref5} has provided an engineering
paradigm for processing spatiotemporal signals on graphs, establishing
time-varying graph signals, based on the spectral graph theory \cite{ref37}.
For analysis and learning, data may be suitably represented by a graph, and the graph Laplacian matrix, 
which is equivalent to graph topology, can be used to solve
many problems including graph signal compression \cite{ref8},
graph signal reconstruction \cite{ref10}, and graph filtering \cite{ref6}.
Although graph-based methods have been successfully applied,
the graph structure is not always available, and straightforward representations
(e.g., geographical $k$-nearest neighbors)
may not adequately capture intrinsic relations among data.
Therefore, efficient graph learning methods should be developed
to improve the quality and efficiency of data analysis (e.g., trend identification).
Extracting underlying relations from observed spatiotemporal
signals is essential for their analysis.

In many cases, the collected spatiotemporal signals are highly redundant and thus strongly correlated. To learn a high-quality graph from these spatiotemporal signals,
their correlation properties must be thoroughly studied.
Recent studies \cite{ref112,ref13,ref35} provided effective ways of characterizing correlation properties by assuming spatiotemporal signals
to be approximately low-rank and have short-term stability. The corresponding results showed methodological superiority in signal processing tasks.
However, most existing graph learning methods neglect the long-term correlation of signals, for example, by modeling the spatiotemporal signals locally \cite{ref24} or by
treating the successive signals independently \cite{ref20,ref21}.
Although the mentioned graph learning methods achieve satisfying results,
there is still much room for improvement.
Therefore, in this paper, we propose an enhanced graph learning method that fully leverages long- and short-term
correlations in spatiotemporal signals.



\vspace{-0.25cm}
\subsection{Related works}
Our work involves joint graph learning and low-rank component estimation. 
Several approaches have been proposed to address these two problems, and detailed surveys are available in \cite{ref16,ref9,ref54}. However, graph learning and low-rank approximation have not been jointly studied.

For low-rank
component estimation, various methods have approximated spatiotemporal
signals as low-rank matrices \cite{ref11}, \cite{ref12}, achieving satisfactory results by 
assuming that the matrix collecting the time sequences is approximately
low-rank.
Recently, GSP approaches were proposed to
recover low-rank components
by using
spectral graph regularization \cite{ref14,ref15,ref16}.
These approaches incorporated graph smoothness on 
low-rank matrices and improved both clustering and recovery performance.
However, the graphs are predefined based on
the geometric distance in these methods, possibly undermining the accuracy for subsequent analyses.

For graph learning,
early studies provided graphical models by
neighborhood selection per node\cite{ref53}. For stability under noise,
graphical lasso methods have been used to
estimate the inverse covariance or precision matrix \cite{ref17,ref18,ref19}.
The fast-growing GSP allows solving graph learning problems by implementing methods related to Gaussian Markov Random Fields (GMRFs) with the precision matrix defined by a graph Laplacian.
For smoothing the graph signals,
smoothness-based methods have been adopted during graph inference.
Dong \emph{et al.} \cite{ref20} first proposed a valid combinatorial graph Laplacian (CGL) learning method
under a smooth graph representation.
Then,
Kalofolias \cite{ref21} reformulated the problem in terms of the adjacency matrix and proposed a
computationally efficient algorithm. 
To generalize the restriction of the precision matrix to be a CGL,
Egilimez \emph{et al.} \cite{ref22} identified a GMRF model whose precision
matrix could be any of multiple types of graph Laplacians. 
Alternative smoothness-based approaches  
have also been effective \cite{ref23,ref24,ref25}, with the methodological implementation in \cite{ref24} and \cite{ref23} being based on space-time modeling and edge selection, respectively, whereas
a theoretical analysis of the reconstruction error was provided in \cite{ref25}.
These methods learned graphs from smooth graph signals,
while a few other works added assumptions on the graph dynamics
for time-varying graph learning.
For instance, dynamic graphs have been learned by assuming that the graph structure changes
smoothly over time \cite{ref26}, whereas a method considering the sparseness of the graph variation was also proposed in \cite{ref27}.

Another family of graph learning approaches adopts a physics perspective for modeling graph signals.
In these cases,
observations are modeled considering a physical process for the graph, such as
diffusion \cite{ref28,ref29,ref30,ref31} and causality \cite{ref32,ref33,ref34}.
Segarra \emph{et al.} \cite{ref28} and Pasdeloup \emph{et al.} \cite{ref29}
identified a graph from stationary observations assumed to be
generated by a diffusion process.
Shafipour \emph{et al.} \cite{ref30} generalized this assumption by exploring
a graph learning method that could be applied to non-stationary
graph signals. Thanou \emph{et al.} \cite{ref31} proposed graph learning considering that the graph signals result from heat
diffusion. Causality-based methods
focus on the asymmetric adjacency matrix corresponding
to a directed graph. Mei \emph{et al.} \cite{ref32} considered a causal graph
process to characterize a time series and applied it to temperature analysis.
Under a structural equation model, Baingana \emph{et al.} \cite{ref33} proposed a
recursive least-squares estimator to track both the signal state and
graph topology. Similarly, Shen \emph{et al.} \cite{ref34} described nonlinear dependencies of signals
via structural vector autoregressive models and developed an efficient estimator to infer a sparse graph.
While these graph learning methods can provide meaningful graphs from time series, long-term correlations (i.e., low rank) have been neglected.

\vspace{-0.73cm}
\subsection{Contributions}
For high-quality graph learning, we propose
a method that considers the low rank and local smoothness of spatiotemporal signals.
Low-rank component estimation allows to improve the quality of the learned graph, with the low-rank component being better estimated from a refined graph.
The main contributions
of this study can be summarized as follows:
\begin{itemize}
\item[1)]
To the best of our knowledge, this is the first model of spatiotemporal signals integrating a low-rank representation and a first-order Gaussian Markov process.
\item[2)]
We introduce a spatiotemporal smooth prior to the time-varying graph signal to facilitate graph learning.
\item[3)]
Graph learning 
is formulated as a joint graph refinement and low-rank component estimation problem
solved using the proposed graph learning method based on low-rank approximation and spatiotemporal smoothness (GL-LRSS), which applies the alternating direction method of multipliers (ADMM) and alternating minimization.
\item[4)]  
We provide visual and quantitative comparisons with state-of-the-art low-rank component estimation and graph learning methods. The extensive experimental results on synthetic and real-world datasets demonstrate the superiority and effectiveness of the proposed GL-LRSS. 
\end{itemize}
\vspace{-0.33cm}
\subsection{Comparison with state-of-the-art methods}
Regarding graph signal representation,
the proposed GL-LRSS extends smoothness-based graph learning. 
Although the methods in \cite{ref35} and \cite{ref20} are the most related to the proposed GL-LRSS, they neglect the local temporal correlations of spatiotemporal signals. Specifically, the smoothness-based method (e.g., \cite{ref20,ref21,ref22}) uses a GMRF model which is mainly suitable for time-independent signals, and the method in \cite{ref35} uses a low-rank signal model which lacks local signal characterization. The proposed GL-LRSS adopts a different model from these similar methods in the following aspects:
\begin{itemize}
\item The novel model for spatiotemporal signals considers local and global correlations. By combining a low-rank representation and a first-order Gaussian Markov process, the proposed model can describe multiple types of time correlations.

\item Although both the method in \cite{ref35} and the proposed GL-LRSS aim to jointly estimate the graph structure and low-rank components, the method in \cite{ref35} obtains its final optimization by directly combining
the objective functions of two estimation subproblems. Our method, on the other hand,
formulates the optimization problem based on Bayesian inference and introduces a new regularization term called spatiotemporal smoothness for graph learning. 
\end{itemize}  

The remainder of this paper is organized as follows.
Section II presents the notation and preliminaries of GSP.
In Section III, we propose the low-rank graph-based model and
the corresponding spatiotemporal smoothness prior.
In Section IV, we formulate the graph learning problem as a
joint low-rank component and graph topology estimation and propose the GL-LRSS to solve the optimization problem alternately.
In Section V, 
the GL-LRSS performance on both synthetic
and real-world datasets is reported and compared with that of baseline methods. 
A discussion is presented in Section VI, and
we draw conclusions in Section VII.
\vspace{-0.23cm}
\section{Notation and Preliminaries}
\subsection{Notations}
Throughout this paper, lowercase letters (e.g., $\alpha$, $\beta$), lowercase boldface letters (e.g., ${\bf{x}}$, ${\bf{u}}$),
and uppercase boldface letters (e.g., ${\bf{X}}$, ${\bf{L}}$)
denote scalars, vectors, and matrices, respectively.
Unless otherwise stated, calligraphic capital letters (e.g., ${\mathcal{E}}$ and ${\mathcal{L}}$)
represent sets.
Additional notation is listed in Table \ref{tab:1}.
\begin{table}[t]
\newcommand{\tabincell}[2]{\begin{tabular}{@{}#1@{}}#2\end{tabular}}
\centering
\caption{\label{tab:1}List of symbols and their meaning}
\begin{tabular}{c c}
\hline
\hline
\specialrule{0em}{2pt}{2pt}
\boldmath{Symbols} & \boldmath{Meaning}  \\
\specialrule{0em}{2pt}{2pt}
\hline
\specialrule{0em}{2pt}{2pt}
$\mathcal{G}$ $|$ $\bf{L}$ $|$ ${{\mathcal {L}}^N}$ & weighted graph $|$ graph Laplacian matrix $|$ set of CGLs\\
${\mathcal{V}}$ $|$ ${\mathcal{E}}$ & vertex set $|$ edge set \\
$N$ $|$ $M$  & number of vertices $|$ number of time instants \\
$\bf{I}$ $|$ $\bf{W}$ $|$ $\bf{D}$ & identity matrix $|$ adjacency matrix $|$ degree matrix\\
${\bf{U}}$ $|$ $\bf{\Lambda}$ & eigenvector matrix $|$ eigenvalue matrix of $\bf{L}$\\
$\bf{0}$ $|$ $\bf{1}$ & column vector of zeros $|$ column vector of ones\\
${{\bf{X}}^{ - 1}}$ $|$ ${{\bf{X}}^\dag}$ & inverse of ${\bf{X}}$ $|$ pseudo-inverse of ${\bf{X}}$\\
${{\bf{X}}^{ T}}$ $|$ ${{\bf{x}}^{T}}$ & transpose of ${\bf{X}}$ $|$ transpose of ${\bf{x}}$\\
${\left( {\bf{X}} \right)_{ij}}$ & entry of ${\bf{X}}$ at $i$-th row and $j$-th column\\
${ {\bf{x}} _{i}}$ & $i$-th entry of ${\bf{x}}$ \\
$ \ge \left(  \le  \right)$ & element-wise greater (less) than or equal to opertor\\
${\bf{X}} \succeq 0$ & ${\bf{X}}$ is a positive semidefinite matrix\\
${\rm{tr}}$ $|$ ${\rm{vec}}$   & trace operator $|$ vectorization operator \\
$ \otimes $ $|$ $\langle  \cdot , \cdot \rangle$ & Kronecker product operator $|$ inner product operator\\
${\rm{diag}}\left( {\bf{x}} \right)$ & diagonal matrix formed by elements of ${\bf{x}}$\\
$p\left({\bf{x}}\right)$ & probability density function of random vector $\bf{x}$\\
${\bf{x}}\sim{\cal N}\left( {{\bf{0}},{\bf{\Sigma}} } \right)$ & zero-mean multivariate Gaussian with covariance ${\bf{\Sigma}}$\\
${\left\| {\bf{X}} \right\|_*}$ & nuclear norm of ${\bf{X}}$ \\
${\left\| {\bf{x}} \right\|_1}$ $|$ ${\left\| {\bf{X}} \right\|_1}$ & sum of absolute values of all elements (${l}_1$-norm)\\
\specialrule{0em}{1pt}{1pt}
$\left\| {\bf{x}} \right\|_2^2$ $|$ $\left\| {\bf{X}} \right\|_F^2$ & sum of squared values of elements\\
\specialrule{0em}{2pt}{2pt}
\hline
\hline
\end{tabular}
\vspace{-0.2cm}
\end{table}

\subsection{Graph Laplacian}
We consider an undirected weighted
graph with non-negative edge weights and no self-loops.
Let $ {\mathcal{G}}=\left( {{\mathcal{V}},{\mathcal{E}},{\bf{W}}} \right)$ be an $N$-vertex weighted graph,
where ${\mathcal{V}}=\left( {{v_1}, \ldots ,{v_N}} \right)$ is the vertex set
and ${\mathcal{E}}$ is the edge set.
The adjacency matrix $\bf{W}$ is an $N \times N$ symmetric matrix. The CGL of ${\mathcal{G}}$ is defined as
$\bf{L} = \bf{D} - \bf{W}$, where diagonal matrix
$\bf{D}$ denotes the degree
matrix with its $i$th diagonal entry indicating the degree of vertex
$i$ (i.e., ${\rm{diag}} {\left( {\bf{D}} \right)}_i = \sum\nolimits_{j = 1}^N {{W_{ij}}}$). 
${{\mathcal {L}}}$ is the set of all valid $N \times N$ CGLs for matrix $\bf{L}$:
\begin{eqnarray}\label{Laplacian defination}
\resizebox{0.93\hsize}{!}{$
\begin{aligned}
{{\mathcal {L}}}=\left\{ {{\bf{L}}|{\bf{L}} \succeq 0,{\kern 1pt} {\left( {\bf{L}} \right)_{ij}} = {\left( {\bf{L}} \right)_{ji}} \le 0{\kern 1pt} ,i \ne j,{\kern 1pt} {\kern 1pt}  {\rm{and}}{\kern 1pt} {\kern 1pt}  {\bf{L}}\cdot{\bf{1}} = {\bf{0}}} \right\}.
\end{aligned} $}
\end{eqnarray}

As the CGL
is a real symmetric positive semidefinite matrix, 
its eigenvalues are non-negative. Let the CGL eigendecomposition be ${\bf{L}} = {\bf{U\Lambda }}{{\bf{U}}^T}$, where ${\bf{\Lambda }}
= {\rm{diag}}\left( {\lambda _1},{\lambda _2}, \ldots ,{\lambda _N}\right)$
and ${\bf{U}} = \left[ {{{\bf{u}}_1},{{\bf{u}}_2}, \ldots ,{{\bf{u}}_N}}\right]$ are matrices containing the eigenvalues and eigenvectors, respectively.
The graph frequency spectrum is defined by the ascending array
of eigenvalues ${0=\lambda_1\le\lambda_2\le\dots\le\lambda_N}$,
referred to as the graph frequency, and orthogonal
eigenvectors ${{{\bf{u}}_1},{{\bf{u}}_2}, \ldots ,{{\bf{u}}_N}}$
are the harmonics associated with the graph frequencies.
In addition, the CGL of a connected graph always has
a zero eigenvalue (i.e., $\lambda_1=0$) 
corresponding
to eigenvector \begin{small}${{\bf{u}}_1}=
1/\sqrt N \cdot{\bf{1}}$\end{small}.

\vspace{-0.3cm}
\subsection{Smooth Graph Signals}
For graph signal ${\bf{x}}={\left[ {{x_1},{x_2}, \ldots ,{x_N}} \right]^T}$,
where ${x_i}$ is attached to vertex ${v_i}$,
its frequency component is defined by the graph Fourier transform denoted as ${\bf{\hat {\bf{x}}}} = {{\bf{U}}^T}{\bf{x}}$.
The frequency components corresponding to higher eigenvalues indicate larger variations between
the signals of the vertices, whereas those corresponding to
small eigenvalues are relatively smooth.
Many real-world datasets have graph signals that change smoothly between the connected vertices.
Such smoothness property indicates the graph signal variation with respect to
the underlying graph. To quantify the smoothness
of signal ${\bf{x}}$, a typical
metric can be written in the graph Laplacian quadratic
form \cite{ref37}:
\begin{eqnarray} \label{sm}
\resizebox{0.63\hsize}{!}{$
\begin{aligned}
S\left({\bf{x}}\right)={{\bf{x}}^T} {\bf{L}}{\bf{x}} = \sum\limits_{\left( {i,j} \right) \in {\mathcal{I}} } {{\left({\bf{W}}\right)}_{i,j}}{{\left[ {x_j - x_i} \right]}^2},
\end{aligned} $}
\end{eqnarray}
where $\mathcal{I}=\left\{(i, j) |\left(v_{i}, v_{j}\right) \in \mathcal{E}\right\}$ is
the set of index pairs of connnected vertices. 
Eq. (\ref{sm})
measures the total variation of connected
vertices associated with edge set ${\mathcal{E}}$.
In the vertex domain, smaller
values of Eq. (\ref{sm}) indicate higher signal smoothness in the graph.

\vspace{-0.3cm}
\subsection{Correlations in Spatiotemporal Signals}
Spatiotemporal signals can be viewed as time-varying graph signals in a graph structure of the observation sites. These signals are usually highly redundant and thus strongly correlated.
Global and local consistency principles have been identified
for data description \cite{ref38}, unveiling long- and short-term correlations.

\emph{Long-term correlation}:
The global consistency indicates that spatiotemporal signals are usually correlated
globally \cite{ref11,ref13}. Such correlation describes the space and time commonalities over a long time and can be interpreted
as temporal sequences of the form ${\bf{x}}_1, {\bf{x}}_2, \dots, {\bf{x}}_{M}$ generated from limited patterns. Hence, spatiotemporal signals $\bf{X}$ can be approximately low-rank \cite{ref16,ref12}.

\emph{Short-term correlation}:
Spatiotemporal
signals can be locally correlated \cite{ref24,ref38}, as observations from a site can be correlated across neighboring time instants for the
temporal sequences to vary smoothly over time.
Likewise, at a given instant, nearby
observation sites can exhibit spatial correlations corresponding to similar values.
These two types of short-term correlations are respectively determined by
temporal smoothness and spatial smoothness.

GSP methods are based upon spatial and temporal smoothness.
Although spatial smoothness has been widely applied \cite{ref10}, \cite{ref21}, \cite{ref16}, few studies have leveraged temporal smoothness \cite{ref51}, \cite{ref52}.
By combining spatial and temporal smoothness,
we introduce the concept of spatiotemporal smoothness
and propose the GL-LRSS, which also considers long-term correlations.
Spatiotemporal smoothness  
describes short-term characteristics
of time-varying graph signals.

\section{Long- and short-term characterization of spatiotemporal signals}

\subsection{Signal Representation}
Spatiotemporal signals exhibit
global and local correlations over the long and short terms, respectively.
To describe these correlations,
we propose a model that characterizes spatiotemporal signals from the local and global perspectives.

Consider an $N$-vertex graph with graph Laplacian matrix \begin{small}${\bf{L}}\in {\mathbb {R}}^{N \times N}$\end{small}.
A spatiotemporal signal can be expressed by matrix \begin{small}${\bf{X}}={\left[ {\bf{x}}_1, {\bf{x}}_2, \dots, {\bf{x}}_{M} \right ]\in {{\mathbb {R}}^{N \times M}}}$\end{small}, where $M$ is the number of time instants. In the proposed model, the observed signal is modeled as
\begin{eqnarray} \label{global representation}
\resizebox{0.2\hsize}{!}{$
\begin{aligned}
{{\bf{Y}}} = {{\bf{X}}} +{{\bf{N}}},
\end{aligned} $}
\end{eqnarray}
where ${{\bf{N}}}$ denotes the additive Gaussian white noise. 

\subsubsection{Short-term signal characterization}
Considering dependencies in the neighboring space and time,
we characterize the observed signal from a local viewpoint as follows:
\begin{eqnarray} \label{representation}
\resizebox{0.24\hsize}{!}{$
\begin{aligned}
{{\bf{y}}_t} = {{\bf{x}}_t} +{{\bf{n}}_t},
\end{aligned} $}
\end{eqnarray}
\vspace{-0.9cm}
\begin{eqnarray} \label{latent}
\resizebox{0.3\hsize}{!}{$
\begin{aligned}
{{\bf{x}}_t} = {\textit{R}}{{\bf{x}}_{t - 1}} + {{\bf{v}}_t},
\end{aligned} $}
\end{eqnarray}
where ${\bf{y}}_t\in{{\mathbb {R}}^{N}}$ is the
observation at the $t$th time instant, and ${\bf{n}}_t\in {{\mathbb {R}}^{N}}$ denotes the multivariate Gaussian noise with zero mean
and covariance ${\sigma _n}^2 {\bf{I}}_{N}$.
The state transition matrix ${\textit{R}}$
is defined as a general diagonal matrix ${\textit{R}} ={\rm{diag}}\left( {{c_1},{c_2}, \ldots ,{c_N}} \right)$, where $c_i$ denotes the coefficient of the $i$th observation site and ranges from 0 to 1. Each $c$ represents the autocorrelation
coefficient that describes the time correlation of data with
a delayed copy (one-time lag in this model) of itself,
and can be obtained in advance.

To represent signals residing on graphs 
and identify structures in data, 
we introduce graph-based process variable \begin{small}${{\bf{v}}_t}$\end{small}:
\begin{eqnarray} \label{latent distribution}
\resizebox{0.2\hsize}{!}{$
\begin{aligned}
{\bf{v}}_t={{\bf{U}}_{(r)}}{{\bf{z}}_t},
\end{aligned} $}
\end{eqnarray}
where \begin{small}${{\bf{U}}_{(r)}}\in {\mathbb {R}}^{N \times r}$\end{small} contains the first $r$ ($r\le N$) eigenvectors of the graph Laplacian matrix, and ${\bf{z}}_{t}\in {{\mathbb {R}}^{r}}$ is assumed to follow a multivariate Gaussian distribution, \begin{small}${{\bf{z}} _t} \sim {\mathcal N}\left( {\bf{0}}, {{{\boldsymbol{\Lambda}}_{\left( r\right)}}^\dag } \right)$\end{small}, with precision matrix \begin{small}${{\boldsymbol{\Lambda}}_{(r)}}^\dag$\end{small}
being the Moore-Penrose pseudoinverse of the matrix that contains the first $r$ eigenvalues. This definition leads to a smooth
graph signal representation and provides an intuitive relationship between the graph structure
and graph signal.
According to Eq. (\ref{latent distribution}),
the assumption about ${\bf{z}}_{t}$ and the basis vector ${{\bf{U}}_{(r)}}$
leads to a multivariate Gaussian distribution for
${\bf{v}}_t$ (i.e., \begin{small}${{\bf{v}} _t} \sim {\mathcal N}\left(
{\bf{0}}, {{{\tilde {\bf{L}}}^\dag }} \right)$\end{small},
with \begin{small}${{{\tilde {\bf{L}}}^\dag }}={{\bf{U}}_{(r)}}{{{ {{\bf{\Lambda }}_{\left( {r} \right)}}}^\dag }}
{{{\bf{U}}}_{(r)}}^T$\end{small}),
such that the representation of time-varying signals reflects the graph topology. 
Furthermore, as is shown in Section III-B-1, 
the short-term characterizations in Eqs. (\ref{latent}) and (\ref{latent distribution}) lead to the local smoothness of spatiotemporal signals.

\subsubsection{Long-term signal characterization} 
As mentioned in Section II-D, spatiotemporal signals are approximately low-rank in practice. Thus, it is realistic and efficient to treat spatiotemporal signals from a global viewpoint.
Considering the spatiotemporal signals across $M$ time instant \begin{small}${\bf{X}}={\left[ {\bf{x}}_1, {\bf{x}}_2, \dots, {\bf{x}}_{M} \right ]}$\end{small} with initialization ${\bf{x}}_0={\bf{v}}_0$, we obtain the matrix form of Eq.  (\ref{representation}) given by Eq. (\ref{global representation}).

For a convenient signal representation of $\bf{X}$,
a normal distribution \begin{small}${\mathcal {N}}\left( {\mu},{\sigma}_{R}^2 \right)$\end{small} is used to model temporal correlation coefficients \begin{small}${c_1}, \ldots ,c_{N}$\end{small}, such that correlation matrix ${\textit{R}}$ can be decomposed as \begin{small}${\textit{R}}={\mu}{\bf{I}}+\Delta{\textit{R}}$\end{small}, 
where ${\mu}{\bf{I}}$ corresponds to the mean and $\Delta{\textit{R}}$ represents the 
fluctuations around ${\mu}{\bf{I}}$. By applying both the decomposition of ${\textit{R}}$ and Eqs. (\ref{latent distribution}) to (\ref{latent}), the spatiotemporal signal $\bf{X}$ in Eq. (\ref{global representation}) becomes
\begin{eqnarray} \label{low rank X}
\resizebox{0.25\hsize}{!}{$
\begin{aligned}
{\bf{X}}={\bf{U}}_{(r)}{\bf{Z}}+{\bf{\Phi}},
\end{aligned} $}
\end{eqnarray}
\vspace{-0.02cm}
where \begin{small}${\bf{Z}}=\left[ {\mu}{\bf{z}}_0+{\bf{z}}_1,\ldots, {\mu}^M{\bf{z}}_0+{\mu}^{M-1}{\bf{z}}_1+\cdots+{\bf{z}}_M\right]$\end{small}, and \begin{small}${\bf{\Phi}}$\end{small} is a complex perturbation term related to $\Delta{\textit{R}}$. Mathematically,
Eq. (\ref{low rank X}) is the matrix expression of Eq. (\ref{latent}) that establishes a relation between the long- and short-term signals. 
As discussed in Section III-B-2, spatiotemporal signals under the long-term characterization are approximately low-rank.

\subsection{Long- and Short-Term Properties in Signal Representation}
Under the local and global signal representations described in Section III-A, we explore the long- and short-term properties (i.e., low-rank property and spatiotemporal smoothness, respectively) of spatiotemporal signals.

\subsubsection{Short-term property}  
As discussed in Section III-A-1,
the model given by Eq. (\ref{latent}) is analogous to a first-order vector autoregressive model. It naturally promotes the temporal smoothness of the signal.
On the other hand, 
as small eigenvalues correspond to smooth eigenvectors on the graph, the selection of \begin{small}${{\bf{U}}_{(r)}}$\end{small} in Eq. (\ref{latent distribution}) as the basis vector supports the spatial smoothness of the signal.
Therefore,
the graph signal under the short-term characterizations in Eqs. (\ref{latent}) and (\ref{latent distribution}) exhibit spatiotemporal smoothness.
Moreover, we show in Section IV-A that our proposed method enforces such spatiotemporal smoothness property in graph learning. The definition of spatiotemporal smoothness is given next.

\begin{definition}[Spatiotemporal smoothness]
The weighted time differences of spatiotemporal signals are smooth
with respect to the graph structure. 
Based on (\ref{sm}), spatiotemporal smoothness can be defined as
\begin{eqnarray} \label{stm}
\resizebox{0.9\hsize}{!}{$
\begin{aligned}
S\left({{\bf{x}} _t}-{\textit{R}}{{\bf{x}} _{t-1}}\right)={\left({{\bf{x}} _t}-{\textit{R}}{{\bf{x}} _{t-1}}\right)^T} {\bf{L}}\left({{\bf{x}} _t{}}-{\textit{R}}{{\bf{x}} _{t-1}}\right)={{\bf{v}}_t}^T{{\bf{L}}}{{\bf{v}}_t}.
\end{aligned} $}
\end{eqnarray}
\end{definition}

Considering the signals across multiple time instants, the corresponding spatiotemporal smoothness in \emph{Definition 1} can be expressed in a matrix form as follows.
\begin{definition}[Weighted difference operator]
The weighted difference operator of graph signal
$\bf{X}$ is \begin{small}${\mathcal{D}}\left( {\bf{X}} \right) = {\bf{X}} - {\textit{R}}{\bf{X}}{\bf{B}}$\end{small},
where ${\bf{B}}$ is
a shift operator given by 
\begin{eqnarray}\label{operator}
\resizebox{0.45\hsize}{!}{$
\begin{aligned}
{\bf{B}} = {\left[ {\begin{array}{*{20}{c}}
0&1&{}&{}&{}\\
{}&0&1&{}&{}\\
{}&{}&0& \ddots &{}\\
{}&{}&{}& \ddots &1\\
{}&{}&{}&{}&0
\end{array}} \right]_{M \times M,}}
\end{aligned} $}
\end{eqnarray}
\end{definition}

The weighted difference signal is equal to
\begin{small}$ \mathcal{D}\left( {\bf{X}} \right)=\left[ {{\bf{x}}_1}, {{\bf{x}}_2}- {\textit{R}}{{\bf{x}}_1},{{{\bf{x}}_3} - {\textit{R}}{{\bf{x}}_2}, \ldots ,{{\bf{x}}_{M}} - {\textit{R}}{{\bf{x}}_{M - 1}}  } \right].$\end{small}
Hence, the matrix expression of spatiotemporal smoothness is 
\begin{eqnarray}\label{Dsmooth}
\resizebox{0.85\hsize}{!}{$
\begin{aligned}
S\left(\mathcal{D}\left( {\bf{X}} \right)\right)=\sum\limits_{t = 1}^{M} S\left({{\bf{x}}_t}-{\textit{R}}{{\bf{x}}_{t-1}}\right) \overset{\left(a\right)}{=} {\rm{tr}}{\kern 1pt} {\kern 1pt}  \left( {{{\mathcal{D}\left( {\bf{X}} \right)}^T}{\bf{L}}\mathcal{D}\left( {\bf{X}} \right)} \right),
\end{aligned} $}
\end{eqnarray}
where (a) follows from \emph{Definition 2} and the smoothness metric $S\left( \cdot\right)$ in (\ref{sm}).

\subsubsection{Long-term property}
Besides the above mentioned short-term properties, the following discussion shows the approximately low-rank property of 
$\bf{X}$ considering Eq. (\ref{low rank X}).

The first term in Eq. (\ref{low rank X}) resembles the formulation of principal component analysis (PCA), which is the most popular
technique for approximating low-rank components. 
According to conventional PCA, the low-rankness of $\bf{X}$ mainly depends on 
the number of basis vectors, i.e., the number of columns of ${\bf{U}}_{(r)}$. However, 
due to the perturbation in Eq. (\ref{low rank X}), the rank of $\bf{X}$ is also affected by ${\bf{\Phi}}$. Specifically, 
a small ${\sigma}_R$ tends to reduce the effect of ${\bf{\Phi}}$, and hence 
$\bf{X}$ is more likely to be low-rank, whereas a large ${\sigma}_R$ tends to weaken the low-rank property of $\bf{X}$. When ${\sigma}_R=0$ (i.e., \begin{small}${\bf{X}}={\bf{U}}_{(r)}{\bf{Z}}$\end{small}), $\bf{X}$ is a low-rank matrix.
Therefore, the proposed model given by (\ref{global representation}) and (\ref{low rank X}) can be viewed as an approximately low-rank representation that roughly characterizes long-term correlations of the signal.
To precisely constrain the low-rank property of $\bf{X}$,
we introduce a nuclear norm for the optimization problem.

\section{Graph Learning based on low rank and spatiotemporal smoothness (GL-LRSS)}
In this section, we propose an efficient graph learning method by jointly exploiting the local smoothness and global correlation of spatiotemporal signals.
We first formulate the graph learning problem and then 
propose an optimization algorithm, GL-LRSS, which
is based on the ADMM and
alternating minimization. Finally, 
the computational complexity of the proposed algorithm is briefly analyzed.
\vspace{-0.3cm}
\subsection{Problem Formulation}
As mentioned in Section III-A, 
the graph structural information is encoded in the covariance of process variable ${{{\bf{v}}}_t}$. In terms of graph structure recovery, our method can be regarded as inverse covariance matrix estimation.
For probabilistic inference,
we first introduce the following weighted difference observation:
\begin{eqnarray} \label{diffsignal}
\resizebox{0.6\hsize}{!}{$
\begin{aligned}
{{\bf{d}}_t} = {{\bf{y}}_t} - {\textit{R}}{{\bf{y}}_{t - 1}}
={{\bf{v}}_t}+ {{\bf{n}}_t} - {\textit{R}}{{\bf{n}}_{t - 1}},
\end{aligned} $}
\end{eqnarray}
with initialization ${{\bf{d}}_1}$=${{\bf{y}}_1}$. Based on the distribution of Gaussian noise, 
the conditional
probability of ${{\bf{d}}_t}$ given ${{\bf{v}} _t}$ satisfies
\begin{eqnarray} \label{condition}
\resizebox{0.55\hsize}{!}{$
\begin{aligned}
{{{\bf{d}}}_t}|{{\bf{v}} _t}\sim{\cal N}\left( {{{\bf{v}} _t},{\sigma _n}^2\left( {{{\bf{I}}_N} + {{\textit{R}}{\textit{R}}^T}} \right)} \right).
\end{aligned} $}
\end{eqnarray}

Given the weighted difference observation ${{{\bf{d}}}_t}$
and the Gaussian prior distribution of ${{{\bf{v}}}_t}$,
we can compute a maximum a posteriori (MAP) estimate of core component ${{{\bf{v}}}_t}$. Specifically, by applying Bayes' rule,
the MAP estimate of ${{{\bf{v}}}_t}$ is given by
\begin{eqnarray} \label{MAPestimate}
\resizebox{0.92\hsize}{!}{$
\begin{aligned}
{{\bf{v}}_t}_{MAP}\left( {{{\bf{d}}_t}} \right)
&: = \mathop {\arg \max }\limits_{{{\bf{v}}_t}\in {{\mathbb {R}}^{N}}} p\left( {{{\bf{v}}_t}|{{\bf{d}}_t}} \right) = \mathop {\arg \max }\limits_{{{\bf{v}}_t}\in {{\mathbb {R}}^{N}}} p\left( {{{\bf{d}}_t}|{{\bf{v}}_t}} \right)p\left( {{{\bf{v}}_t}} \right)\\
& = \mathop {\arg \min }\limits_{{{\bf{v}}_t}\in {{\mathbb {R}}^{N}}} \left( { - \log {p_E}\left( {{{\bf{d}}_t} - {{\bf{v}}_t}} \right) - \log {p_V}\left( {{{\bf{v}}_t}} \right)} \right)\\
& = \mathop {\arg \min }\limits_{{{\bf{v}}_t}\in {{\mathbb {R}}^{N}}} {\left( {{{\bf{d}}_t} - {{\bf{v}}_t}} \right)^T}{{\bf{W}}^{ - 1}}\left( {{{\bf{d}}_t} - {{\bf{v}}_t}} \right) +\alpha {{\bf{v}}_t}^T{\tilde {\bf{L}}}{{\bf{v}}_t},
\end{aligned} $}
\end{eqnarray}
where \begin{small}${\bf{W}}={{{\bf{I}}_N} + {{\textit{R}}{\textit{R}}^T}} $\end{small}and
$\alpha$ is a constant parameter proportional to the noise variance, ${{ \sigma }_n}^2$.
However, the objective function in Eq. (\ref{MAPestimate}) is difficult to process \cite{ref24}, especially for the unknown correlation matrix ${\textit{R}}$.
To obtain a solution, a relaxation procedure can be adopted for the problem. 
By leveraging the diagonal property of matrix ${\textit{R}}$ and 
inequality 
\begin{eqnarray} \label{inequality}
\resizebox{0.83\hsize}{!}{$
\begin{aligned}
{{\left( {{{\bf{d}}_t} -   {{\bf{v}} _t}} \right)}^T}{\bf{W}}^{-1}{\left( {{{\bf{d}}_t} -   {{\bf{v}} _t}} \right)}
\geq  \lambda_{min}\left({\bf{W}}^{-1}\right) \left\| {{{\bf{d}}_t} - {{\bf{v}}_t}} \right\|_2^2,
\end{aligned} $}
\end{eqnarray}
we obtain a relaxed MAP estimation:
\begin{eqnarray} \label{MAP}
\resizebox{0.75\hsize}{!}{$
\begin{aligned}
{{\bf{v}}_t}_{MAP}\left( {{{\bf{d}}_t}} \right){: = } \mathop {\arg \min }\limits_{{{\bf{v}}_t}\in {{\mathbb {R}}^{N}}} \left\| {{{\bf{d}}_t} - {{\bf{v}}_t}} \right\|_2^2 + \alpha{\kern 1pt}{\kern 1pt} {{\bf{v}}_t}^T{\tilde {\bf{L}}}{{\bf{v}}_t}.
\end{aligned} $}
\end{eqnarray}

In Eq. (\ref{MAP}), the Laplacian quadratic term is the same as that of Eq. (\ref{stm}).
Therefore, it verifies that the proposed method enforces the 
spatiotemporal smoothness in graph learning.

Considering observations at $M$ time instants
\begin{small}${\bf{Y}}={\left[ {\bf{y}}_1, {\bf{y}}_2, \dots, {\bf{y}}_{M} \right ]\in {{\mathbb {R}}^{N \times M}}}$\end{small}, 
we focus on 1)
learning the graph Laplacian matrix that is equivalent
to the graph structure and 2) improving the
low-rank component estimation. By imposing additional constraints
on the graph Laplacian $\bf{L}$ and the low-rank component $\bf{X}$,
we propose to solve the problem of (\ref{MAP}) using the following objective function, given in a matrix form:
\begin{equation} \label{problem}
\resizebox{1\hsize}{!}{$
\begin{aligned}
&\left( {\rm{P1}} \right) {\kern 1pt} {\kern 1pt} {\kern 1pt} {\kern 1pt} {\kern 1pt} \mathop {\min }\limits_{{\bf{X }}\in {{\mathbb {R}}^{N \times M}},{\bf{L}}} {\kern 1pt} {\kern 1pt} {\kern 1pt} {\kern 1pt} {\kern 1pt} {\kern 1pt} {\kern 1pt} {\kern 1pt} {\kern 1pt} {\kern 1pt} {\kern 1pt} Q_1\left( {{\bf{L}},{\bf{X}}} \right)\\
&{\kern 1pt} {\kern 1pt} {\kern 1pt} {\kern 1pt} {\kern 1pt} {\kern 1pt} {\kern 1pt} {\kern 1pt} {\kern 1pt} {\kern 1pt} {\kern 1pt} {\kern 1pt} {\kern 1pt} {\kern 1pt} {\kern 1pt} {\kern 1pt} {\kern 1pt} {\kern 1pt} {\kern 1pt} {\kern 1pt} {\kern 1pt} {\kern 1pt} {\kern 1pt} {\kern 1pt} {\kern 1pt} {\kern 1pt} {\kern 1pt} {\rm{s}}.{\rm{t}}.{\kern 1pt} {\kern 1pt} {\kern 1pt} {\kern 1pt} {\kern 1pt} Q_1\left( {{\bf{L}},{\bf{X}}} \right) = {\kern 1pt} \left\| \mathcal{D}{\left( { {\bf{X}}- {\bf{Y}} } \right)} \right\|_F^2 + \alpha  {\kern 1pt} {\rm{tr }}\left( {{{\mathcal{D}\left( {\bf{X}} \right)}^T}{\bf{L}}\mathcal{D}\left( {\bf{X}} \right)} \right){\kern 1pt} \\
&  {\kern 1pt} {\kern 1pt} {\kern 1pt} {\kern 1pt} {\kern 1pt}  {\kern 1pt} {\kern 1pt} {\kern 1pt} {\kern 1pt} {\kern 1pt} {\kern 1pt} {\kern 1pt} {\kern 1pt} {\kern 1pt} {\kern 1pt} {\kern 1pt} {\kern 1pt} {\kern 1pt} {\kern 1pt} {\kern 1pt} {\kern 1pt} {\kern 1pt} {\kern 1pt} {\kern 1pt} {\kern 1pt} {\kern 1pt} {\kern 1pt} {\kern 1pt} {\kern 1pt} {\kern 1pt} {\kern 1pt} {\kern 1pt} {\kern 1pt} {\kern 1pt} {\kern 1pt} {\kern 1pt} {\kern 1pt} {\kern 1pt} {\kern 1pt} {\kern 1pt} {\kern 1pt} {\kern 1pt} {\kern 1pt} {\kern 1pt} {\kern 1pt} {\kern 1pt} {\kern 1pt} {\kern 1pt} {\kern 1pt} {\kern 1pt} {\kern 1pt} {\kern 1pt} {\kern 1pt} {\kern 1pt} {\kern 1pt} {\kern 1pt} {\kern 1pt} {\kern 1pt} {\kern 1pt} {\kern 1pt} {\kern 1pt} {\kern 1pt} {\kern 1pt} {\kern 1pt} {\kern 1pt} {\kern 1pt} {\kern 1pt} {\kern 1pt} {\kern 1pt} {\kern 1pt}{\kern 1pt}{\kern 1pt}{\kern 1pt}{\kern 1pt}{\kern 1pt}{\kern 1pt}{\kern 1pt}{\kern 1pt} {\kern 1pt} {\kern 1pt} {\kern 1pt} {\kern 1pt} {\kern 1pt}{\kern 1pt}{\kern 1pt}{\kern 1pt}{\kern 1pt}{\kern 1pt}{\kern 1pt}{\kern 1pt}{\kern 1pt}{\kern 1pt}{\kern 1pt}{\kern 1pt} {\kern 1pt} {\kern 1pt}{\kern 1pt} {\kern 1pt} {\kern 1pt}{\kern 1pt} {\kern 1pt} {\kern 1pt}  + \beta {\kern 1pt} \left\| {\bf{L}} \right\|_F^2+ \gamma {\kern 1pt} {\left\| {\bf{X}} \right\|_*}, \\
&{\kern 1pt} {\kern 1pt} {\kern 1pt} {\kern 1pt} {\kern 1pt} {\kern 1pt} {\kern 1pt} {\kern 1pt} {\kern 1pt} {\kern 1pt} {\kern 1pt} {\kern 1pt} {\kern 1pt} {\kern 1pt} {\kern 1pt} {\kern 1pt} {\kern 1pt} {\kern 1pt} {\kern 1pt} {\kern 1pt} {\kern 1pt} {\kern 1pt} {\kern 1pt} {\kern 1pt} {\kern 1pt} {\kern 1pt} {\kern 1pt} {\kern 1pt} {\kern 1pt} {\kern 1pt} {\kern 1pt} {\kern 1pt} {\kern 1pt} {\kern 1pt} {\kern 1pt} {\kern 1pt} {\kern 1pt} {\kern 1pt} {\kern 1pt} {\kern 1pt} {\kern 1pt} {\kern 1pt} {\kern 1pt} {\kern 1pt} {\bf{L}} \in {{\cal L}},{\kern 1pt} {\kern 1pt} {\kern 1pt} {\kern 1pt} {\kern 1pt}{\rm{tr}}\left( {\bf{L}} \right) = N, \nonumber
\end{aligned} $}
\end{equation}
where $\alpha$, $\beta$, and $\gamma$ are positive
regularization parameters corresponding to the regularization
terms.
The first regularization term, \begin{small}${\rm{tr }}\left( {{{\mathcal{D}\left( {\bf{X}} \right)}^T}{\bf{L}}\mathcal{D}\left( {\bf{X}} \right)} \right)$\end{small}, induces the spatiotemporal smoothness
encoded in Eq. (\ref{Dsmooth}).
Together with the trace constraint that aims to avoid
trivial solutions, the second regularization term, $\left\| {\bf{L}} \right\|_F^2$, controls the sparsity of the off-diagonal entries
in $\bf{L}$ (i.e., the edge weights of the graph).
To promote long-term correlations, we impose
nuclear norm
${\left\| {\bf{X}} \right\|_*}$, which is defined as
the sum of the singular values of ${\bf{X}}$ and corresponds to the convex
envelope of ${\rm{rank}}\left( {\bf{X}} \right)$.
The last Laplacian constraint guarantees that 
the learned graph Laplacian
is a valid CGL that satisfies Eq. (\ref{Laplacian defination}).

Note that in problem (P1), we particularly introduce two regularization terms, i.e.,
\begin{small}${\rm{tr }}\left( {{{\mathcal{D}\left( {\bf{X}} \right)}^T}{\bf{L}}\mathcal{D}\left( {\bf{X}} \right)} \right)$\end{small}
and ${\left\| {\bf{X}} \right\|_*}$
to characterize the correlation properties of spatiotemporal signals.
Although these two terms promote
the correlation of spatiotemporal signals from the local and global 
perspectives, respectively,
they compensate each other to deduce a meaningful graph, as detailed below. 


\begin{itemize}
\item
Regularization term \begin{small}${\rm{tr }}\left( {{{\mathcal{D}\left( {\bf{X}} \right)}^T}{\bf{L}}\mathcal{D}\left( {\bf{X}} \right)} \right)$\end{small} is derived from the proposed signal representation for graph learning.
This term encodes spatial and temporal correlations of $\bf{X}$ in graph
Laplacian ${\bf{L}}$ and weighted difference operator ${\mathcal{D}}$, respectively, while enforcing the weighted difference signal
to be smooth on the graph.
Unlike differential smoothness \cite{ref13},
this term contains a general correlation matrix ${\textit{R}}$
that considers the
varying temporal evolution of data
at distinct observation sites.
As demonstrated in real experiments,
when proper matrix ${\textit{R}}$ is known a priori,
the graph learning performance can be further improved.

\item
As data from many applications have the low-rank property,
we utilized ${\left\| {\bf{X}} \right\|_*}$ to improve the low-rank approximation. 
The nuclear norm directly forces spatiotemporal signal 
${\bf{X}}$ to achieve a low rank, thus compensating for the limitation of 
the proposed model in terms of long-term signal characterization, where 
the low-rank property of the spatiotemporal signal is partially depicted  
through the term resembling PCA in Eq. (\ref{low rank X}).
Moreover, nuclear norm ${\left\| {\bf{X}} \right\|_*}$ can increase the
graph learning performance, as verified experimentally.
\end{itemize}

In the optimization of (P1),
the graph Laplacian interacts with the low-rank component. Our hypothesis is that accurate low-rank component estimation
improves the quality of the learned graph, which in turn 
improves the low-rank component estimation.
Therefore, we adopted an alternating minimization framework that
iteratively refines the graph topology and estimates the low-rank components.
This hypothesis was validated through experiments on synthetic dataset.

\begin{algorithm}[t]
\renewcommand{\algorithmicrequire}{\textbf{Input:}}
\renewcommand\algorithmicensure {\textbf{Output:} }
\caption{\textbf{: Graph learning based on low rank and spatiotemporal smoothness (GL-LRSS)}}
\begin{algorithmic}[1]
\REQUIRE
Observations ${\bf{Y}}$, local correlation ${\textit{R}}$,
regularization parameters $\alpha$, $\beta$, $\gamma$, maximum iteration $K$, threshold $\varepsilon$.
\STATE Initialization: ${\bf{X}}^0 = {\bf{Y}}$, $k=1$;
\REPEAT
\STATE 1) Graph topology refinement: \\
\quad \quad\begin{small}${\bf{L}}^{k+1}=G\left({\bf{X}}^{k},{\bf{Y}}\right)$
\end{small} by (\ref{adm21})-(\ref{adm23})\\
\STATE 2) Low-rank component estimation:\\
\quad\quad
\begin{small}${\bf{X}}^{k+1}=C\left({\bf{L}}^{k+1},{\bf{Y}}\right)$
\end{small} by (\ref{admm21})-(\ref{admm23})
\STATE 3) $\left( {\hat {\bf{L}},\hat {\bf{X}}} \right) = \left( {{{\bf{L}}^k},{{\bf{X}}^k}} \right)$, $k=k+1$;
\UNTIL 
 $k=K$ or \begin{small}$\left| {{Q_1}\left( {{{\bf{L}}^k},{{\bf{X}}^k}} \right) - {Q_1}\left( {{{\bf{L}}^{k + 1}},{{\bf{X}}^{k + 1}}} \right)} \right| < \varepsilon $\end{small}
\ENSURE
Refined graph $\hat{\bf{L}}$, low-rank component $\hat{\bf{X}}$.
\end{algorithmic}
\end{algorithm}
\subsection{Optimization algorithm}
The formulation in (P1) establishes a biconvex optimization problem, that is, it is a convex problem with respect to ${\bf{L}}$ when ${\bf{X}}$ is fixed and vice versa.
We propose the GL-LRSS to solve the optimization problem
via alternating optimization.
At each step, one variable is optimized while keeping the
other variables constant.
The iterative procedure is given by
\begin{equation} \label{up}
\resizebox{0.8\hsize}{!}{$
\begin{aligned}
& 1. {\kern 1pt} {\kern 1pt}{\kern 1pt} {\kern 1pt}{\kern 1pt}{\kern 1pt}{\kern 1pt}{ G\left({\bf{X}},{\bf{Y}}\right) {\kern 1pt}{\kern 1pt}\triangleq {\kern 1pt}{\kern 1pt}\arg \mathop {\min }\limits_{\bf{L}} {\kern 1pt} {\kern 1pt}{\kern 1pt} {\kern 1pt}Q_1\left( {{\bf{L}},{\bf{X}}} \right)}, {\kern 1pt} {\kern 1pt}{\kern 1pt} {\kern 1pt}{\kern 1pt}{\kern 1pt}{\kern 1pt}{\kern 1pt} {\kern 1pt}{\kern 1pt} {\kern 1pt}{\kern 1pt}{\kern 1pt}{\kern 1pt}{\kern 1pt} {\kern 1pt}{\kern 1pt} {\kern 1pt}{\kern 1pt}{\kern 1pt}{\kern 1pt}{\kern 1pt} {\kern 1pt}{\kern 1pt} {\kern 1pt} {\kern 1pt}{\kern 1pt} {\kern 1pt}{\kern 1pt}{\kern 1pt}{\kern 1pt}{\kern 1pt}{\kern 1pt}{\kern 1pt}{\kern 1pt}{\kern 1pt}{\kern 1pt}{\kern 1pt}{\kern 1pt}{\kern 1pt}{\kern 1pt}\left({\mathcal{S}}_L\right)\\
&  {\kern 1pt} {\kern 1pt} {\kern 1pt} {\kern 1pt} {\kern 1pt} {\kern 1pt} {\kern 1pt} {\kern 1pt} {\kern 1pt} {\kern 1pt} {\kern 1pt} {\kern 1pt} {\kern 1pt} {\kern 1pt} {\kern 1pt} {\kern 1pt} {\kern 1pt} {\kern 1pt} {\kern 1pt} {\kern 1pt} {\kern 1pt} {\kern 1pt} {\kern 1pt} {\kern 1pt} {\kern 1pt} {\kern 1pt} {\kern 1pt} {\kern 1pt} {\kern 1pt} {\kern 1pt} {\kern 1pt} {\kern 1pt} {\kern 1pt} {\kern 1pt} {\kern 1pt}{\kern 1pt}{\kern 1pt}{\kern 1pt}{\kern 1pt}{\kern 1pt}{\kern 1pt}{\kern 1pt}{\kern 1pt}{\kern 1pt} {\rm{s}}{\rm{.t}}{\rm{.}}{\kern 1pt} {\kern 1pt} {\kern 1pt} {\kern 1pt} {\kern 1pt} {\kern 1pt} {\kern 1pt} {\kern 1pt}{\bf{L}} \in {{\mathcal {L}}}, {\kern 1pt} {\kern 1pt} {\kern 1pt} {\kern 1pt} {\rm{tr}}\left( {\bf{L}} \right) = N.\\
& 2. {\kern 1pt} {\kern 1pt}{\kern 1pt} {\kern 1pt}{\kern 1pt}{\kern 1pt}{\kern 1pt}C\left({\bf{L}},{\bf{Y}}\right) {\kern 1pt}{\kern 1pt}\triangleq {\kern 1pt}{\kern 1pt} \mathop {\arg \min }\limits_{{\kern 1pt}{\kern 1pt}{\bf{X}}\in {{\mathbb {R}}^{N \times M}} } Q_1\left( {{\bf{ L}},{\bf{X}}} \right).{\kern 1pt}  {\kern 1pt}{\kern 1pt} {\kern 1pt}{\kern 1pt}{\kern 1pt}{\kern 1pt}{\kern 1pt} {\kern 1pt}{\kern 1pt} {\kern 1pt} {\kern 1pt}{\kern 1pt} {\kern 1pt}{\kern 1pt}{\kern 1pt} {\kern 1pt}{\kern 1pt} {\kern 1pt}{\kern 1pt}{\kern 1pt}{\kern 1pt} {\kern 1pt}{\kern 1pt}{\kern 1pt}{\kern 1pt} {\kern 1pt}{\kern 1pt}{\kern 1pt}{\kern 1pt} {\kern 1pt}{\kern 1pt}{\kern 1pt}{\kern 1pt} {\kern 1pt}{\kern 1pt}{\kern 1pt}{\kern 1pt}{\kern 1pt}{\kern 1pt}\left({\mathcal{S}}_{X}\right)\nonumber
\end{aligned} $}
\end{equation}

By iteratively refining the graph from the low-rank
representation and estimating the low-rank
component using the learned graph,
we obtain the final solution of (P1) by alternating
minimization.
The detailed 
procedure for solving (P1) is shown as follows.

\subsubsection{Graph refinement in subproblem $\left({\mathcal{S}}_{L}\right)$}
Notice that $\left({\mathcal{S}}_{L}\right)$ is
a strictly convex problem under convex
constraints, because the Hessian matrix
of the objective function, $2\beta {\bf{I}}_N$, is positive definite.
To solve this constrained convex problem, we apply the ADMM method \cite{ref39}.
Specifically, 
we reformulate problem (P1) with respect to the graph Laplacian $\bf{L}$
as follows:
\begin{eqnarray} \label{updateLL}
\resizebox{0.55\hsize}{!}{$
\begin{aligned}
&{  \mathop {\min }\limits_{\bf{L}\in {{\mathcal {L}}^*}} \alpha {\kern 1pt} {\kern 1pt} {\rm{tr }}\left( {{{\mathcal{D}\left( {\bf{X}} \right)}^T}{\bf{L}}\mathcal{D}\left( {\bf{X}} \right)} \right) + \beta \left\| {\bf{L}} \right\|_F^2},\\
&{{\kern 1pt}   {\kern 1pt} {\kern 1pt} {\kern 1pt} {\kern 1pt} {\kern 1pt} {\kern 1pt} {\kern 1pt} {\kern 1pt} {\kern 1pt} {\kern 1pt} {\kern 1pt} {\kern 1pt} {\kern 1pt} {\kern 1pt} {\kern 1pt} {\kern 1pt} {\kern 1pt} {\kern 1pt} {\kern 1pt} {\kern 1pt} {\kern 1pt} {\kern 1pt} {\kern 1pt} {\kern 1pt} {\kern 1pt} {\kern 1pt} {\kern 1pt} {\kern 1pt} {\kern 1pt} {\rm{s}}{\rm{.t}}{\rm{.}}{\kern 1pt} {\kern 1pt} {\kern 1pt} {\kern 1pt} {\kern 1pt} {\kern 1pt} {\kern 1pt} {\kern 1pt} {\bf{L}}-{\bf{Z}}=0}, {\kern 1pt}{\kern 1pt}{\kern 1pt}{\kern 1pt}{\bf{Z}}\in {{\mathcal {L}}^*},
\end{aligned} $}
\end{eqnarray}
where ${\bf{Z}}$ is an auxiliary variable matrix and \begin{small}${{\mathcal {L}}^*}$\end{small}is expressed as
\begin{eqnarray}\label{Laplacian1}
\resizebox{0.9\hsize}{!}{$
\begin{aligned}
{{\mathcal {L}}^*}=\left\{ {{\bf{L}}| {\bf{L}}\succeq 0, {L_{ji}}={L_{ij}} \le 0{\kern 1pt} ,i \ne j,{\kern 1pt} {\kern 1pt}  {\rm{and}}{\kern 1pt} {\kern 1pt}  {\bf{L}}\cdot{\bf{1}} = {\bf{0}},  {\rm{tr}}\left( {\bf{L}} \right) = N} \right\}.
\end{aligned} $}
\end{eqnarray}
Therefore, the augmented Lagrangian of (\ref{updateLL}) is given by
\begin{eqnarray} \label{updateLa}
\resizebox{0.69\hsize}{!}{$
\begin{aligned}
{\mathcal {L}}_{\rho}\left({\bf{L}}, {\bf{Z}}, {\bf{\Xi }} \right)
&=  \alpha {\kern 1pt} {\kern 1pt} {\rm{tr }}\left( {{{\mathcal{D}\left( {\bf{X}} \right)}^T}{\bf{L}}\mathcal{D}\left( {\bf{X}} \right)} \right) + \beta \left\| {\bf{L}} \right\|_F^2\\
& + \left\langle {{\bf{\Xi }},{\bf{Z}} - {\bf{L}}} \right\rangle
+\frac{\rho }{2}\left\| {{\bf{Z}} - {\bf{L}}} \right\|_F^2,
\end{aligned} $}
\end{eqnarray}
where ${\bf{\Xi }}$ is the Lagrange multiplier, $\left\langle { \cdot , \cdot } \right\rangle $ denotes the inner
product of matrices, and $\rho>0$ is a prescribed penalty parameter.
We use the following formulas to update \begin{small}${\bf{L}}$, ${\bf{Z}}$, and ${\bf{\Xi }}$\end{small} to find a saddle
point of (\ref{updateLa})
\begin{eqnarray} \label{adm21}
\resizebox{0.55\hsize}{!}{$
\begin{aligned}
{{\bf{L}}^{k + 1}} = \mathop {\arg \min }\limits_{\bf{L}\in {{\mathcal {L}}^*}} {\kern 1pt}{\kern 1pt}{\kern 1pt}{\kern 1pt} {{\mathcal {L}}_\rho }\left( {{\bf{L}},{{\bf{Z}}^k},{{\bf{\Xi }}^k}} \right),
\end{aligned} $}
\end{eqnarray}
\vspace{-0.72cm}
\begin{eqnarray} \label{adm22}
\resizebox{0.56\hsize}{!}{$
\begin{aligned}
{{\bf{Z}}^{k + 1}} = \mathop {\arg \min }\limits_{\bf{Z}\in {{\mathcal {L}}^*}} {\kern 1pt}{\kern 1pt}{\kern 1pt}{\kern 1pt}{{\mathcal {L}}_\rho }\left( {{{\bf{L}}^{k + 1}},{\bf{Z}},{{\bf{\Xi }}^k}} \right),
\end{aligned} $}
\end{eqnarray}
\vspace{-0.72cm}
\begin{eqnarray} \label{adm23}
\resizebox{0.53\hsize}{!}{$
\begin{aligned}
{{\bf{\Xi }}^{k + 1}} = {{\bf{\Xi }}^k} + \rho \left( {{{\bf{Z}}^{k + 1}} - {{\bf{L}}^{k + 1}}} \right).
\end{aligned} $}
\end{eqnarray}

Setting the derivatives of Eqs. (\ref{adm21}) and (\ref{adm22}) with respect to $\bf{L}$ and $\bf{Z}$, respectively, equal to zero,
we obtain the following solutions:
\begin{eqnarray} \label{admmsolution}
\resizebox{0.93\hsize}{!}{$
\begin{aligned}
{{\bf{L}}^{k + 1}} = \frac{{\rho {{\bf{Z}}^k} + {{\bf{\Xi }}^k} - \alpha {{\mathcal{D}\left( {\bf{X}} \right)}}{{\mathcal{D}\left( {\bf{X}} \right)}^T}}}{{2\beta  + \rho }},{\kern 1pt}{\kern 1pt}{\kern 1pt}{\kern 1pt}{\kern 1pt}
{{\bf{Z}}^{k + 1}} = P_{{\mathcal {L}}^*} {\left( {{{\bf{L}}^{k + 1}} - \frac{1}{\rho }{{\bf{\Xi }}^k}} \right)},
\end{aligned} $}
\end{eqnarray}
where \begin{small}$P_{{\mathcal {L}}^*}(\cdot)$\end{small} denotes the Euclidean
projection onto set ${{\mathcal {L}}^*}$.

\subsubsection{Low-rank component estimation in subproblem $\left({\mathcal{S}}_{X}\right)$}
The first two terms of (P1) are differentiable, and the third term of ${\bf{X}}$ is proximable. We apply ADMM to solve problem $\left({\mathcal{S}}_{X}\right)$.
First, we provide an equivalent formulation of (P1)
with respect to $\bf{X}$: 
\begin{eqnarray} \label{updateLX}
\resizebox{0.85\hsize}{!}{$
\begin{aligned}
\mathop {\min }\limits_{{\bf{X}},{\bf{P}}\in {{\mathbb {R}}^{N \times M}}} {\kern 1pt} {\kern 1pt} {\kern 1pt} \left\| {{\cal D}\left( {{\bf{X}} - {\bf{Y}}} \right)} \right\|_F^2 +
& \alpha {\kern 1pt}{\kern 1pt}{\rm{tr}}\left( {{\cal D}{{\left( {\bf{X}}\right)}^T}{\bf{L}}{\cal D}\left( {\bf{X}} \right)} \right) + \gamma {\left\| {\bf{P}}\right\|_*},\\
&{\rm{s}}{\rm{.t}}{\rm{.}} {\kern 1pt}{\kern 1pt}{\kern 1pt}{\kern 1pt}{\bf{X}}={\bf{P}}.
\end{aligned} $}
\end{eqnarray}

The objective function is split into two parts
by introducing the linear equality constraint.
Then, the augmented Lagrangian of (\ref{updateLX}) is
given by
\begin{eqnarray} \label{Lagrangian-X}
\resizebox{0.83\hsize}{!}{$
\begin{aligned}
{\mathcal {L}}_{\rho}\left({\bf{X}},{\bf{P}},{\bf{Q}}\right)
&=\left\| {{\cal D}\left( {{\bf{X}} - {\bf{Y}}} \right)} \right\|_F^2
+\alpha {\kern 1pt}{\kern 1pt}{\rm{tr}}\left( {{\cal D}{{\left( {\bf{X}}\right)}^T}{\bf{L}}{\cal D}\left( {\bf{X}} \right)} \right)\\
&+\gamma {\left\| {\bf{P}}\right\|_*}+\left\langle {{\bf{Q}},{\bf{X}} - {\bf{P}}} \right\rangle+\frac{\rho }{2}\left\| {{\bf{X}} - {\bf{P}}} \right\|_F^2,
\end{aligned} $}
\end{eqnarray}
where ${\bf{Q}}$ is the Lagrange multiplier. Based on the augmented Lagrangian
in (\ref{Lagrangian-X}),
the solution is obtained iteratively, as follows:
\begin{eqnarray} \label{admm21}
\resizebox{0.55\hsize}{!}{$
\begin{aligned}
{{\bf{X}}^{k + 1}} = \mathop {\arg \min }\limits_{{\bf{X}}\in {{\mathbb {R}}^{N \times M}}} {\kern 1pt}{\kern 1pt}{\kern 1pt}{\kern 1pt} {{\mathcal {L}}_\rho }\left( {{\bf{X}},{{\bf{P}}^k},{{\bf{Q}}^k}} \right),
\end{aligned} $}
\end{eqnarray}
\vspace{-0.72cm}
\begin{eqnarray} \label{admm22}
\resizebox{0.56\hsize}{!}{$
\begin{aligned}
{{\bf{P}}^{k + 1}} = \mathop {\arg \min }\limits_{{\bf{P}}\in {{\mathbb {R}}^{N \times M}}} {\kern 1pt}{\kern 1pt}{\kern 1pt}{\kern 1pt}{{\mathcal {L}}_\rho }\left( {{{\bf{X}}^{k + 1}},{\bf{P}},{{\bf{Q}}^k}} \right),
\end{aligned} $}
\end{eqnarray}
\vspace{-0.72cm}
\begin{eqnarray} \label{admm23}
\resizebox{0.53\hsize}{!}{$
\begin{aligned}
{{\bf{Q}}^{k + 1}} = {{\bf{Q}}^k} + \rho \left( {{{\bf{X}}^{k + 1}} - {{\bf{P}}^{k + 1}}} \right).
\end{aligned} $}
\end{eqnarray}
According to (\ref{Lagrangian-X}), the subproblem of (\ref{admm21}) can be
rewritten as
\begin{eqnarray} \label{admm21rewitten}
\resizebox{0.83\hsize}{!}{$
\begin{aligned}
{{\bf{X}}^{k + 1}}  = \mathop{\arg \min}\limits_{{\bf{X}}\in {{\mathbb {R}}^{N \times M}}} &{\kern 1pt} {\kern 1pt} {\kern 1pt} \left\| {{\cal D}\left( {{\bf{X}} - {\bf{Y}}} \right)} \right\|_F^2 +
\alpha {\kern 1pt}{\kern 1pt}{\rm{tr}}\left( {{\cal D}{{\left( {\bf{X}}\right)}^T}{\bf{L}}{\cal D}\left( {\bf{X}} \right)} \right)\\
&  + \frac{\rho }{2}\left\| {{\bf{X}} - {{\bf{P}}^k} + {{{{\bf{Q}}^k}} \mathord{\left/{\vphantom {{{Q^k}} \rho }} \right.\kern-\nulldelimiterspace} \rho }} \right\|_F^2.
\end{aligned} $}
\end{eqnarray}

The expression in (\ref{admm21rewitten})
is a differentiable convex optimization problem
that admits a closed-form solution. 
Using \begin{small}${\rm{vec}}\left({\bf{A}}{\bf{X}}{\bf{B}}\right)=
\left({\bf{B}}^T \otimes {\bf{A}} \right){\rm{vec}}\left({\bf{X}}\right)$\end{small}, 
the optimal update of ${{\bf{X}}^{k + 1}}$ is
given by
\begin{eqnarray}\label{Uniquesolution}
\resizebox{0.93\hsize}{!}{$
\begin{aligned}
{\rm{vec}}\left( {{{\bf{X}}^{k + 1}}} \right) = {\left( {2{{\bf{T}}_d}{{\bf{T}}_d}^T + 2\alpha \mathord{\buildrel{\lower3pt\hbox{$\scriptscriptstyle\smile$}}
\over {\bf{L}}}  + \rho {{\bf{I}}_{MN}}} \right)^{ - 1}}\left( {{\rm{vec}}\left( {\rho {{\bf{P}}^k} - {{\bf{Q}}^k}} \right) + \mathord{\buildrel{\lower3pt\hbox{$\scriptscriptstyle\smile$}}
\over {\bf{Y}}} } \right),
\end{aligned} $}
\end{eqnarray}
where the operator \begin{small}${\rm{vec}}\left(  \cdot  \right)$\end{small} stacks the columns of an $M \times N$ matrix into a vector of dimension $MN$, and parameters $\mathord{\buildrel{\lower3pt\hbox{$\scriptscriptstyle\smile$}}
\over {\bf{L}}} $ and $\mathord{\buildrel{\lower3pt\hbox{$\scriptscriptstyle\smile$}}
\over {\bf{Y}}}$ are represented by
\begin{small}${{\bf{T}}_d}\left( {{{\bf{I}}_M} \otimes {\bf{L}}} \right){{\bf{T}}_d}^T$\end{small} and \begin{small}$ 2{{\bf{T}}_d}{{\bf{T}}_d}^T{\rm{vec}}\left( {\bf{Y}} \right)$\end{small}, respectively, with ${{\bf{T}}_d}$ expressed as
\begin{eqnarray}\label{Difference}
\resizebox{0.6\hsize}{!}{$
\begin{aligned}
{{\bf{T}}_d} = \left[ {\begin{array}{*{20}{c}}
{\bf{I}}_N&-{\textit{R}}&{}&{}&{}\\
{}&{\bf{I}}_N&-{\textit{R}}&{}&{}\\
{}&{}&{\bf{I}}_N& \ddots &{}\\
{}&{}&{}& \ddots &-{\textit{R}}\\
{}&{}&{}&{}&{\bf{I}}_N
\end{array}} \right]_{NM \times NM.}
\end{aligned} $}
\end{eqnarray}

A detailed derivation of (\ref{Uniquesolution}) is outlined
in Appendix A.
\begin{algorithm}[t]
\renewcommand{\algorithmicrequire}{\textbf{Input:}}
\renewcommand\algorithmicensure {\textbf{Output:} }
\caption{\textbf{: Method for solving subproblem (\ref{admm21rewitten})}}
\begin{algorithmic}[1]
\REQUIRE
\begin{small}${\bf{Y}}$, ${\textit{R}}$, ${\bf{B}}$, ${\bf{L}}^{k+1}$, ${{\bf{P}}^k}$, ${{\bf{Q }}^k}$, $\alpha$, $\rho$, $K$,\end{small} error tolerance $\delta$.
\STATE Initialization: \begin{small}${\bf{X }}_0 = {\bf{0}}$; $\Delta {{\bf{X }} _0}=- \nabla {f_X}\left( {{{\bf{X }} _{0}}} \right)$\end{small};
\REPEAT
\STATE 1) Dynamic stepsize selection:
\STATE  \quad $\mu  =  - \frac{{{\rm{tr}}\left\{ {{{\left( {\Delta {{\bf{X }} _m}} \right)}^T}\nabla {f_X}\left( {{{\bf{X }} _m}} \right)} \right\}}}{{{\rm{tr}}\left\{ {{{\left( {\Delta {{\bf{X }} _m}} \right)}^T}\left[ {\nabla {f_X}\left( {\Delta {{\bf{X }} _m}} \right) +{\bf{\psi }}} \right]} \right\}}}$,\\
\quad with \begin{small}$ {\bf{\psi }}=2\mathcal{D}\left( {\bf{Y}} \right) - 2{\bf{R}}\mathcal{D}\left( {\bf{Y}} \right){{\bf{B}}^T}+\rho{{\bf{P}}^k}-{{\bf{Q}}^k};$\end{small}
\STATE 2) Conjugate direction update:
\STATE \quad \, \begin{small}${{\bf{X }} _{m + 1}} = {{\bf{X }} _m} + \mu \Delta {{\bf{X }} _m}$\end{small};
\STATE \quad \, \begin{small}$\Delta {{\bf{X }} _{m + 1}} =  - \nabla {f_X}\left( {{{\bf{X }} _{m + 1}}} \right) + \theta \Delta {{\bf{X }} _m}$\end{small};
\STATE $m=m+1$;
\UNTIL $m$ reaches maximum number of iterations
\ENSURE
Recovered  ${\bf{X }}$.
\end{algorithmic}
\end{algorithm}
The solution in (\ref{Uniquesolution}) requires the calculation of the inverse of an
\begin{small}$MN \times MN$\end{small} matrix. Thus, for a large number of vertices or time instants, this procedure is expected to be time-consuming.
Instead, 
the conjugate gradient method \cite{ref40} can be adopted to efficiently obtain a solution. Let \begin{small}${f_X}\left(  \cdot  \right)$\end{small} represent
the objective function in (\ref{admm21rewitten}).
The algorithm mainly updates the stepsize and searching direction in each iteration.
Denoting the search direction of the
$m$th iteration as \begin{small}$\Delta {{\bf{X}} ^m}$\end{small},
the optimal stepsize ${\mu }$ at the $m$th step
can be obtained
by an exact line search \cite{ref41} given as 
\begin{small}$\mathop {\min }\limits_{\mu } f_X \left({{\bf{X}} ^m} + \mu \Delta {{\bf{X}} ^m}\right)$\end{small}.
By setting the derivative of ${f_X}$ with respect to ${\mu}$ equal to zero, we obtain
\begin{eqnarray} \label{resu}
\resizebox{0.6\hsize}{!}{$
\begin{aligned}
{\rm{tr}}\left[ {{{\left( {\Delta {{\bf{X }}^m}} \right)}^T}{\nabla} {f_X}\left( {{{\bf{X }}^m} + \mu \Delta {{\bf{X }}^m}} \right)} \right]=0,
\end{aligned} $}\nonumber
\end{eqnarray}
with the gradient of ${f_X}$ calculated as
\begin{eqnarray} \label{grad}
\resizebox{0.85\hsize}{!}{$
\begin{aligned}
\nabla {f_X} =
& 2{\cal D}\left( {{\bf{X }} - {\bf{Y }}} \right)  - 2{\bf{R }}{\cal D}\left( {{\bf{X }} - {\bf{Y }}} \right){{\bf{B }}^T}+
\rho \left( {{\bf{X }} - {{\bf{P }}^k}} \right) + {{\bf{Q }}^k}\\
& + 2\alpha \left( {{\bf{L }}{\cal D}\left( {\bf{X }} \right) - {\bf{R }}{\bf{L }}{\bf{X }}{{\bf{B}}^T} + {\bf{L}}{\bf{X }}{\bf{B }}{{\bf{B }}^T}} \right).
\end{aligned} $}
\end{eqnarray}
Therefore, we can determine optimal stepsize $\mu$ and
update the searching direction by introducing the Fletcher-Reeves
parameter given by \begin{small}$\theta={{\left\| {\nabla {f_X}\left( {{{\bf{X }} ^{m + 1}}} \right)} \right\|_F^2} \mathord{\left/
 {\vphantom {{\left\| {\nabla {f_X}\left( {{{\bf{X }} ^{m + 1}}} \right)} \right\|_F^2} {\left\| {\nabla {f_X}\left( {{{\bf{X }} ^m}} \right)} \right\|_F^2}}} \right.
 \kern-\nulldelimiterspace} {\left\| {\nabla {f_X}\left( {{{\bf{X }} ^m}} \right)} \right\|_F^2}}$\end{small}. 
The corresponding iterative optimization is detailed in
\emph{Algorithm}2.

Similar to the subproblem of (\ref{admm21}), by adding a
constant term \begin{small}$\frac{1}{2}{\rm{tr}}\left( {\frac{{{{\left( {{{\bf{Q}}^k}} \right)}^T}{{\bf{Q}}^k}}}{{{\rho ^2}}}} \right)$\end{small},
the subproblem of (\ref{admm22}) is
equivalent to the following optimization problem
\begin{eqnarray} \label{admm22rewitten}
\resizebox{0.75\hsize}{!}{$
\begin{aligned}
{{\bf{P}}^{k + 1}}  = \mathop{\arg \min} \limits_{{\bf{P}}\in {{\mathbb {R}}^{N \times M}}}&{\kern 1pt} {\kern 1pt} {\kern 1pt} \frac{1}{2}\left\| {{\bf{P}}- {{\bf{X}}^{k + 1}} - \frac{{{{\bf{Q}}^k}}}{\rho }} \right\|_F^2 + \frac{\gamma }{\rho }{\left\| {\bf{P}} \right\|_*},
\end{aligned} $}
\end{eqnarray}
which has closed-form solution
\begin{eqnarray} \label{admm22solution}
\resizebox{0.45\hsize}{!}{$
\begin{aligned}
{{\bf{P}}^{k + 1}}  = {\Gamma _{{\gamma  \mathord{\left/
 {\vphantom {\gamma  \rho }} \right.
 \kern-\nulldelimiterspace} \rho }}}\left( {{{\bf{X}}^{k + 1}} + \frac{{{{\bf{Q}}^k}}}{\rho }} \right),
\end{aligned} $}
\end{eqnarray}
where $\Gamma $ is singular value thresholding operator
\cite{ref42}
that is the proximity operator associated with the
nuclear norm. For each $\tau  \ge 0$, the $\Gamma $ is defined
as follows:
\begin{eqnarray} \label{SVT}
\resizebox{0.4\hsize}{!}{$
\begin{aligned}
{\Gamma _\tau }\left( {\bf{X}} \right) = {\bf{U}}{\Theta _\tau }\left( {\bf{\Sigma }} \right){{\bf{V}}^T},
\end{aligned} $}
\end{eqnarray}
where ${\bf{U}}$, ${\bf{V}}$ and ${\bf{\Sigma }}$ are obtained
from the singular value decomposition (SVD) of ${\bf{X}}$,
that is, ${\bf{X}} = {\bf{U}}{\bf{\Sigma}} {{\bf{V}}^T}$, with
${\sigma _i}$ denoting the $i$th singular value and
\begin{eqnarray} \label{softsholding}
\resizebox{0.6\hsize}{!}{$
\begin{aligned}
{\Theta _\tau }\left( {{\sigma _i}} \right) = {\rm{sign}}\left( {{\sigma _i}} \right)\max \left( {\left| {{\sigma _i}} \right| - \tau ,0} \right).
\end{aligned} $}
\end{eqnarray}
The operator (\ref{softsholding}) applies a soft-thresholding rule to the singular values of ${\bf{X}}$, effectively shrinking
these towards zero.

The overall graph learning framework is presented in \emph{Algorithm} 1.
It should be noted that the optimization problem in (P1) is not jointly convex in $\bf{L}$ and $\bf{X}$, the solution therefore corresponds to a local optimum rather than a global optimum. Besides,
our empirical results suggest
that after only eight iterations or less, the objective $Q_1\left( {{\bf{L}},{\bf{X}}} \right)$ does not change more than the predefined threshold.
\vspace{-0.3cm}
\subsection{Complexity analysis}
We next provide a brief complexity analysis of the proposed GL-LRSS.
For problem $\left({\mathcal{S}}_{L}\right)$,
the computation
is dominated by the update of ${\bf{L}}$ in (\ref{admmsolution}), 
and the update is in turn dominated by \begin{small}${{\mathcal{D}\left( {\bf{X}} \right)}}{{\mathcal{D}\left( {\bf{X}} \right)}^T}$\end{small}, where
the matrix-matrix
product costs \begin{small}$\mathcal{O}(N^2M+M^2N+N^3)$\end{small}.
For problem $\left({\mathcal{S}}_{X}\right)$, there are
two main steps having the highest computation burden.
For the first step of updating ${\bf{X}}^k$,
we utilize the conjugate gradient method instead of
the calculation in (\ref{Uniquesolution}).
In \emph{Algorithm} 2, the computation
is dominated by the gradient calculation according to
(\ref{grad}), which is mainly determined
by the matrix-matrix product, that is, ${\bf{R}}{\bf{L }}{\bf{X }}{{\bf{B}}^T}$,
which costs \begin{small}$\mathcal{O}(N^2M+M^2N+N^3)$\end{small}
flops. 
When updating ${\bf{P}}^k$ in the second
step of (\ref{admm22}), the computation of $\Gamma $
dominates the computation consumption.
The SVD of ${\bf{X}}$ takes the computational cost of 
\begin{small}$\mathcal{O}(\min ( {{M^2}N,{N^2}M} ))$\end{small} \cite{ref43}.
The last step of updating ${\bf{\Xi }}$ and ${\bf{Q}}$ 
involves the product of scalars and matrices with a cost of
\begin{small}$\mathcal{O}(MN)$\end{small}.
Overall, the
GL-LRSS is dominated by the updates of ${\bf{X}}$ in (\ref{admm21})
and ${\bf{L}}$ in (\ref{admmsolution}).

\begin{figure*}[htb]
\begin{minipage}[b]{0.18\linewidth}
  \centering
  \centerline{\includegraphics[width=3.3cm]{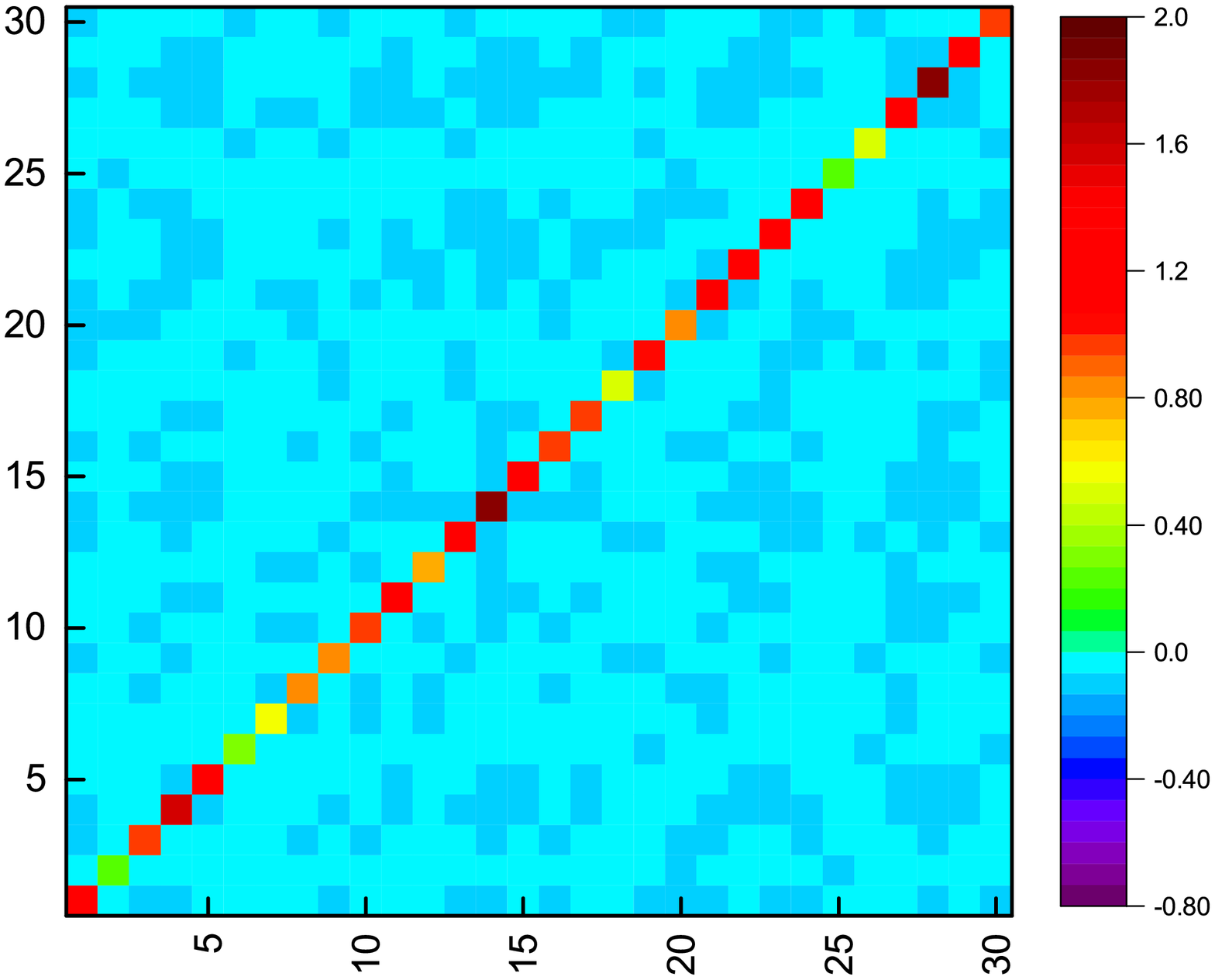}}
  \centerline{ Groundtruth}
\end{minipage}
\hfill
\begin{minipage}[b]{0.18\linewidth}
  \centering
  \centerline{\includegraphics[width=3.3cm]{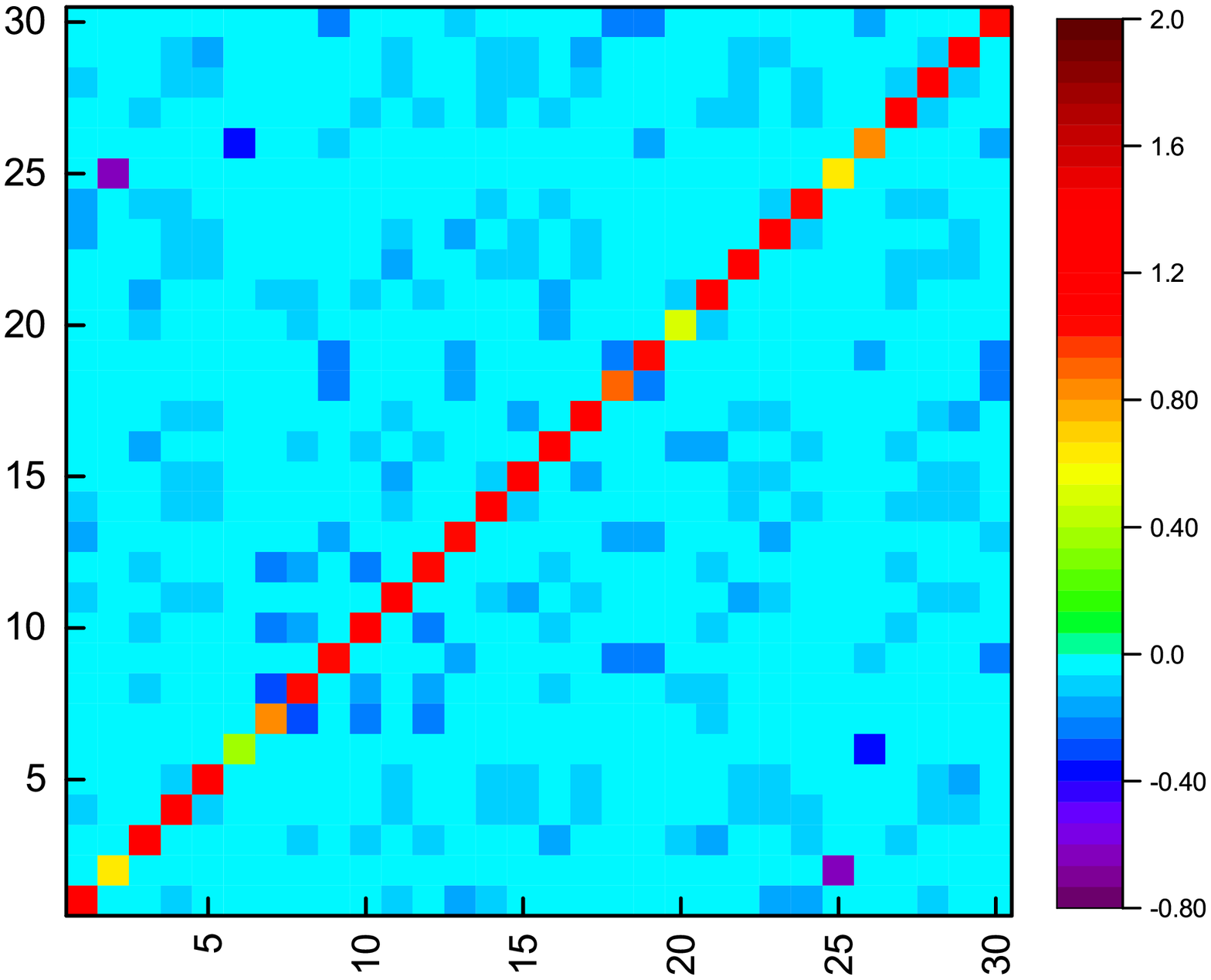}}
  \centerline{ (a) GL-LRSS}
\end{minipage}
\hfill
\begin{minipage}[b]{0.18\linewidth}
  \centering
  \centerline{\includegraphics[width=3.3cm]{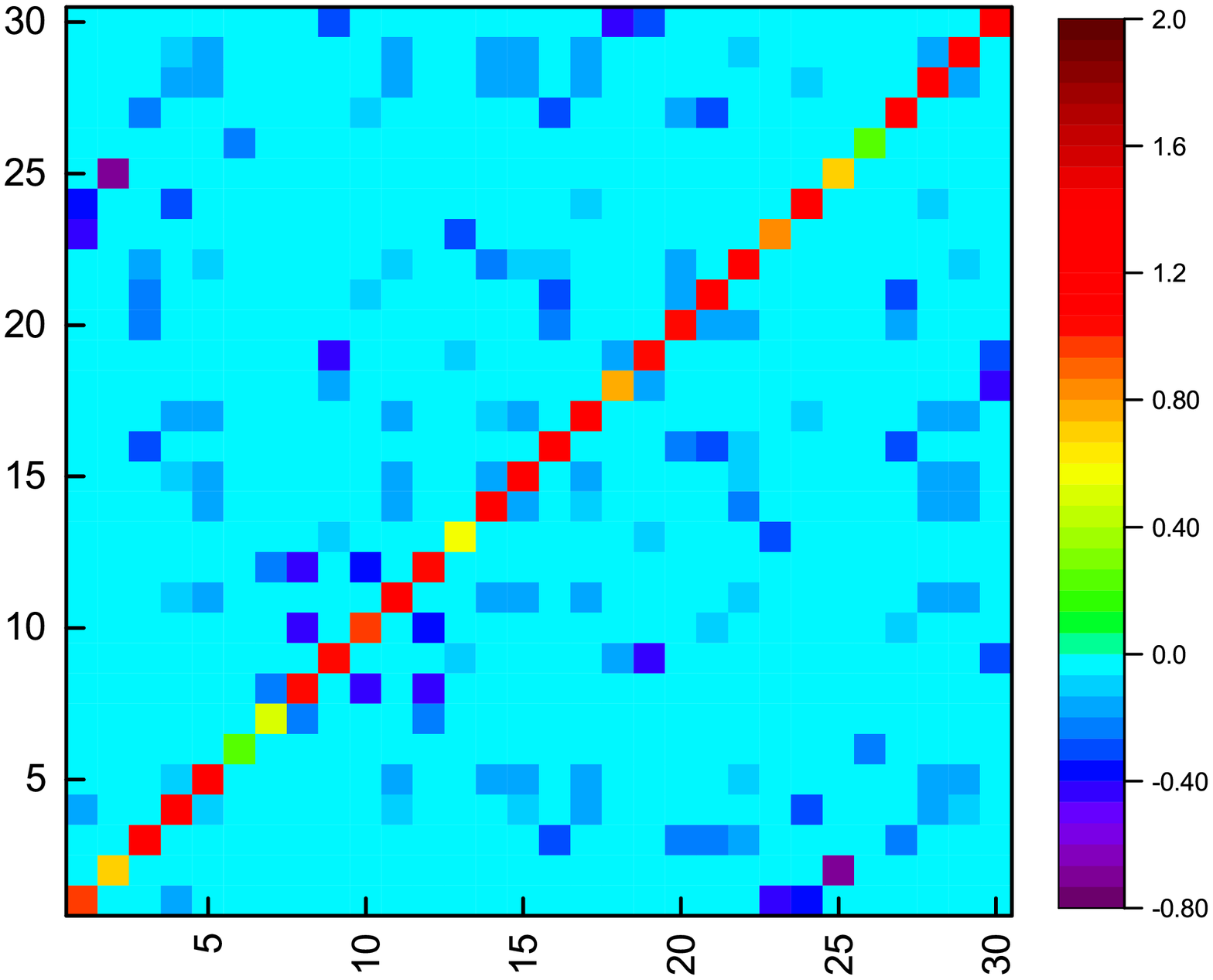}}
  \centerline{ (b) GL-Sigrep}
\end{minipage}
\hfill
\begin{minipage}[b]{0.18\linewidth}
  \centering
  \centerline{\includegraphics[width=3.3cm]{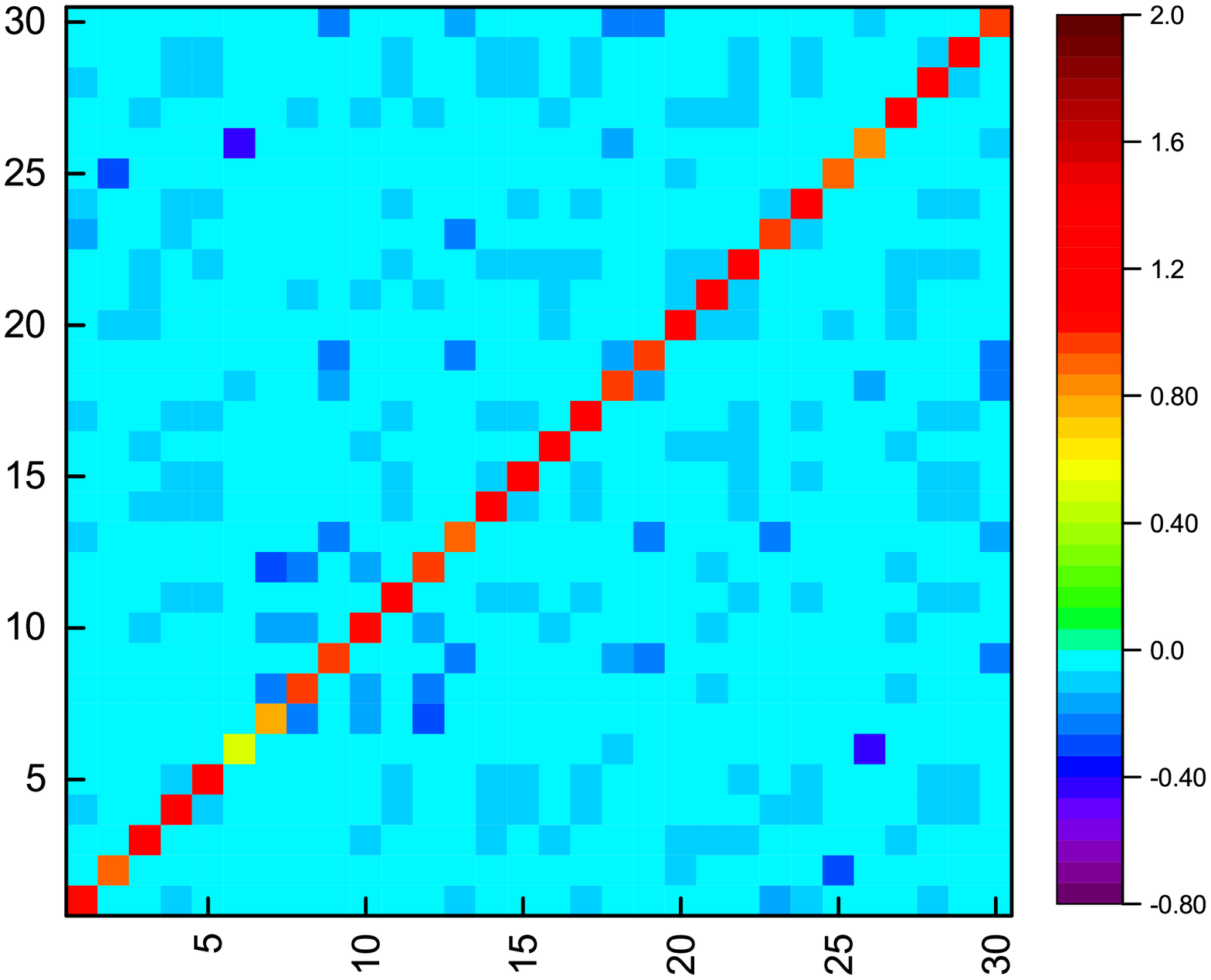}}
  \centerline{ (c) LGE}
\end{minipage}
\hfill
\begin{minipage}[b]{0.18\linewidth}
  \centering
  \centerline{\includegraphics[width=3.3cm]{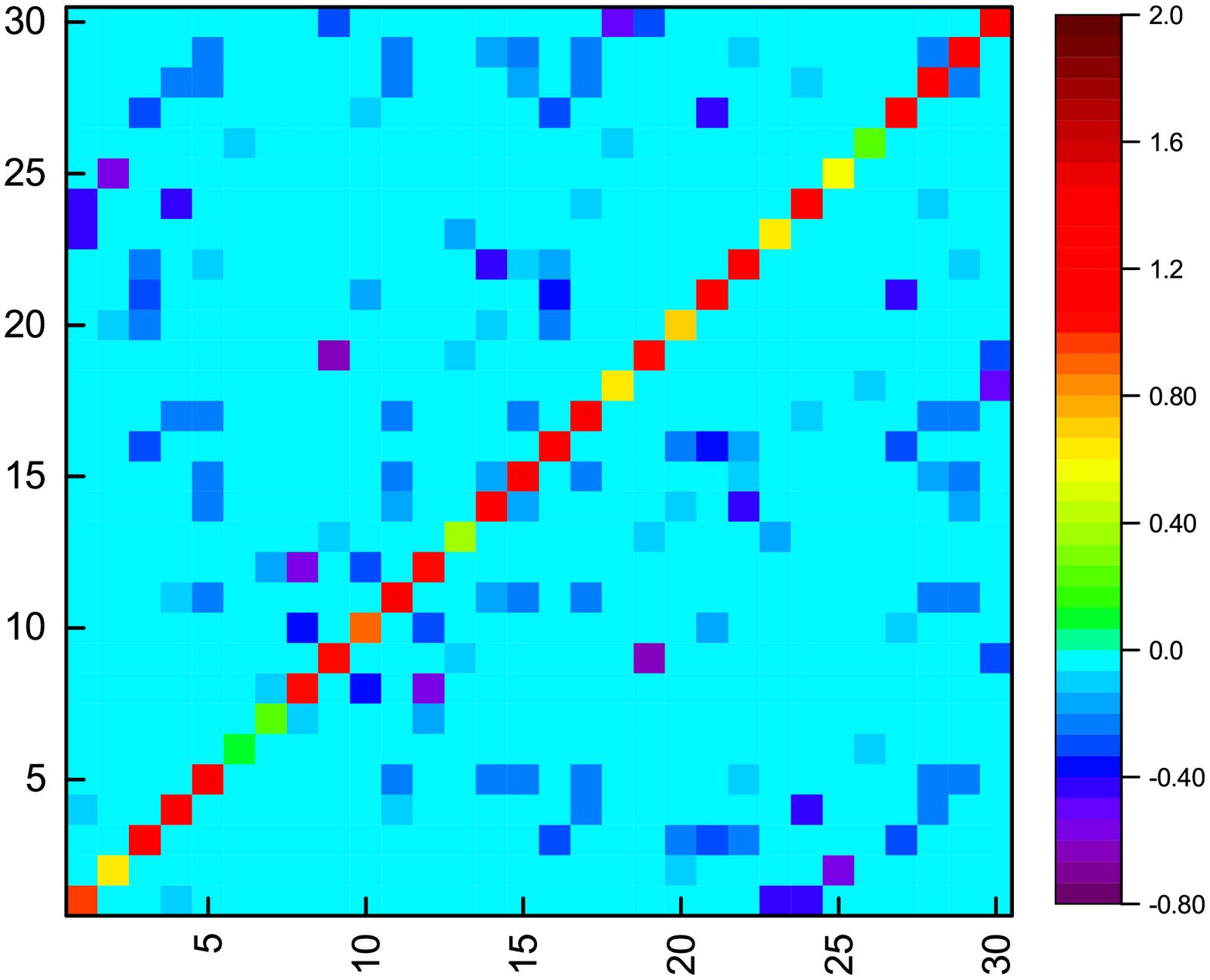}}
  \centerline{ (d) GL-logdet}
\end{minipage}
\vspace{0.1cm}
\vfill
\begin{minipage}[b]{0.18\linewidth}
  \centering
  \centerline{\includegraphics[width=3.3cm]{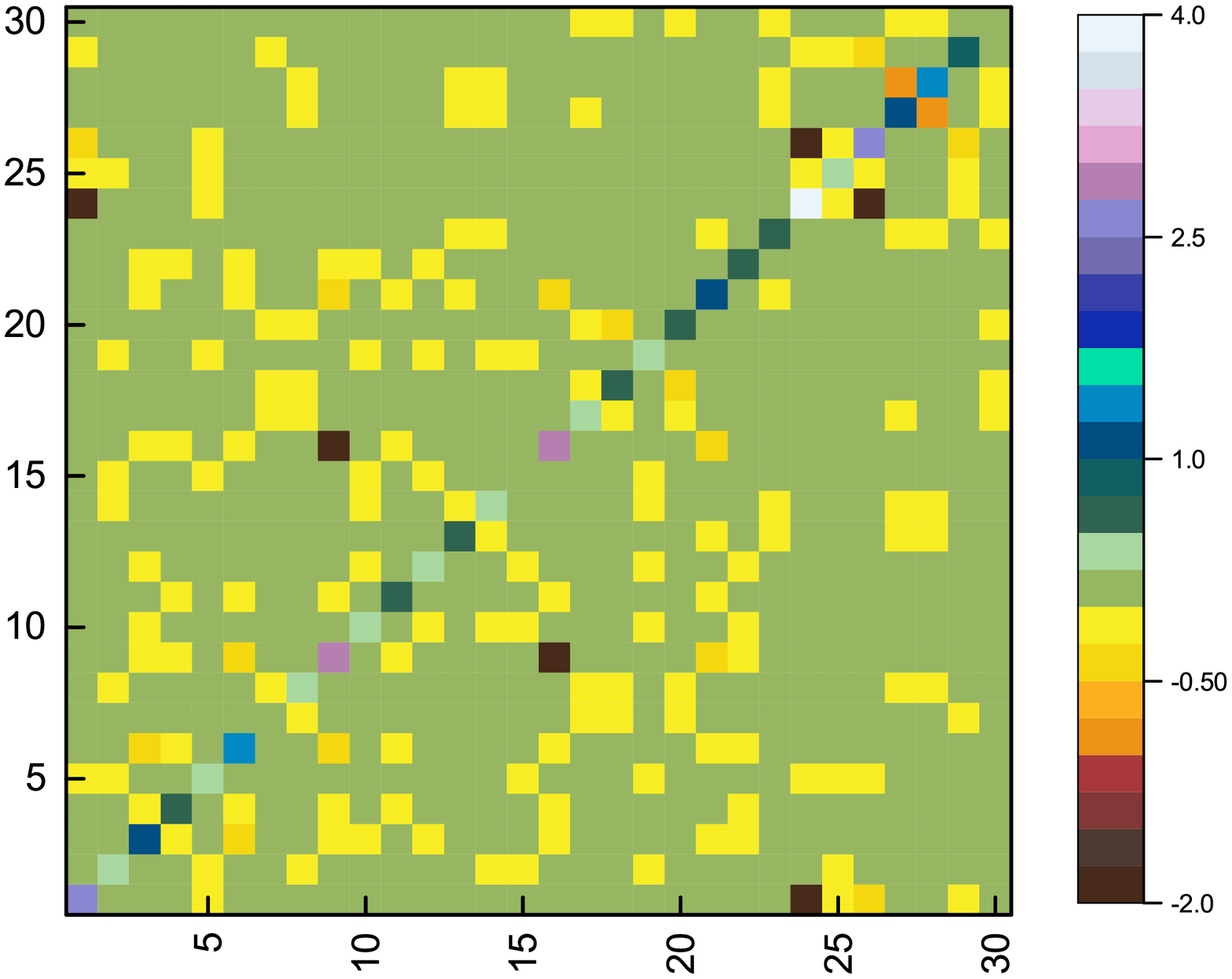}}
  \centerline{ Groundtruth}
\end{minipage}
\hfill
\begin{minipage}[b]{0.18\linewidth}
  \centering
  \centerline{\includegraphics[width=3.3cm]{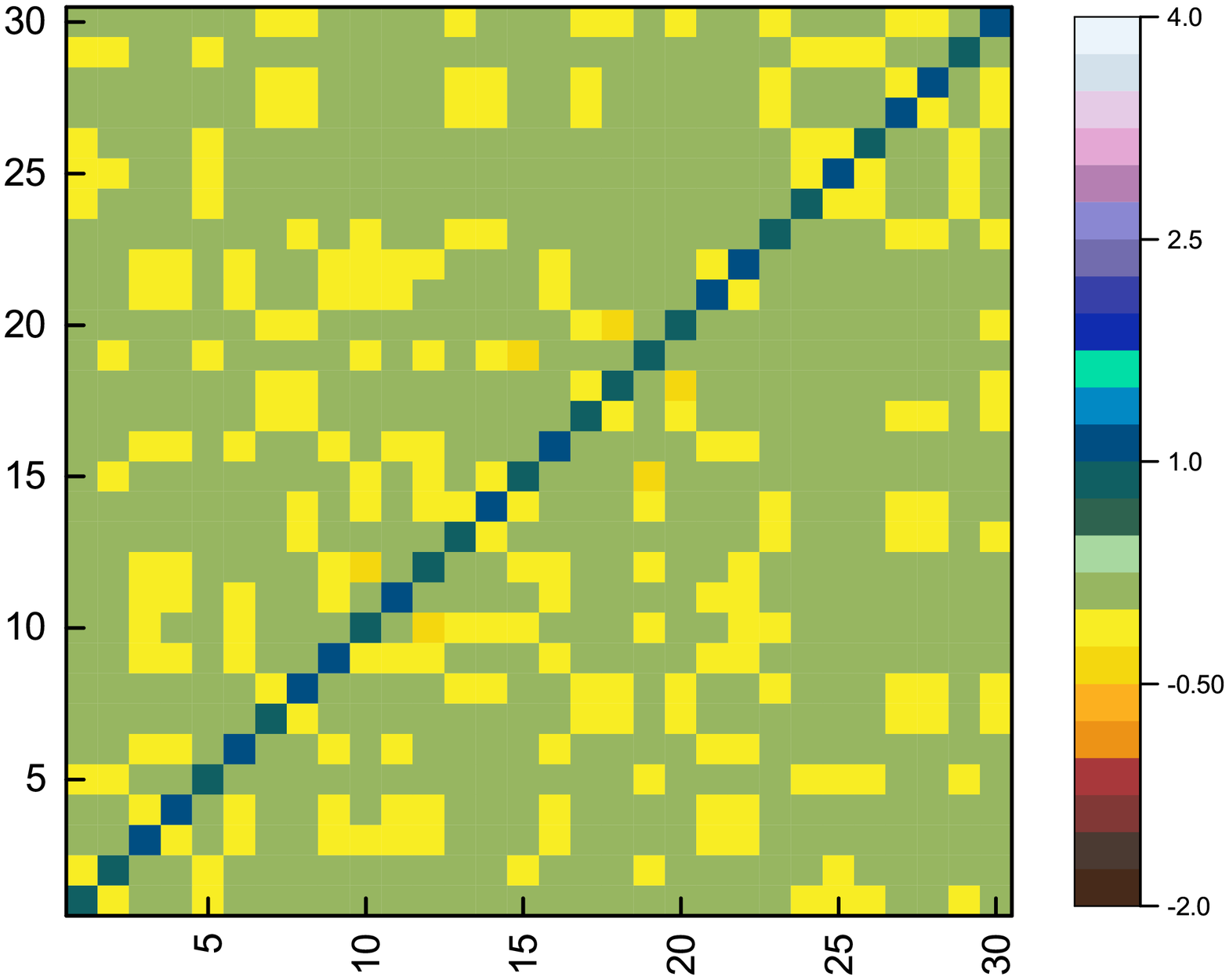}}
  \centerline{ (e) GL-LRSS}
\end{minipage}
\hfill
\begin{minipage}[b]{0.18\linewidth}
  \centering
  \centerline{\includegraphics[width=3.3cm]{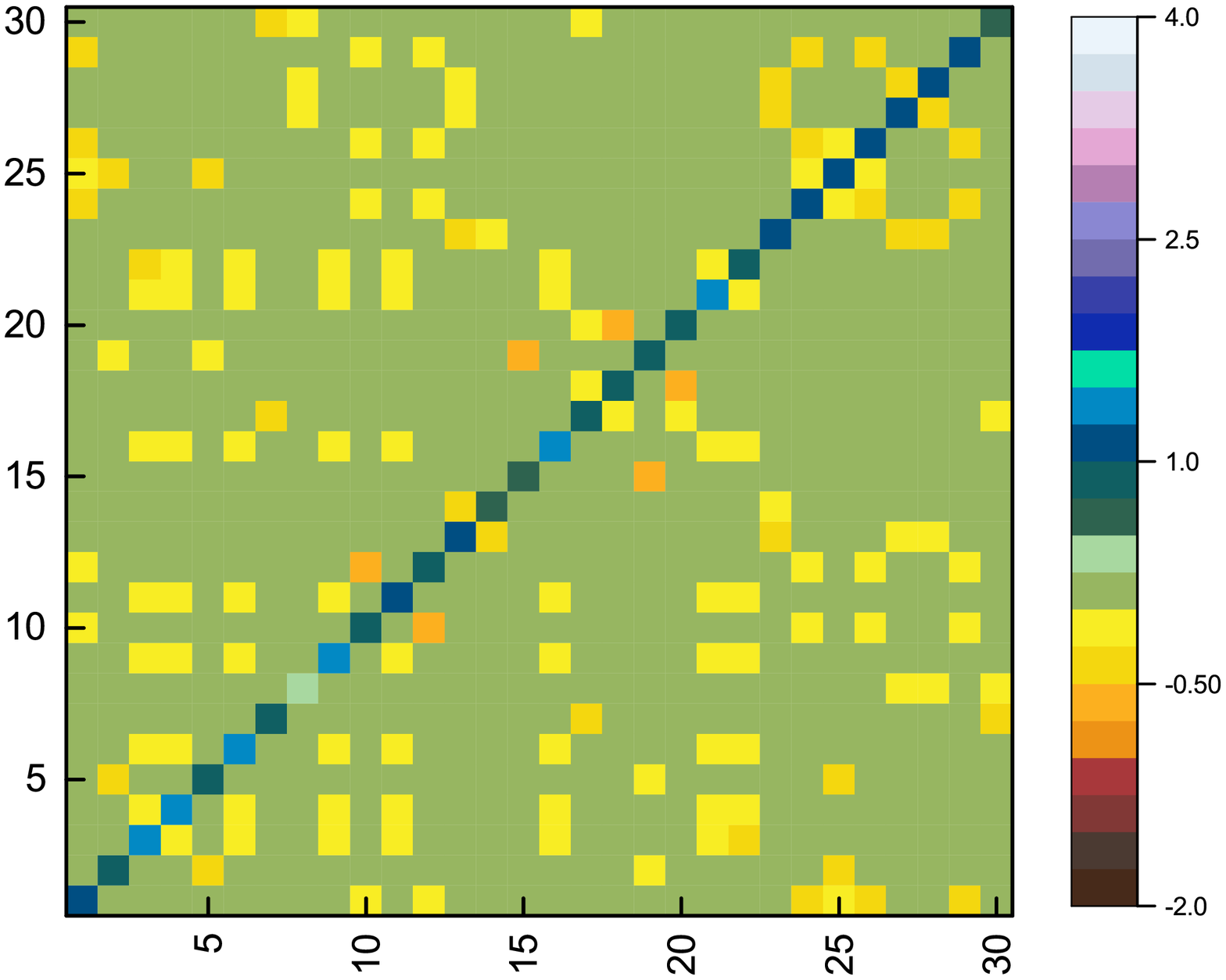}}
  \centerline{ (f) GL-Sigrep}
\end{minipage}
\hfill
\begin{minipage}[b]{0.18\linewidth}
  \centering
  \centerline{\includegraphics[width=3.3cm]{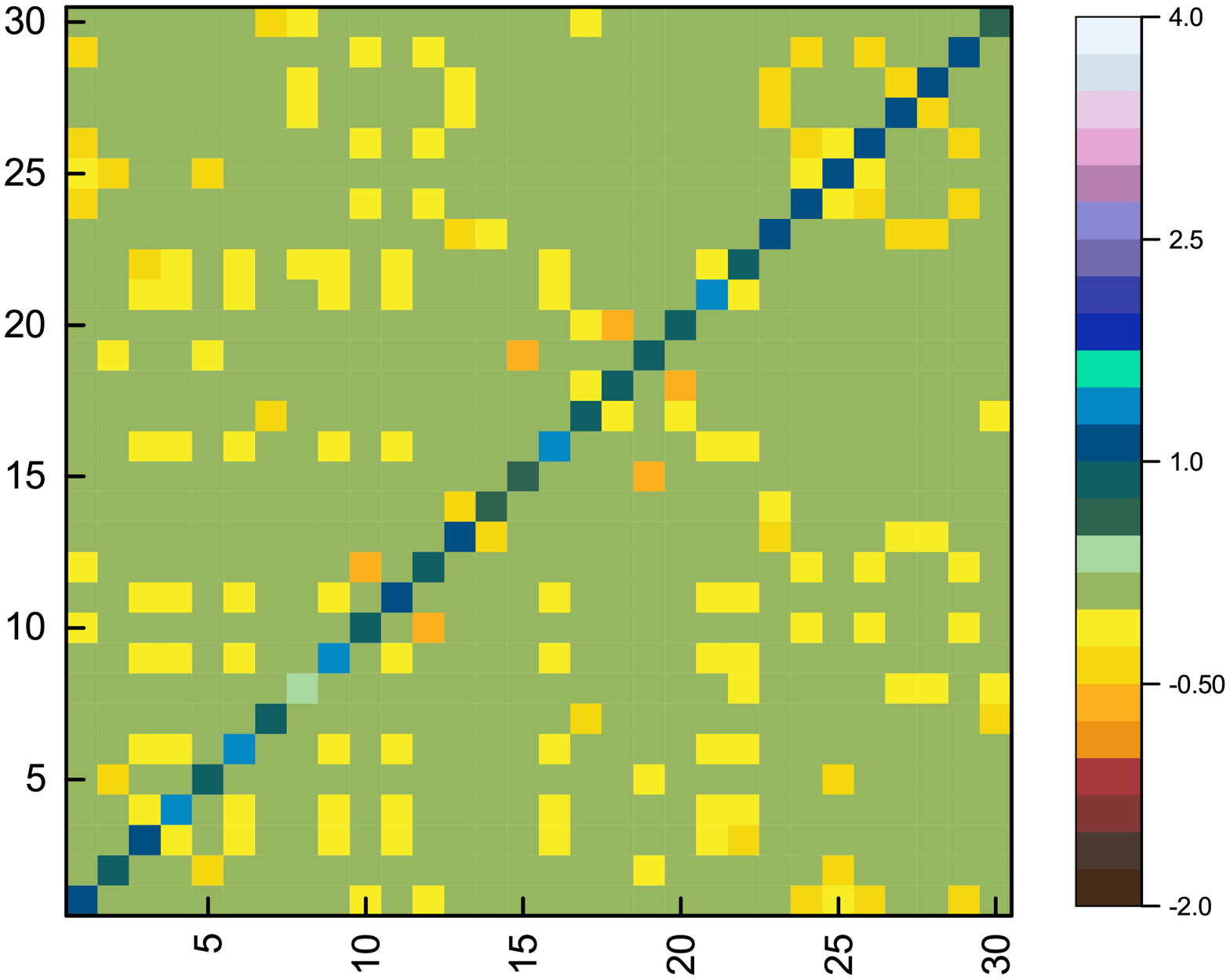}}
  \centerline{ (g) LGE}
\end{minipage}
\hfill
\begin{minipage}[b]{0.18\linewidth}
  \centering
  \centerline{\includegraphics[width=3.3cm]{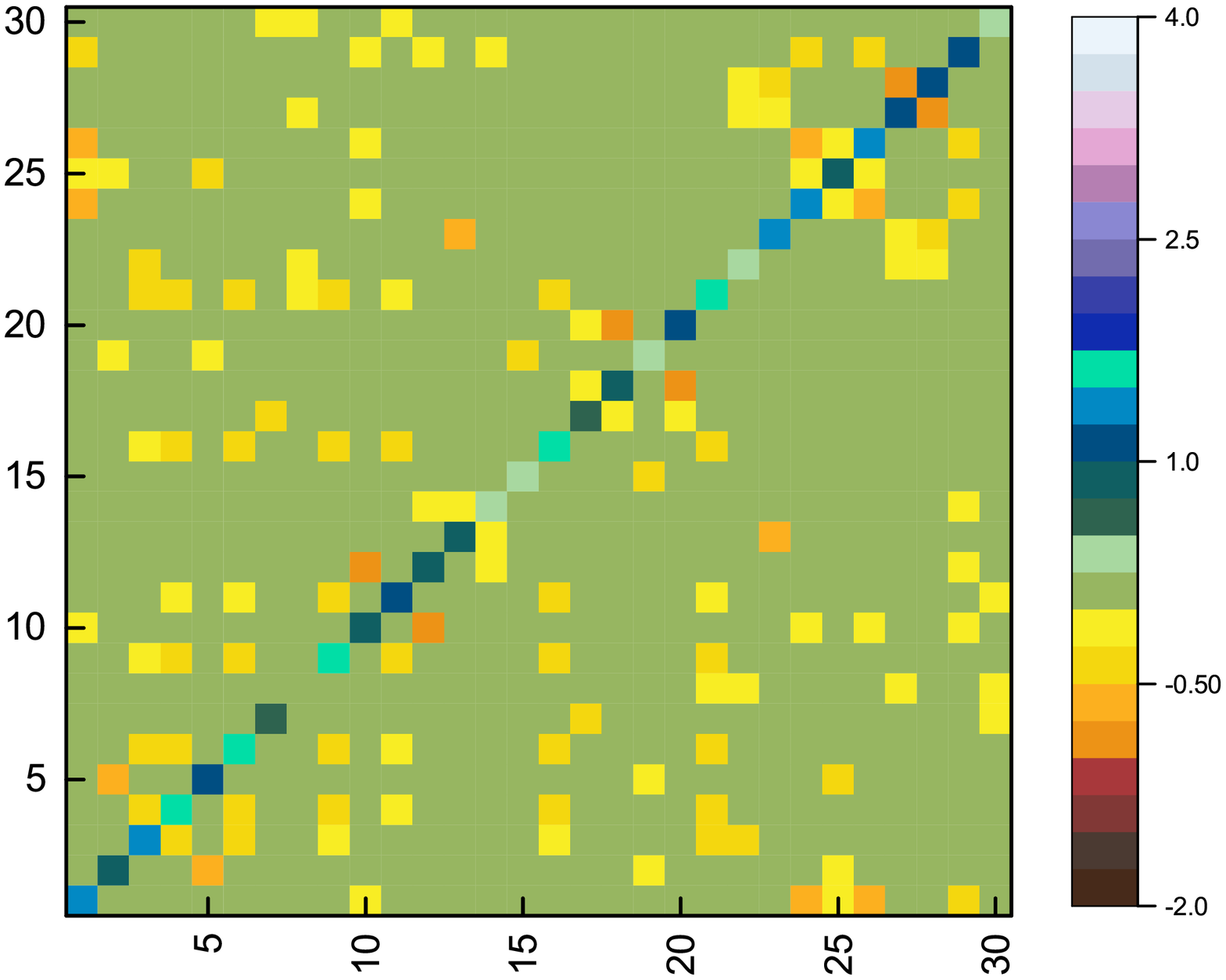}}
  \centerline{ (h) GL-logdet}
\end{minipage}
\caption{Visual comparison between the learned graph Laplacian matrices and the ground-truth Laplacian.
The columns from the left to the right are the ground-truth Laplacian,
the Laplacians recovered by GL-LRSS, GL-Sigrep, LGE and GL-logdet.
The rows from the top to the bottom are the learning results for the random geometric graph ${\mathcal{G}}_{RGG}$ and grid graph ${\mathcal{G}}_{grid}$, respectively.}
\label{fig2}
\end{figure*}

\begin{table*}[!t]
\centering
\caption{\label{tab:2}Graph learning performance from different types of time-varying graph signal in the proposed and baseline methods.}
\begin{tabular}{l c c c c p{20pt}<{\centering} p{20pt}<{\centering} c c c c p{20pt}<{\centering} p{20pt}<{\centering}}
\hline
\hline
\specialrule{0em}{2pt}{2pt}
\multirow{2}{*}{}
& \multicolumn{6}{c}{Random geometirc graph ${\mathcal{G}}_{RGG}$ } & \multicolumn{6}{c}{Grid graph ${\mathcal{G}}_{grid}$}\\
\cmidrule(lr){2-7}  \cmidrule(lr){8-13}
&\boldmath{GL-LRSS} & \boldmath{GL-Sigrep} & \boldmath{LGE} & \boldmath{GL-logdet} &\boldmath{PCAG} & \boldmath{RPCAG}&\boldmath{GL-LRSS} & \boldmath{GL-Sigrep} & \boldmath{LGE} & \boldmath{GL-logdet} &\boldmath{PCAG} & \boldmath{RPCAG}  \\
\specialrule{0em}{2pt}{2pt}
\hline
\specialrule{0em}{2pt}{2pt}
\boldmath{F-measure}& \bf{0.8201} & 0.7087& 0.7196 & 0.6861 & - & - & \bf{0.7832} & 0.6913& 0.7029 & 0.6764 & - & -\\
\boldmath{Precision}& \bf{0.8709} & 0.7834 & 0.6469 & 0.8565 & - & -& \bf{0.7633} & 0.6547 & 0.6593 & 0.7517 & - & -\\
\boldmath{Recall}& \bf{0.7984} & 0.6561 & 0.8212 & 0.5793 & - & - & \bf{0.8117} & 0.7554 & 0.7575 & 0.6456 & - & -\\
\boldmath{NMI}& \bf{0.5096} & 0.2330 & 0.2761 & 0.2138 & - & - & \bf{0.4198} & 0.3282 & 0.3339 & 0.3033 & - & -\\
\boldmath{GSE}& \bf{0.3315} & 0.3814 & 0.3445 & 0.5375 & - & - & \bf{0.7068} & 0.7229 & 0.7234 & 0.9664 & - & - \\
\boldmath{LCE}&\bf{0.0545}& 0.2446 & 0.1424 & - & 0.4220 & 0.2432 &\bf{0.0665}& 0.2465 & 0.1452 & - &0.2223  & 0.1221\\
\specialrule{0em}{2pt}{2pt}
\hline
\hline
\end{tabular}
\end{table*}
\vspace{-0.25cm}
\section{Experiments}
We verified the effectiveness and performance of the proposed method 
on a variety of datasets: 1) two synthetic
datasets under different graph structures, 2) meshes representing a dancing man \cite{ref44}, 3)
a daily temperature dataset of China obtained from the National Oceanic and Atmospheric Administration \cite{ref45},
and 4) a daily evaporation
dataset of California from the Department of Water Resources \cite{ref46}.
Moreover, we compare the proposed GL-LRSS with several state-of-the-art methods,
including GMS \cite{ref53}, GL-Logdet \cite{ref19},
GL-Sigrep \cite{ref20}, SpecTemp \cite{ref28}, LGE \cite{ref35}, PCAG \cite{ref14} and RPCAG \cite{ref15}.
GMS, GL-Logdet and SpecTemp are graph learning methods
that only infer the graph structure from observations, whereas
PCAG and RPCAG estimate low-rank components
under a $k$-nearest-neighbor graph. In contrast,
GL-LRSS, GL-Sigrep, and LGE simultaneously estimate the graph and low-rank components.
For real-world data, we evaluated two types of ${\textit{R}}$ matrices for the proposed method. Specifically, we either set ${\textit{R}}$ to identity matrix $\bf{I}$ or considered 
prior information of ${\textit{R}}$.

We provided visual and quantitative results of 
the edges from the learned graph and the ground-truth graph.
We conducted 20 independent Monte Carlo simulations to test
the average performance of the proposed and baseline methods.
To measure the estimation performance,
we used the low-rank component estimation error (LCE):
\begin{small}${{{{{\left\| {\hat {\bf{X}} - {{\bf{X}}_0}} \right\|}_F}} \mathord{\left/
 {\vphantom {{{{\left\| {\hat {\bf{X}} - {{\bf{X}}_0}} \right\|}_F}} {\left\| {{X_0}} \right\|}}} \right.
 \kern-\nulldelimiterspace} {\left\| {{{\bf{X}}_0}} \right\|}_F}}$\end{small} and graph structure estimation error (GSE):
\begin{small}${{{{{\left\| {\hat {\bf{L}} - {{\bf{L}}_0}} \right\|}_F}} \mathord{\left/
 {\vphantom {{{{\left\| {\hat {\bf{L}} - {{\bf{L}}_0}} \right\|}_F}} {\left\| {{{\bf{L}}_0}} \right\|}}} \right.
 \kern-\nulldelimiterspace} {\left\| {{{\bf{L}}_0}} \right\|}_F}}$\end{small}.
Additionally, to measure 
the recovery performance of the edge position in the ground-truth graphs,
we obtained the \emph{Precision},
\emph{Recall}, \emph{F-measure} and \emph{Normalized Mutual Information (NMI)} \cite{ref47}.
These four measures take values between 0 and 1, where values
close to 1 indicate higher learning performance.
For a fair comparison, we used a grid search to set the regularization parameters that maximize the performance for each method. 

\begin{figure*}
\centering
\subfigure[]{ \label{Fig3:a}
\includegraphics[width=2.2in]{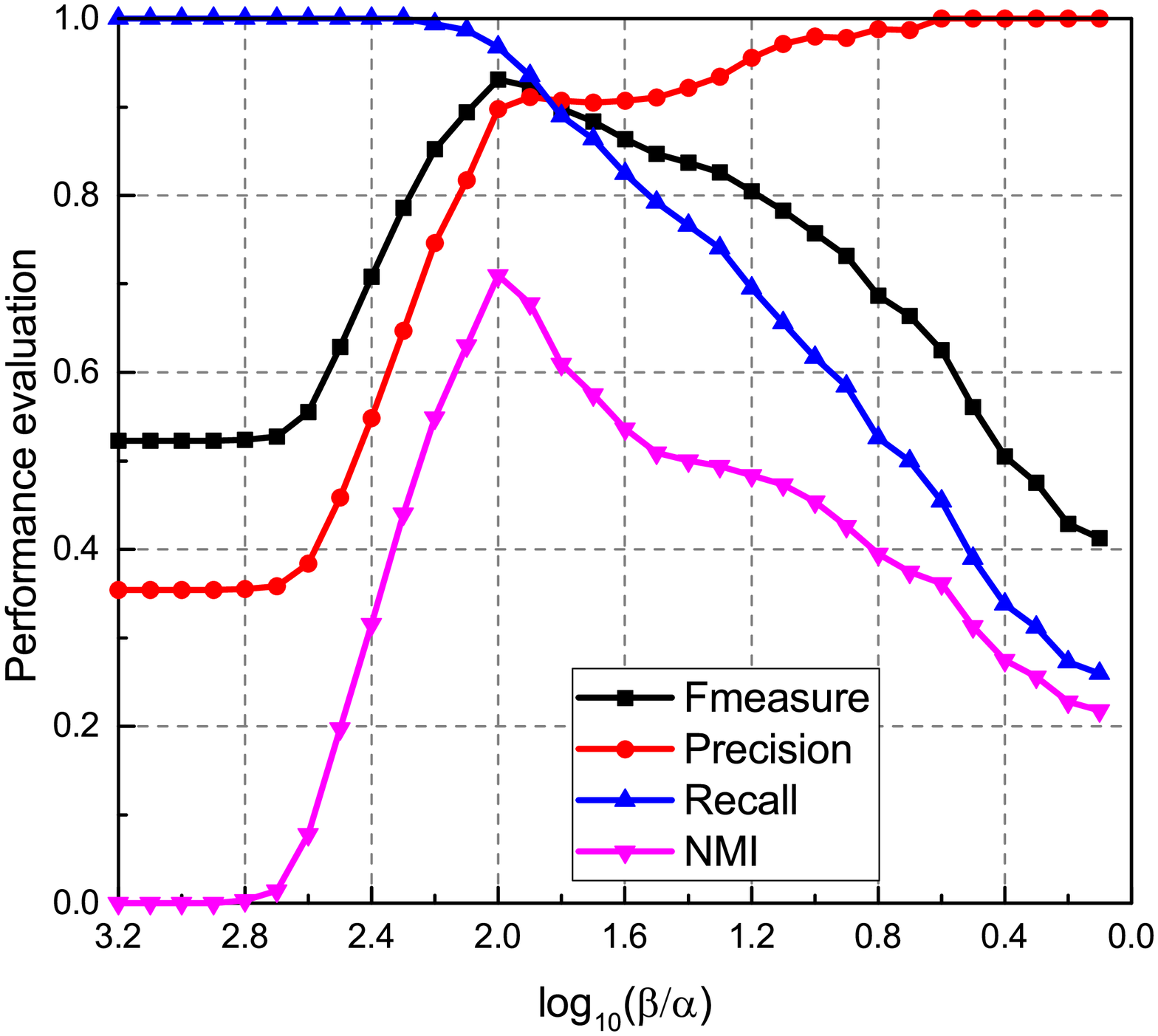}}
\hspace{0.05in}
\subfigure[]{ \label{Fig3:b}
\includegraphics[width=2.2in]{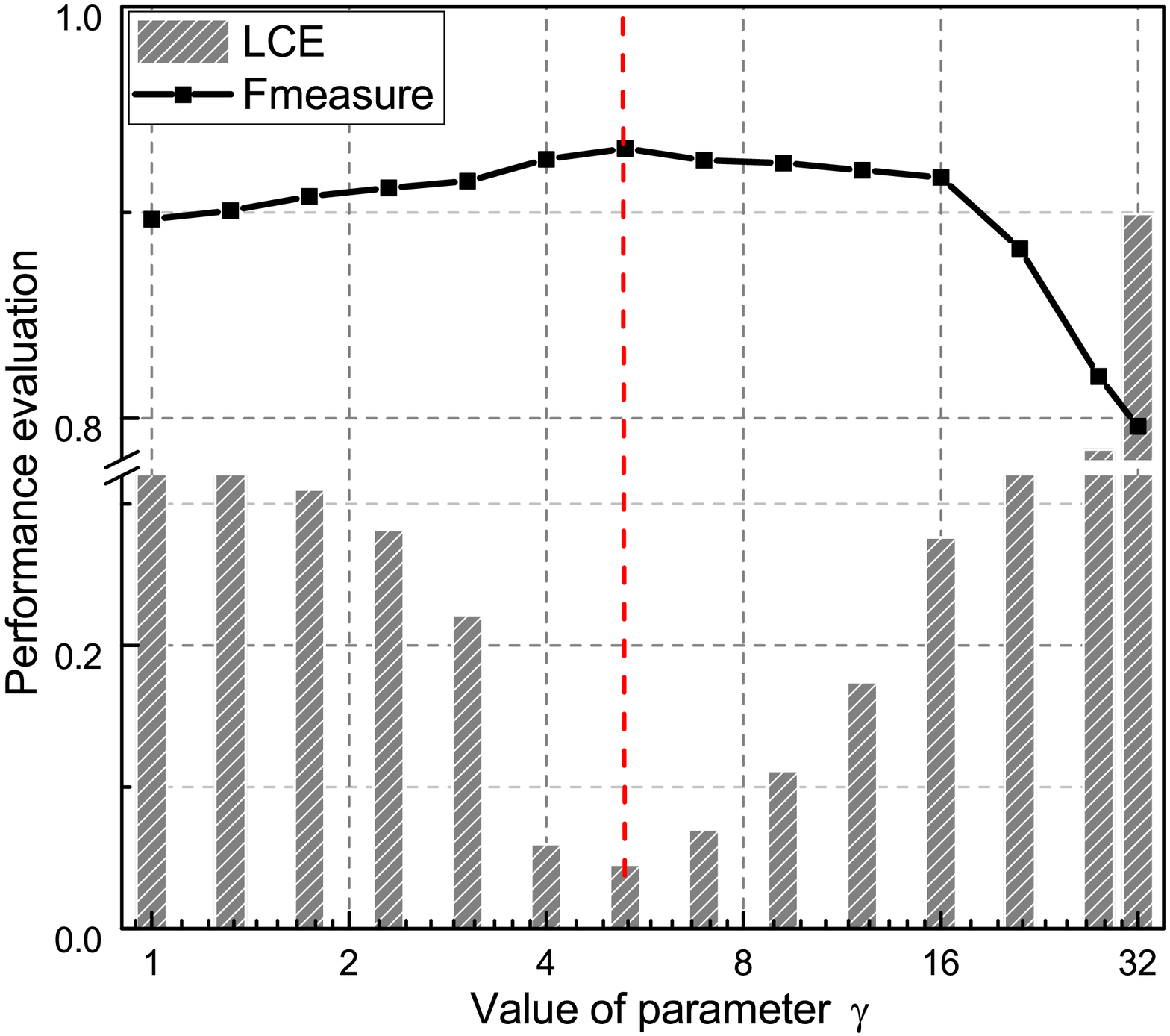}}
\hspace{0.05in}
\subfigure[]{ \label{Fig3:c}
\includegraphics[width=2.2in]{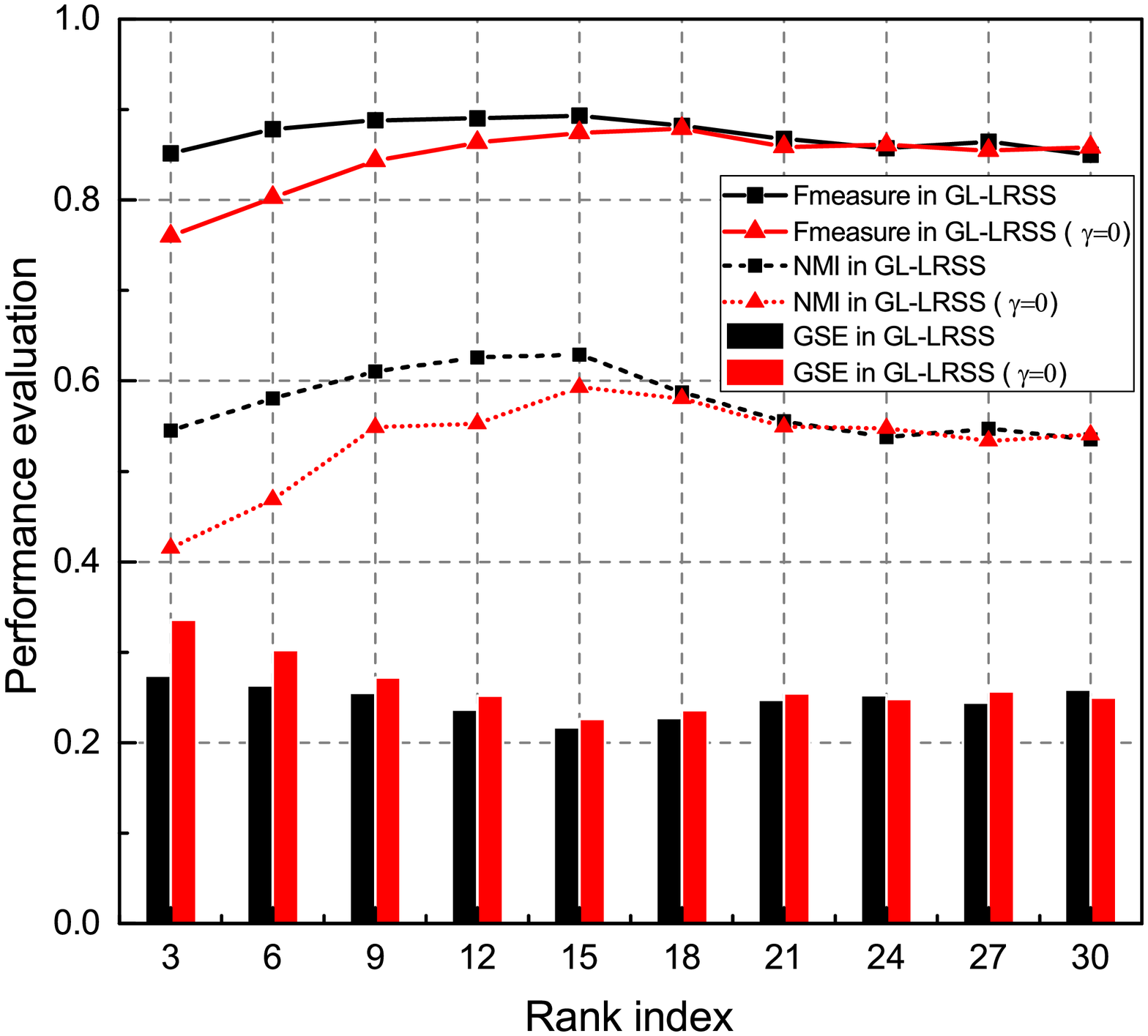}}
\caption{For a random instance of ${\mathcal{G}}_{RGG}$, (a) performance of the GL-LRSS under different ratios of $\beta$ to $\alpha$, with $\gamma=5.278 $, (b) performance of the GL-LRSS under different value of $\gamma$, where $\alpha$ and $\beta$ are chosen to maximize the \emph{F-measure} for each $\gamma$ and (c) the performance comparison of the proposed GL-LRSS and the
GL-LRSS ($\gamma=0$) without nuclear norm under the
different rank index.}
\label{Fig3}
\end{figure*}

\vspace{-0.3cm}
\subsection{Experiments on synthetic data}
We evaluated the GL-LRSS performance on synthetic datasets, which were created considering a 30-vertex undirected graph and two different graph connectivity models: grid graph ${\mathcal{G}}_{grid}$
and random geometric graph ${\mathcal{G}}_{RGG}$.
For the grid graph, each vertex with a random coordinate was connected to its five nearest neighbors,
and the edge weight between two vertices was inversely proportional to their distance.
For the random geometric graph, the vertex coordinates were uniformly
random in a unit square, and each edge weight was determined from a
Gaussian function,
$W\left( {i,j} \right) = \exp \left( { - \frac{{d{{\left( {i,j} \right)}^2}}}{{2{\sigma ^2}}}} \right)$,
where $\sigma  = 0.5$, considering threshold weights below $0.7$.
After graph construction, we computed the graph Laplacian matrix and
normalized its trace to 30.

Given a ground-truth graph, we generated $30 \times 100$ time-varying graph signals
based on the proposed model in Eqs. (\ref{representation}) and (\ref{latent}).
Unless otherwise stated, state transition ${\textit{R}}$ was set as the identity matrix. The case of a general diagonal ${\textit{R}}$ was also considered, as reported below.
We selected the eigenvectors
corresponding to the smallest $r=3$ eigenvalues as the basis
vectors, that is, the columns of $\bf{U}$. 
Zero-mean Gaussian noise having a standard deviation of 0.5 was set as the perturbation.
Initial signal ${\bf{x}}_1$ and weighted difference signal
\begin{small}${{\bf{x}}_t}- {\textit{R}}{{\bf{x}}_{t - 1}}$\end{small} were smooth graph signals residing
on the subspace corresponding to the three smallest eigenvalues of the graph Laplacian
$\bf{L}$. 
Hence, the time-varying graph signals were approximately low-rank
and presented spatiotemporal smoothness.

We applied GL-LRSS, GL-Sigrep, LGE, and GL-Logdet to learn the graph Laplacian matrices
given only observation \begin{small}$\bf{Y}$\end{small}. 
We used GL-LRSS, GL-Sigrep, LGE, 
PCAG, and RPCAG to estimate the low-rank components and obtained their 
average performance across 20 random instances of two graphs with the associated graph signals.

\subsubsection{Performance comparison} 
A visual comparison of the evaluated methods is shown
in Fig. \ref{fig2}, which depicts
the ground-truth graph Laplacian and the Laplacian matrices learned by
GL-LRSS, GL-Sigrep, LGE, and GL-Logdet from left to right. The first and second rows show
the results under graph models ${\mathcal{G}}_{RGG}$ and ${\mathcal{G}}_{grid}$,
respectively.
In both cases, the graph Laplacian provided by GL-LRSS is visually more consistent with
the ground truth than the Laplacians obtained from the baseline methods.
For further performance analysis, we obtained the quantitative results listed in
Table \ref{tab:2}.
Compared with the other four graph learning methods, 
the \emph{F-measure} increases with the decreasing score of the \emph{LCE} for the proposed GL-LRSS. Hence, better low-rank component
estimation improves the graph estimation accuracy.
For the five low-rank component estimation methods applied to ${\mathcal{G}}_{grid}$,
the \emph{LCE} decreases with the increasing \emph{F-measure} scores.
Specifically, the performance of PCAG and RPCAG on
${\mathcal{G}}_{grid}$ is better than that on ${\mathcal{G}}_{RGG}$
because the predefined graph is 
closer to the ground truth in ${\mathcal{G}}_{grid}$.
These results suggest that a better graph inference improves low-rank component estimation. 
Thus, because the two estimation
steps are alternately optimized, the proposed
method outperforms GL-logdet, PCAG, and RPCAG.

The proposed GL-LRSS outperforms the baseline methods in 
graph inference and low-rank component estimation.
For ${\mathcal{G}}_{RGG}$, GL-LRSS
achieves the highest \emph{F-measure} of 0.8201 and an \emph{NMI} of 0.5096, as well as the lowest \emph{GSE} of 0.3315 and an \emph{LCE} of 0.0545.
The improvement of GL-LRSS, compared with GL-Sigrep,
is caused by the exploitation of long-term correlations, that is, the low-rank components.
The improvement of GL-LRSS over LGE is due to the proper
modeling of short-term correlations in Eq. (\ref{latent}), which demonstrate 
the benefits of applying
spatiotemporal smoothness during graph learning.
For ${\mathcal{G}}_{grid}$,
the superior GL-LRSS performance is less obvious,
possibly due to the low-rank assumption that limits the
graph information encoded in the low-rank component, which varies depending on graph types. 

\begin{figure}
\centering
\subfigure[]{ \label{Fig4:a}
\includegraphics[width=1.65in]{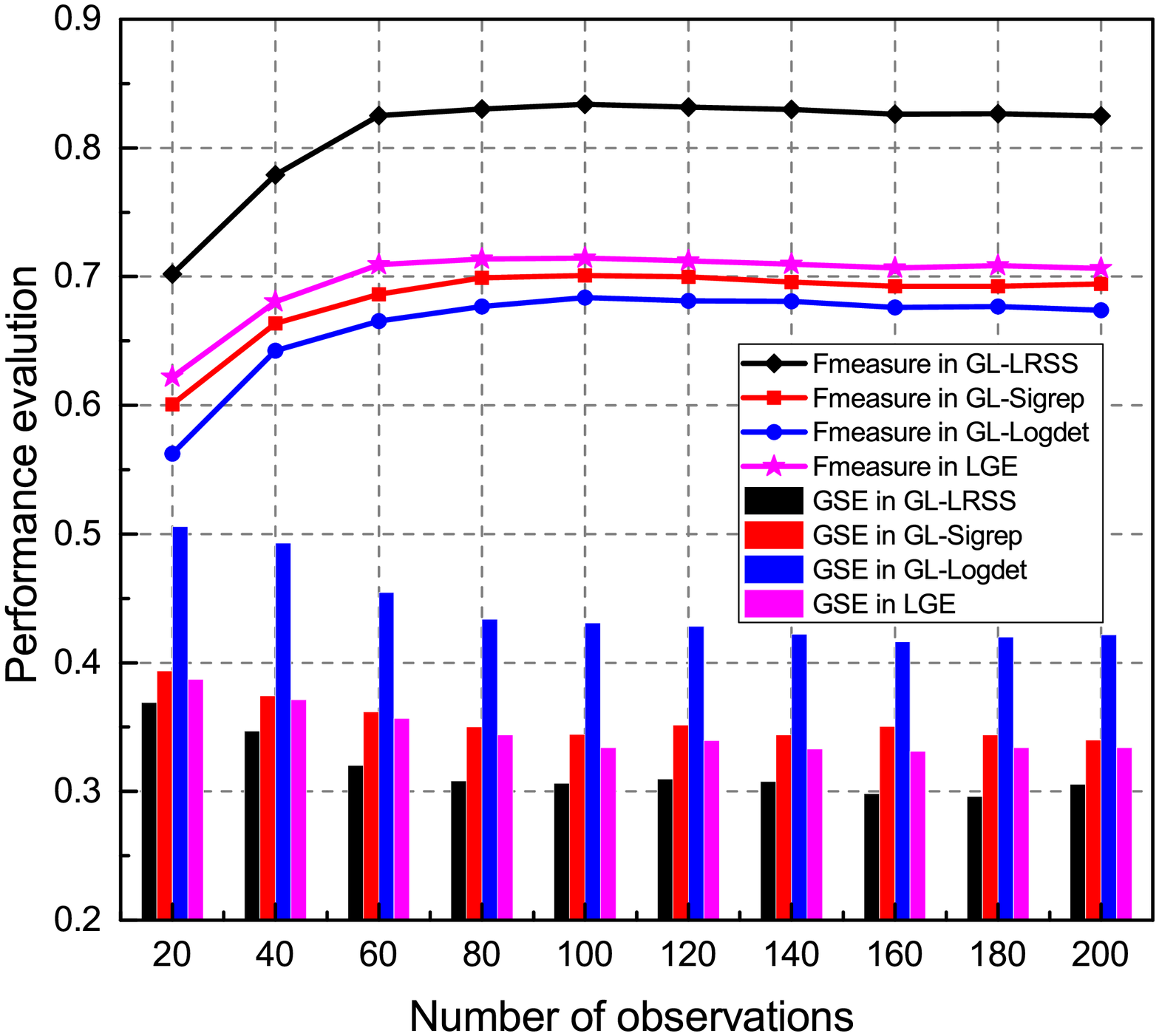}}
\subfigure[]{ \label{Fig4:b}
\includegraphics[width=1.65in]{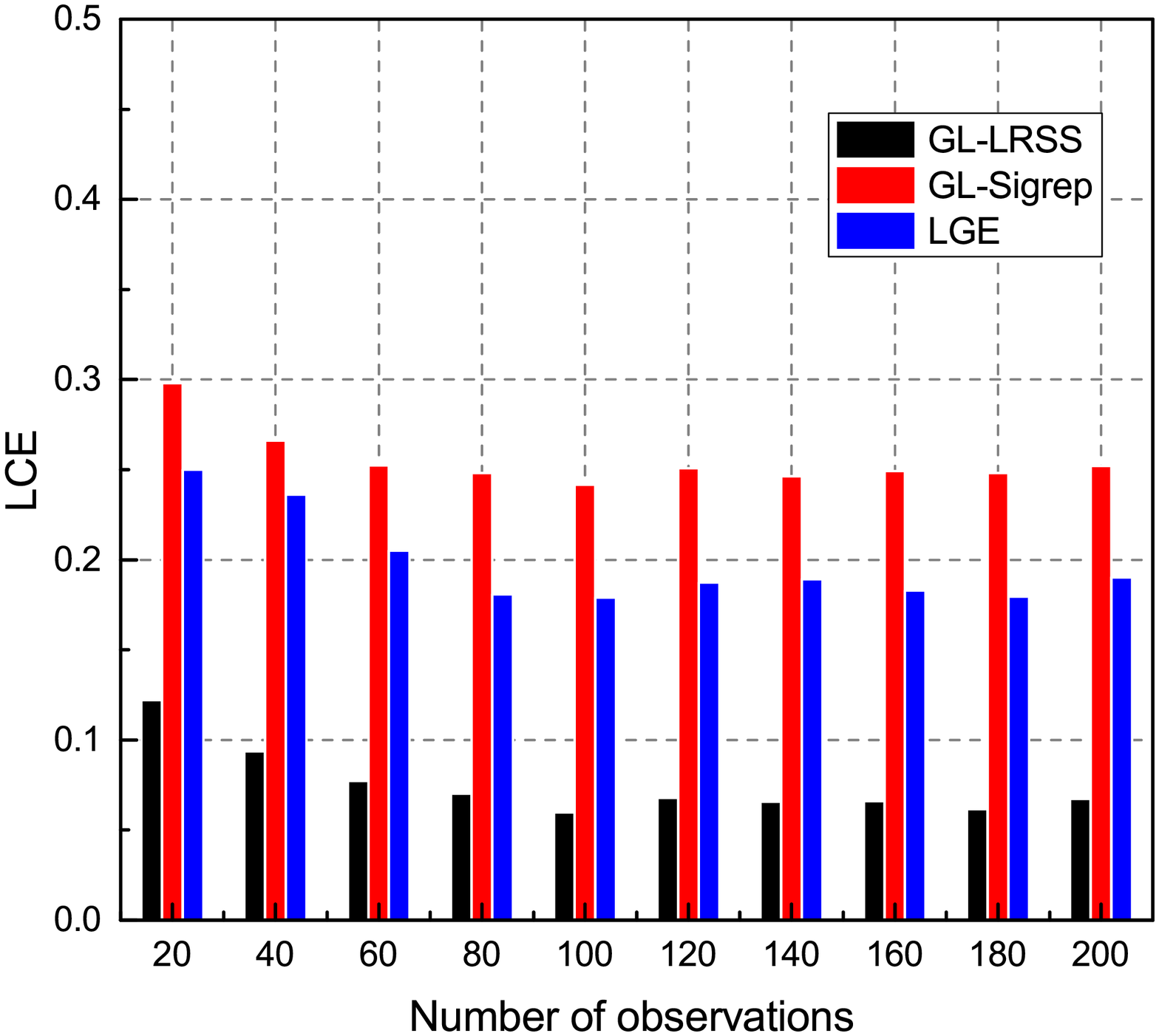}}
\caption{(a) Graph learning performance of the baseline and proposed methods under different number of signals, and (b) low-rank component estimation performance of the baseline and proposed methods under different number of signals,
for a random instance of ${\mathcal{G}}_{RGG}$. }
\label{Fig4}
\end{figure}

\subsubsection{Effect of regularization parameters}
To better understand the behavior of the proposed GL-LRSS under different
regularization parameter settings,
we chose different powers of 2 ranging from
0 to 5 with variations of 0.4 to set $\gamma$,
and
different powers of 10 ranging from
0 to -2 with variations of 0.1 for $\alpha$ and from
2 to 0 with variations of 0.1 for $\beta$.
For the same ${\mathcal{G}}_{RGG}$,
Fig. \ref{Fig3:a} shows
the learning performance for fixed $\gamma$ and varying
ratios of $\beta$ to $\alpha$.
Because the learned graph approaches the
ground truth, the \emph{recall}-\emph{precision} curve gradually interacts,
leading to an \emph{F-measure} peak. Thus, 
an appropriate ratio of $\beta$ to $\alpha$
can maximize the graph learning performance
of the proposed GL-LRSS. A similar trend can be observed in the \emph{NMI} curve.

To investigate the effect of parameter $\gamma$,
we fixed $\alpha$ and $\beta$ to their best values (Fig. \ref{Fig3:a}) while varying $\gamma$.
The GL-LRSS performance according to $\gamma$ is shown in Fig. \ref{Fig3:b}.
The \emph{F-measure} initially
increases with increasing $\gamma$, possibly due to the action of the unclear norm in (P1) on the low-rank component estimation.
After the \emph{F-measure} reaches its peak of 0.93 and the
\emph{LCE} reaches its minimum, the performance decreases,
because the influence of the unclear norm weakens. Hence,
an appropriate value of $\gamma$ improves the low-rank component estimation and the overall
graph inference.

To verify the effectiveness of the nuclear norm ${\left\| {\bf{X}} \right\|_*}$, we generated graph signals
for a random instance of ${\mathcal{G}}_{RGG}$ under varying $r$.
Then, we
inferred the graph by solving (P1)
with $\gamma >0$ and $\gamma=0$.
The GL-LRSS performance with and without ($\gamma=0$) the nuclear norm according to the rank index is shown in Fig. \ref{Fig3:c}.
For $\gamma=0$, the GL-LRSS performance is not affected by
the nuclear norm regularization term.
For the \emph{F-measure},
GL-LRSS with ${\left\| {\bf{X}} \right\|_*}$ outperforms that without ${\left\| {\bf{X}} \right\|_*}$ for the low rank index. The superior
GL-LRSS performance with ${\left\| {\bf{X}} \right\|_*}$ is less obvious
as the rank index increases.
This is possibly due to the introduction of the nuclear norm, which is more effective at a lower rank index, whereas its influence declines as the index approaches 30.
Similar results were obtained for \emph{NMI} and \emph{GSE}.
These results verify the correctness
of the optimization problem (P1).

\subsubsection{Effect of number of observations}
For one random instance of a
random geometric graph,
we investigated the influence of the number of
signals, from 20 to 200 in increments of 20.
The graph learning performance is shown
in
Fig. \ref{Fig4:a} in terms of \emph{F-measure} and \emph{GSE}.
We also report
the performance of GSP-based methods as
baselines for Laplacian recovery.
The performance
of all methods initially increases with the availability of an increasing number of signals for graph learning, until convergence
after approximately 80 signals.
Moreover, among all the evaluated methods, the proposed GL-LRSS
attains the highest \emph{F-measure} of approximately 0.82 and the lowest \emph{GSE} of approximately 0.28, indicating its higher graph learning performance.
The errors of the low-rank components recovered by GL-LRSS, GL-Sigrep,
and LGE are shown in Fig. \ref{Fig4:b}. The \emph{LCE} trend is similar to that of the \emph{F-measure}.
Figs. \ref{Fig4:a} and \ref{Fig4:b} verify that
GL-LRSS outperforms the other
methods in terms of both graph learning and low-rank component estimation, possibly because our formulation
utilizes long- and short-term
correlations in spatiotemporal signals to facilitate learning.

\subsubsection{Effect of general diagonal matrix ${\textit{R}}$}
To examine the GL-LRSS performance when ${\textit{R}}$ is a general diagonal matrix, we generated ${\textit{R}}$ from a normal distribution \begin{small}${\mathcal {N}}\left( 0.5,0.25^2 \right)$\end{small} and
guaranteed that every entry in ${\textit{R}}$ was less than 1. We considered two cases of known and unknown ${\textit{R}}$. For the unknown ${\textit{R}}$, we assumed an incorrect ${\textit{R}}$ (${\textit{R}}={\bf{I}}_N$). For a random instance of ${\mathcal{G}}_{RGG}$, the results of the evaluated method for the two cases are listed in Table \ref{tab:6}. 
The GL-LRSS performance with unknown ${\textit{R}}$ is much worse than that for the known matrix, possibly due to the mismatching ${\textit{R}}$. 
From Tables \ref{tab:2} and \ref{tab:6} for known ${\textit{R}}$, the GL-LRSS performance for a general diagonal ${\textit{R}}$ is not as good as that for ${\textit{R}}$ being the identity matrix. 
Regarding the \emph{LCE}, the advantage of GL-LRSS for ${\textit{R}}$ being the identity matrix is obvious, possibly because ${\textit{R}}={\bf{I}}_N$ is the best case for low-rank component recovery during graph learning.

\begin{table}[t]
\newcommand{\tabincell}[2]{\begin{tabular}{@{}#1@{}}#2\end{tabular}}
\centering
\caption{\label{tab:6} The GL-LRSS performance in two cases of general diagonal $\textit{R}$.}
\setlength{\tabcolsep}{1.4mm}{
\begin{tabular}{c c c c c c c}
\hline
\hline
\specialrule{0em}{2pt}{2pt}
\tabincell{c}{} & F-measure& Precision &
Recall &NMI &GSE & LCE \\
\specialrule{0em}{2pt}{2pt}
\hline
\specialrule{0em}{2pt}{2pt}
${\textit{R}}_{\rm{known}}$ & 0.7657 & 0.8203 & 0.7346 & 0.3817 & 0.3632 &0.1390\\
${\textit{R}}_{\rm{unknown}}$& 0.6707 & 0.7873 & 0.5874 & 0.2833 & 0.6249 &0.2870\\
\specialrule{0em}{2pt}{2pt}
\hline
\hline
\end{tabular}}
\vspace{-0.2cm}
\end{table}

\begin{figure}[t]
\centering
\epsfig{figure=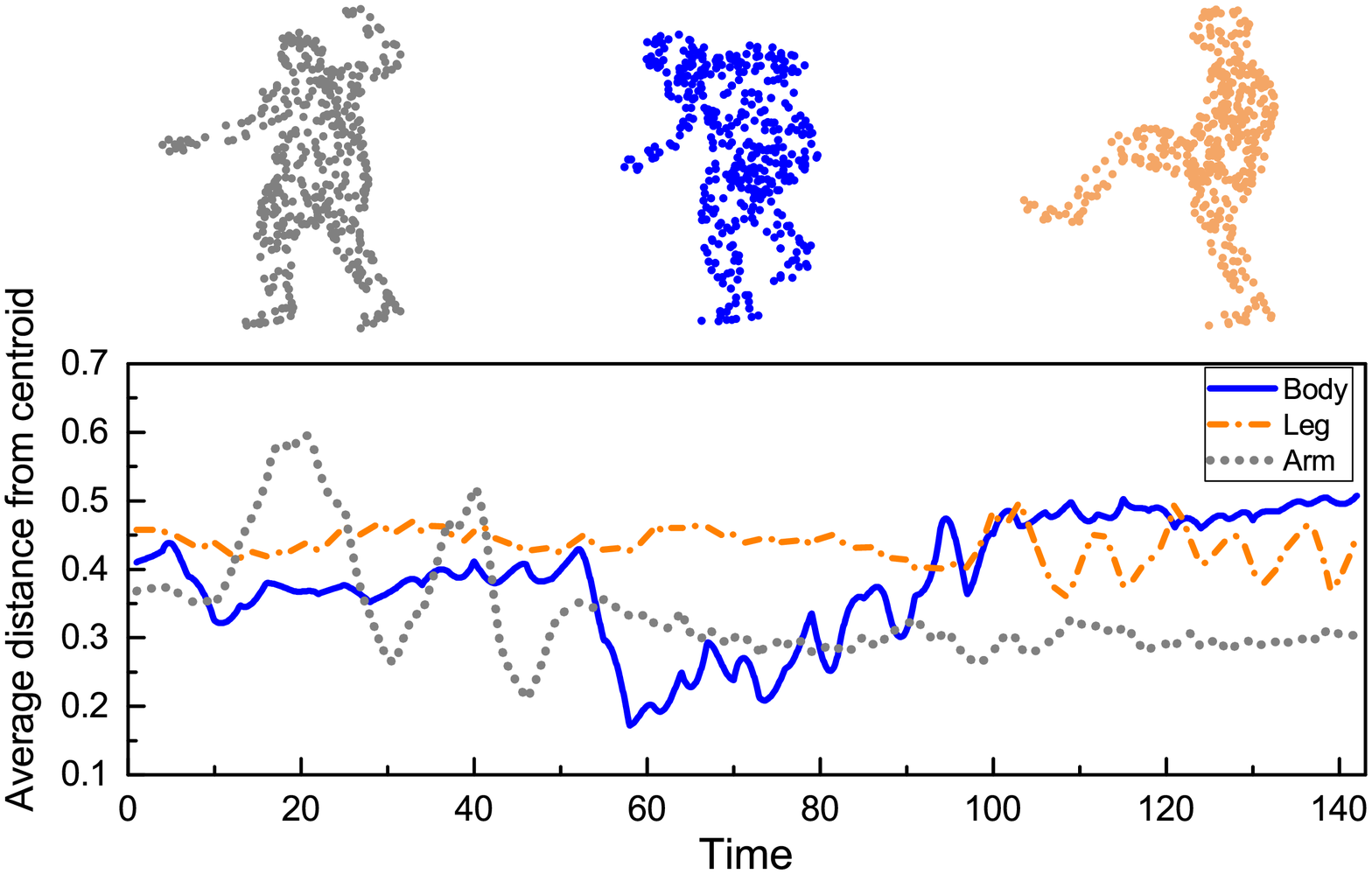,height=2in}
\caption{Clustering of the dancer mesh: the plot (below) shows for
each line the average distance between the points in different part of
body and the centroid. We observe that each frame belongs to different
phase of the dance, named "Arm", "Leg", "Body". The classification of
the motion depends on the main fluctuation of the lines, that is, the
part of body mainly involved in the dance.
}
\label{fig5}
\end{figure}
\begin{figure*}
\centering
\subfigure[]{ \label{Fig6:a}
\includegraphics[width=1.57in]{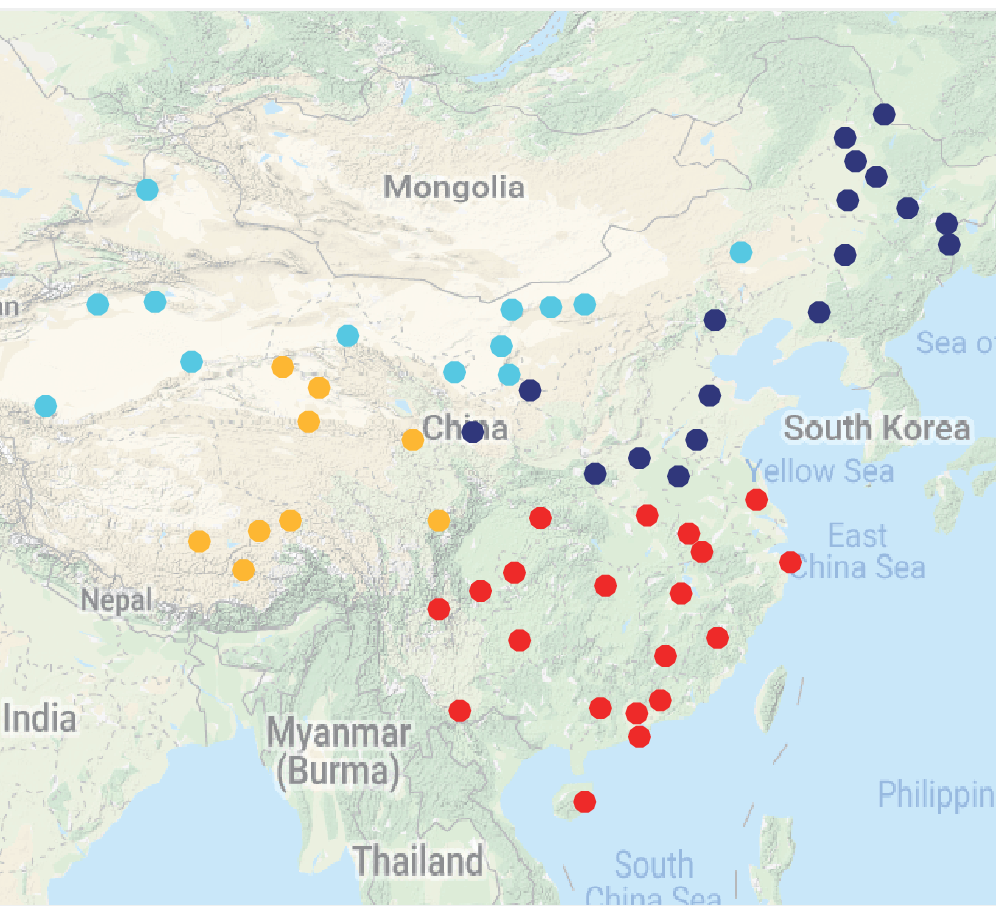}}
\hspace{-0.1in}
\subfigure[]{ \label{Fig6:b}
\includegraphics[width=1.57in]{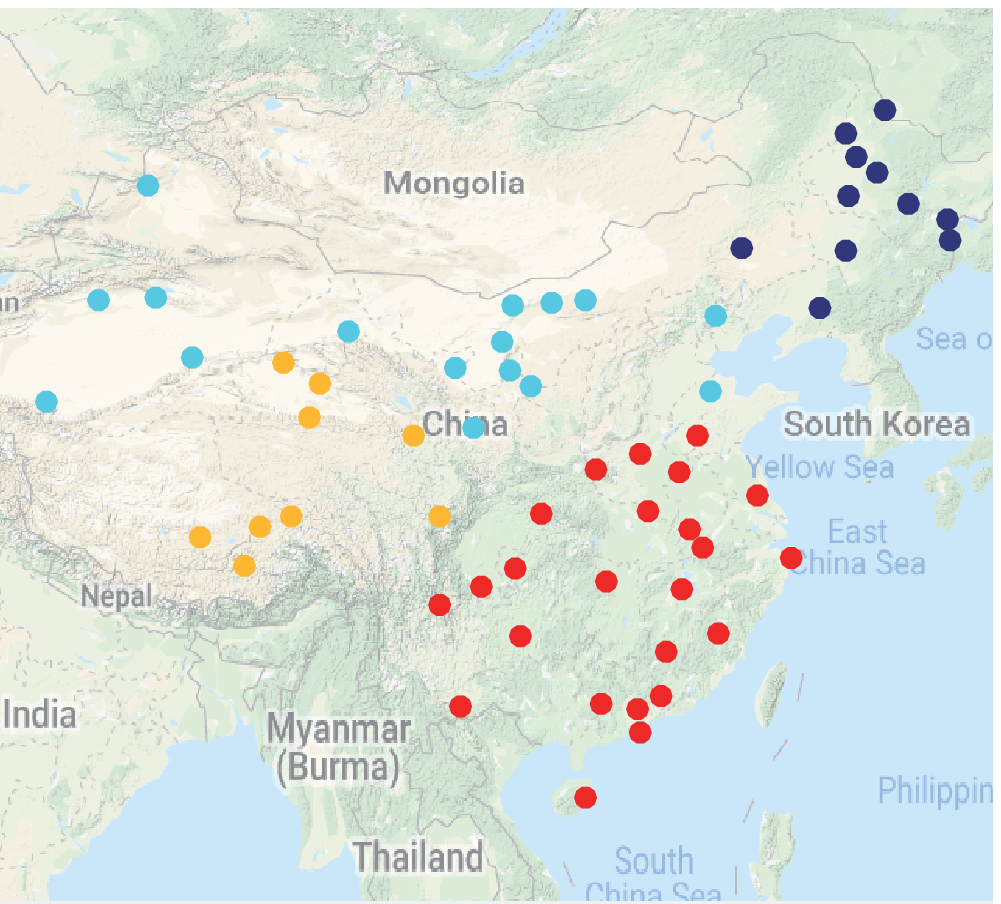}}
\hspace{-0.1in}
\subfigure[]{ \label{Fig6:c}
\includegraphics[width=1.9in]{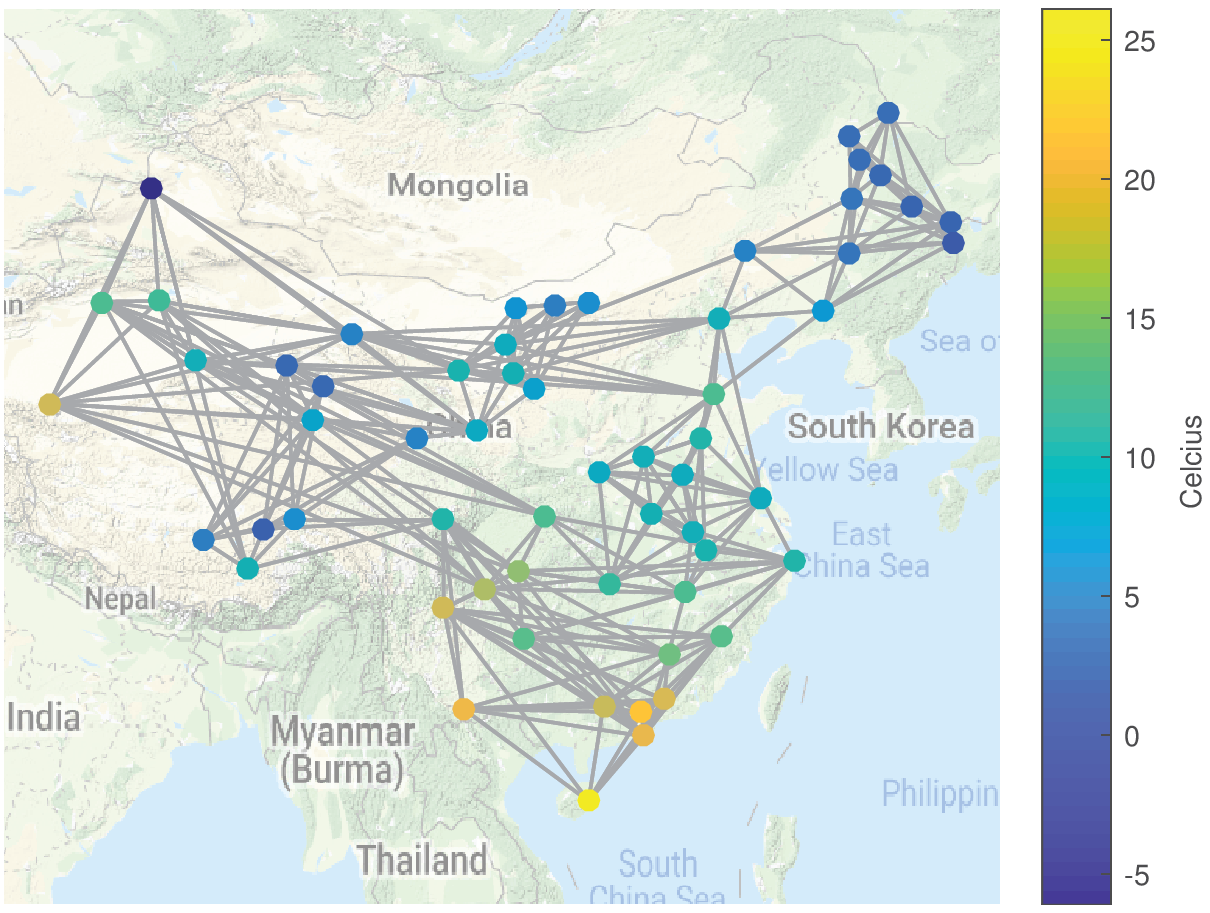}}
\hspace{-0.2in}
\subfigure[]{ \label{Fig6:d}
\includegraphics[width=1.9in]{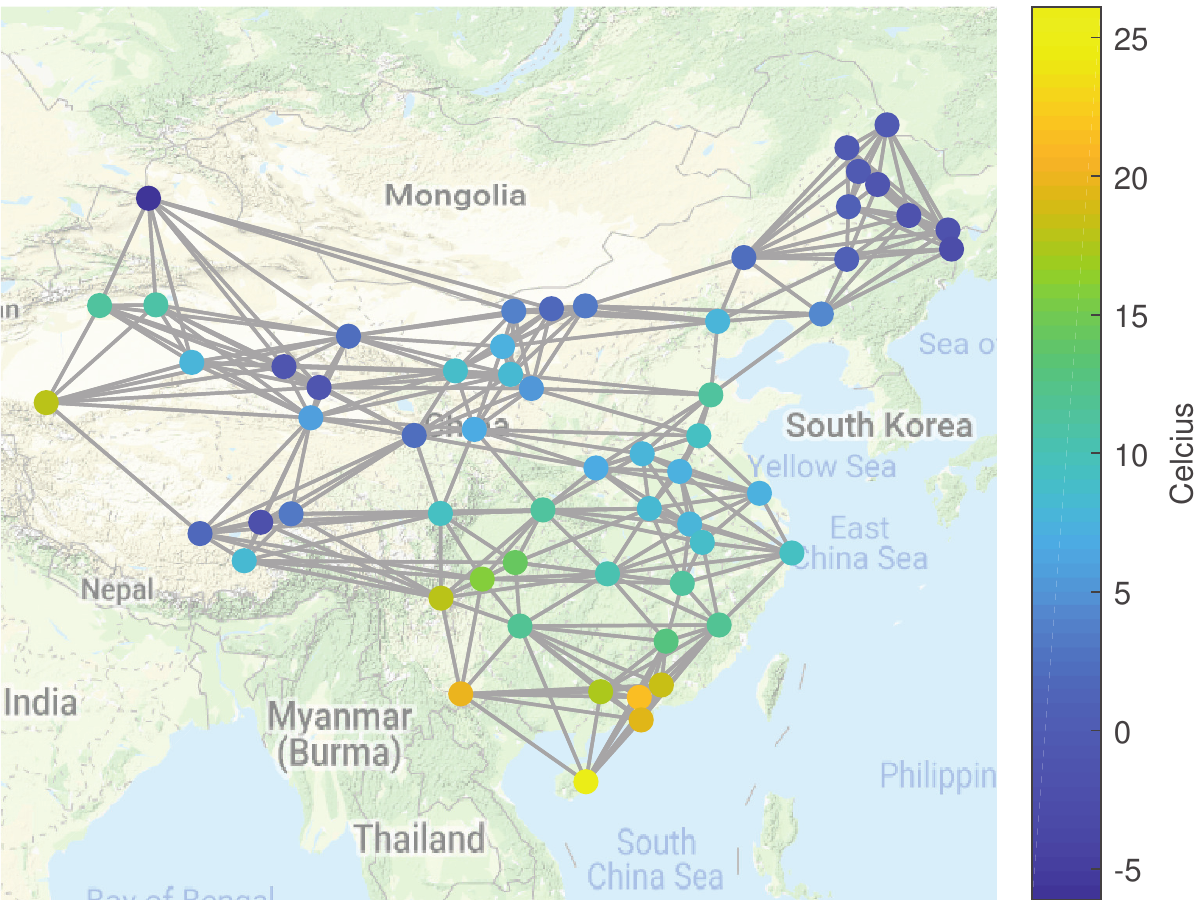}}
\caption{(a) The locations of 60 measuring stations in China. Different colors represent the ground-truth 4 clusters that correspond to 4 geographical regions. (b) The clustering results utilizing the graph Laplacian obtained by the GL-LRSS(${\textit{R}}_I$). (c) Graph structure learned by the GL-LRSS(${\textit{R}}_I$), which achieves the best \emph{RI} score in clustering performance. (d) Graph structure established by eight nearest neighbors according to the physical location of measuring stations.  The color code in (c) and (d) represents the
realistic temperature in Celcius scale on the 20th day.}
\label{Fig6}
\end{figure*}
\subsection{Graph learning from dancer mesh dataset}
We also evaluated the proposed GL-LRSS on real-world
data. We first considered the dancer mesh dataset containing
143 frames that represent different dancing postures.
At each frame,
we considered
the distance of 300 mesh vertices from each coordinate
to the centroid
as the observed signals, 
thus obtaining 143 time-varying graph signals with 
dimension of 300.
During the whole sequence, the graph
between the mesh vertices is unknown but is assumed to be fixed.
GL-LRSS aimed to
determine the intrinsic graph by capturing
the body connectivity between the mesh vertices in terms
of distances during the dancing sequence.

\begin{table}[t]
\newcommand{\tabincell}[2]{\begin{tabular}{@{}#1@{}}#2\end{tabular}}
\centering
\caption{\label{tab:3} Comparison of the motion classification performance
between different methods in dancer mesh data.}
\setlength{\tabcolsep}{1.4mm}{
\begin{tabular}{c c c c c c c}
\hline
\hline
\specialrule{0em}{2pt}{2pt}
\tabincell{c}{} & GL-LRSS& GL-Sigrep &
LGE &PCAG &RPCAG & \tabincell{c}{K-means on\\original data} \\
\specialrule{0em}{2pt}{2pt}
\hline
\specialrule{0em}{2pt}{2pt}
\boldmath{RI} & 0.8385 & 0.7271 & 0.7835 & 0.7340 & 0.7455 &0.6698\\
\boldmath{Purity}  & 0.8671 & 0.7203 & 0.8015 & 0.7343 & 0.7343 &0.5874\\
\boldmath{NMI}  & 0.6422 & 0.5040 & 0.6095 & 0.5412& 0.5651 &0.4519\\
\specialrule{0em}{2pt}{2pt}
\hline
\hline
\end{tabular}}
\vspace{-0.2cm}
\end{table}
As depicted in Fig. \ref{fig5}, we obtained the ground-truth clusters of frames labeled by three dancing postures (i.e.,
moving arms, stretching legs, and bending body). 
For performance evaluation, we performed $k$-means clustering on the recovered low-rank component and compared the classification results.
According to our experimental results on synthetic data, 
the effectiveness of the low-rank component estimation 
depends on the quality of the learned graph. Therefore,
clustering performance on the low-rank component
reflects the graph learning performance.
We used 
\emph{Purity}, \emph{NMI}, and \emph{RI} \cite{ref48} to make a quantitative evaluation of clustering results.



We compared the clustering performance of the proposed GL-LRSS with two
GSP-based methods (i.e., PCAG and RPCAG) both having a predefined
five-nearest-neighbor graph.
We also applied $k$-means clustering to the original data for reference.
The classification results are listed
in Table \ref{tab:3}. 
The proposed
GL-LRSS achieves the highest \emph{RI} score of 0.8385,
greater than the scores of 0.7271, 0.7835, 0.7340, and 0.7455 obtained by
GL-Sigrep, LGE, PCAG, and RPCAG, respectively.
Similar results were obtained in terms of \emph{Purity} and \emph{NMI}.
As expected,
the performance of k-means clustering on the original dataset is the worst,
possibly due to
its susceptibility to noise.
These results
demonstrate that the proposed GL-LRSS provides superior performance
compared with the comparison methods on the dancer mesh dataset.

\vspace{-0.3cm}
\subsection{Graph learning from temperature dataset}
The daily average temperature data is collected
from 60 observation sites in China \cite{ref45} over 150 days starting from
January 1, 2017
and have a size of $60 \times 150$.
We aimed at
learning a graph structure for uncovering
the inherent relationship between these observation
sites in terms of daily temperature variations. 
In the experiment, we did not have an available ground-truth
graph.
Additionally, a $k$-nearest-neighbor graph was inappropriate. 
However, four climate zones of China (i.e., northern, southern, northwest, Qinghai-Tibet)
could be regarded as a ground-truth clustering of the observation sites,
which are differentiated by colors in Fig. \ref{Fig6:a}.
For performance evaluation, 
we compared the clustering results of the proposed and baseline methods by applying spectral clustering \cite{ref49}, 
which utilizes
the learned graph Laplacian to divide the observation sites
into four disjoint clusters. The clustering performance can reflect the quality of a graph.

\begin{table}[t]
\newcommand{\tabincell}[2]{\begin{tabular}{@{}#1@{}}#2\end{tabular}}
\centering
\caption{\label{tab:4}The performance of graph learning methods in recovering ground-truth clusters of temperature measuring stations.}
\setlength{\tabcolsep}{3.9mm}{
\begin{tabular}{c c c c }
\hline
\hline
\specialrule{0em}{2pt}{2pt}
\tabincell{c}{} & RI & Purity &
NMI  \\
\specialrule{0em}{2pt}{2pt}
\hline
\specialrule{0em}{2pt}{2pt}
KNN & 0.7567 & 0.6667 &  0.4855\\
GMS & 0.7667 & 0.5833 &  0.5037\\
GL-logdet  & 0.7411 & 0.6667 & 0.4701 \\
SpecTemp  & 0.7832 & 0.5833 & 0.5201 \\
GL-Sigrep  & 0.79 & 0.7167 & 0.5397 \\
LGE  & 0.7833 & 0.75 & 0.5236 \\
GL-LRSS (${\textit{R}}_I$)  & 0.8633 & 0.85 & 0.7203 \\
GL-LRSS (${\textit{R}}_{prior}$)  & 0.8656 & 0.8333 & 0.7352 \\
\specialrule{0em}{2pt}{2pt}
\hline
\hline
\end{tabular}}
\vspace{-0.3cm}
\end{table}

Fig. \ref{Fig6:b} and \ref{Fig6:c} show the four-cluster
partitions and graph topology obtained from
the proposed GL-LRSS(${\textit{R}}_I$). The clusters are clearly
distinguishable and close to the ground truth in Fig. \ref{Fig6:a}.
For comparison, we also show the natural choice of the graph constructed
using eight nearest neighbors\footnote{For the KNN baseline,
we choose $k=8$ for both the temperature and ETo datasets, which leads
to the best performance (i.e., \emph{RI} score).}
in Fig. \ref{Fig6:d}. 
The resulting graph seems inaccurate because it only considers the
physical distance regardless of other influencing factors (e.g., altitude).
Consequently,
observation sites that are geometrically
close may be geographically distant.
Table \ref{tab:4} lists the best \emph{RI}, \emph{Purity}, and \emph{NMI} values achieved by the evaluated graph learning algorithms. Compared with the baseline
methods, the proposed GL-LRSS achieves the highest scores for all three evaluation measures.
Moreover,
by properly using the prior information of ${\textit{R}}$\footnote{The parameter $c_i$ in matrix ${\textit{R}}$ can be viewed as autocorrelation coefficient of data
in the $i$th observation site. Here,
we obtain $c_i$ in advance through the autocorrelation function (ACF) (i.e., function [acf,lags]=autocorr(x)). },
GL-LRSS (${\textit{R}}_{prior}$) outperforms
GL-LRSS(${\textit{R}}_I$).
Hence, the proposed GL-LRSS
outperforms the baseline methods for
learning the graph topology on the temperature data from China.
\vspace{-0.4cm}
\subsection{Graph learning from evapotranspiration dataset}
\begin{figure}[h]
\centering
\subfigure[]{ \label{Fig7:a}
\includegraphics[width=1.6in]{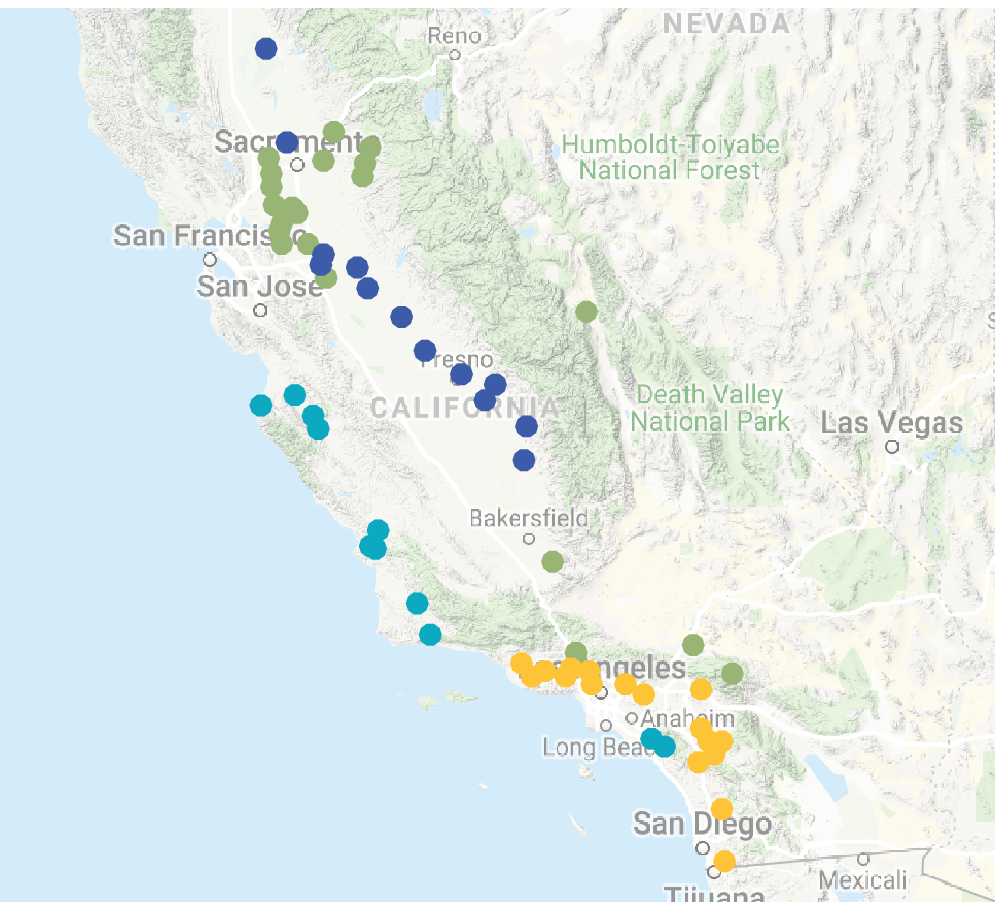}}
\hspace{-0.05in}
\subfigure[]{ \label{Fig7:b}
\includegraphics[width=1.6in]{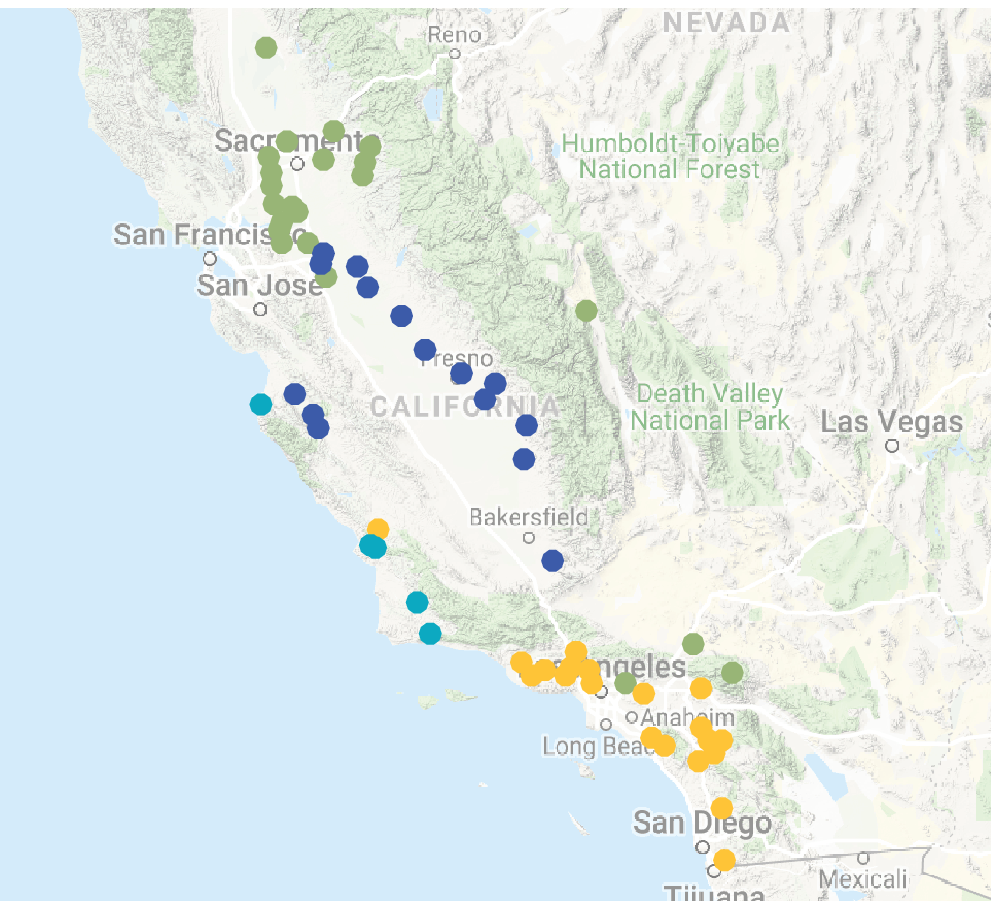}}
\caption{(a) The ground-truth clusters of 62 observation sites in California.
The colors from green, blue, cyan-blue to yellow represent
ETo zone 14, zone 12, zone 6 and zone 9, respectively. (b) The
resulting clusters obtained by the proposed GL-LRSS(${\textit{R}}_I$).}
\label{Fig7}
\vspace{-0.2cm}
\end{figure}

We now move onto the final real-world dataset,
California daily evapotranspiration (ETo) dataset,
published by the California Department of Water Resources \cite{ref46}.
It is collected from 62 active observation sites over
150 days starting from February 1, 2018,
and it contains data with a size
of $62 \times 150$.
We aimed at 
inferring a graph that captures the similarity in the evapotranspiration evolution of different observation stations.
In this experiment, 
because we did not have an obvious ground-truth graph topology,
we used an ETo Zone Map \cite{ref50} as reference,
which divides the 62 observation sites into one of four zones.
The ground-truth clusters are shown in Fig. \ref{Fig7:a}.
Similar to the previous example, we applied spectral clustering to the learned graphs and compared
the resulting clusters with the ground truth.
\begin{table}[t]
\newcommand{\tabincell}[2]{\begin{tabular}{@{}#1@{}}#2\end{tabular}}
\centering
\caption{\label{tab:5}The performance of graph learning methods in recovering ground-truth clusters of ETo measuring stations.}
\setlength{\tabcolsep}{3.9mm}{
\begin{tabular}{c c c c }
\hline
\hline
\specialrule{0em}{2pt}{2pt}
\tabincell{c}{} & RI & Purity &
NMI  \\
\specialrule{0em}{2pt}{2pt}
\hline
\specialrule{0em}{2pt}{2pt}
KNN & 0.7644 & 0.6613 &  0.4805\\
GMS & 0.7685 & 0.6774 &  0.5113\\
GL-logdet  & 0.7612 & 0.6290 & 0.4613 \\
SpecTemp  & 0.7653 & 0.6451 & 0.4799 \\
GL-Sigrep  & 0.8065 & 0.7419 & 0.5865 \\
LGE  & 0.8153 & 0.7903 & 0.5945 \\
GL-LRSS (${\textit{R}}_I$)  & 0.8496 & 0.8225 & 0.6544 \\
GL-LRSS (${\textit{R}}_{prior}$)  & 0.8486 & 0.8064 & 0.6462 \\
\specialrule{0em}{2pt}{2pt}
\hline
\hline
\end{tabular}}
\vspace{-0.2cm}
\end{table}

Fig. \ref{Fig7:b} shows the clustering results of the proposed
GL-LRSS(${\textit{R}}_I$). 
The clusters obtained from the proposed GL-LRSS are visually similar to the ground-truth clusters.
The corresponding
quantitative results are listed in Table \ref{tab:5}.
Compared with the GSP-based methods (i.e., GL-Sigrep, GL-Logdet,
SpecTemp, and LGE) and the other baseline methods,
the proposed GL-LRSS(${\textit{R}}_I$) achieves the highest scores of
0.8496 for \emph{RI}, 0.8255 for \emph{Purity}, and
0.6544 for \emph{NMI}.
Nevertheless, the superior performance of GL-LRSS(${\textit{R}}_{prior}$)
is not obvious, possibly because the correlation
coefficients are inaccurate for the ETo data. Therefore,
the proposed method exhibits higher
performance than the comparison methods
on the ETo dataset.
\section{discussion}
In this section, we first clarify the difference of the proposed method with methods for vector autoregressive models. Then, we discuss the
applicability of the proposed method for
a nondiagonal case of ${\textit{R}}$ in our model.
\vspace{-0.4cm}
\subsection{Comparison with methods for vector autoregressive models}
The model given by Eqs. (\ref{representation}) and (\ref{latent}) is analogous to a noisy version of a first-order vector autoregressive model. The differences between these models are twofold.

First, the driving noise in vector autoregression is assumed to be Gaussian white noise, whereas ${{\bf{v}}_t}$ in (\ref{latent}) follows a GMRF encoding  graph structure in the covariance matrix. Benefitting from the vector autoregressive model with driving noise having a covariance following the GMRF,
the proposed model in (\ref{latent}) can thus characterize the spatial and temporal structures of the spatiotemporal signal through ${{\bf{v}}_t}$ and ${\textit{R}}$, respectively.

Second, the transition matrices in the vector autoregressive model can be viewed as weighted graphs that show Granger causal connections between nodes, but they are usually unknown. Thus, vector autoregression methods (e.g., those proposed in \cite{ref32} and \cite{ref60}) aim to recover multiple transition matrices from observations. In contrast, transition matrix ${\textit{R}}$ in (\ref{latent}) is known. As mentioned, because the graph structure is encoded in the covariance of ${{\bf{v}}_t}$, our graph learning problem can be regarded as inverse covariance matrix estimation.

\subsection{ 
Applicability of our method for a nondiagonal ${\textit{R}}$}
In this work, we assume all measuring nodes to be independent for simplicity, that is, considering the dependencies of data in its own measuring node. Thus, we use a diagonal transition matrix ${\textit{R}}$ in the proposed model. In some cases, there exists dependencies between measuring nodes and hence the transition matrix is nondiagonal. 
However, 
under certain conditions of nondiagonal ${\textit{R}}$, we can also derive
the same optimization problem in (\ref{MAP}) by Proposition 1 below; thus, the proposed method can be also applied in those conditions.
\begin{proposition}
If state transition matrix
${\textit{R}}$ in (\ref{latent}) is real and symmetric but not necessarily diagonal, 
the eigendecomposition of ${\textit{R}}$ is denoted as ${\textit{R}}={\bf{Q\Lambda }}{{\bf{Q}}^T}$, and
the model given by (\ref{global representation}) and (\ref{latent}) can be transformed into an equivalent model by multiplying ${{\bf{Q}}^T}$ on both sides of the equation. The solution to (\ref{MAP}) still represents the MAP estimate of the process variable in the equivalent model.
\end{proposition}
\begin{myproof}
As matrix ${\textit{R}}$ is real and symmetric,
the eigendecomposition of ${\textit{R}}$ is denoted as
\begin{small}${\textit{R}} = {\bf{Q\Lambda }}{{\bf{Q}}^T}$\end{small}.
We reformulate the model in (\ref{global representation}) and (\ref{latent})
by multiplying matrix \begin{small}${{\bf{Q}}^T}$\end{small} as
\begin{eqnarray} \label{Representation}
\resizebox{0.23\hsize}{!}{$
\begin{aligned}
{\tilde{{\bf{y}}_t}} = {\tilde{{\bf{x}}_t}} + {\tilde{{\bf{n}}_t}},
\end{aligned} $}
\end{eqnarray}
\vspace{-0.85cm}
\begin{eqnarray} \label{Latent}
\resizebox{0.28\hsize}{!}{$
\begin{aligned}
{\tilde{{\bf{x}}_t}} = {\bf{\Lambda}}{\tilde{{\bf{x}}}_{t-1}} + {\tilde{{\bf{v}}_t}},
\end{aligned} $}
\end{eqnarray}
where \begin{small}${\tilde{{\bf{y}}_t}}={{\bf{Q}}^T}{{\bf{y}}_t}$\end{small},
\begin{small}${\tilde{{\bf{x}}_t}}={{\bf{Q}}^T}{{\bf{x}}_t}$\end{small},
\begin{small}${\tilde{{\bf{n}}_t}}={{\bf{Q}}^T}{{\bf{n}}_t}$\end{small} and
\begin{small}${\tilde{{\bf{v}}_t}}={{\bf{Q}}^T}{{\bf{v}}_t}$\end{small}.
Based on the definition of ${{\bf{n}}_t}$ and ${{\bf{v}}_t}$,
we have that \begin{small}${\tilde{\bf{n}}_t} \sim {\mathcal {N}}\left( {{\bf{0}},{\sigma _n}^2{{\bf{I}}_N}} \right)$\end{small} and
\begin{small}${\tilde{\bf{v}} _t} \sim {\mathcal N}\left(
{\bf{0}}, {{\bf{Q}}^T}{{{\tilde {\bf{L}}}^\dag }}{{\bf{Q}}} \right)$\end{small}.

Note that model (\ref{latent})
with a non-diagonal transition matrix
can be transformed into an equivalent model (\ref{Latent}) with a diagonal
transition matrix.

Given the weighted difference signal
\begin{small}${\tilde{\bf{d}}_t} = {\tilde{\bf{y}}_t} - {\bf{\Lambda}}{\tilde{\bf{y}}_{t - 1}}
={\tilde{\bf{v}} _t}+{\tilde{\bf{n}}_t}-{\bf{\Lambda}}{\tilde{\bf{n}}_{t - 1}}$\end{small} and the multivariate Gaussian distribution on
${{\tilde{\bf{v}}}_t}$,
we can compute
the MAP estimate of ${{\tilde{\bf{v}}}_t}$ as follows: 
\begin{eqnarray} \label{2}
\resizebox{0.85\hsize}{!}{$
\begin{aligned}
\tilde{{\bf{v}}_t}_{MAP}\left( {{\tilde{\bf{d}}_t}} \right)
&: = \arg \mathop {\max }\limits_{{\tilde{\bf{v}}_t}} p\left( {{\tilde{\bf{v}}_t}|{\tilde{\bf{d}}_t}} \right) = \arg \mathop {\max }\limits_{{\tilde{\bf{v}}_t}} p\left( {{\tilde{\bf{d}}_t}|{\tilde{\bf{v}}_t}} \right)p\left( {{\tilde{\bf{v}}_t}} \right)\\
& = \arg \mathop {\min }\limits_{{\tilde{\bf{v}}_t}}  { - \log {p_E}\left( {{\tilde{\bf{d}}_t} - {\tilde{\bf{v}}_t}} \right) - \log {p_V}\left( {{\tilde{\bf{v}}_t}} \right)}\\
& = \arg \mathop {\min }\limits_{{{\bf{v}}_t}} {\kern 1pt} {\kern 1pt} {\kern 1pt} {\kern 1pt} {\left( {{\tilde{\bf{d}}_t} - {\tilde{\bf{v}}_t}} \right)^T}{{\bf{W}}^{ - 1}}\left( {{\tilde{\bf{d}}_t} - {\tilde{\bf{v}}_t}} \right) + \alpha {\tilde{\bf{v}}_t}^T{{\bf{Q}}^T}{\tilde {\bf{L}}}{{\bf{Q}}}{\tilde{\bf{v}}_t},
\end{aligned} $}\nonumber
\end{eqnarray}
where \begin{small}${\bf{W}}={{{\bf{I}}_N} + {{\bf{\Lambda}}{\bf{\Lambda}}^T}}$\end{small}.
By leveraging the inequality in (\ref{inequality}), we obtain a relaxed optimization problem:
\begin{eqnarray}\label{4}
\resizebox{0.5\hsize}{!}{$
\begin{aligned}
\mathop {\min }\limits_{{\tilde{\bf{v}}_t}} \left\| {{\tilde{\bf{d}}_t} - {\tilde{\bf{v}}_t}} \right\|_2^2 + \alpha {\tilde{\bf{v}}_t}^T{{\bf{Q}}^T}{\tilde {\bf{L}}}{{\bf{Q}}}{\tilde{\bf{v}}_t}.
\end{aligned} $}
\end{eqnarray}

Specifically, the first and second
terms in (\ref{4}) can be rewritten as \begin{small}${{\bf{Q}}^T}\left({{{\bf{d}}_t} - {{\bf{v}}_t}}\right)$\end{small} and \begin{small}${{\bf{v}}_t}^T{\tilde {\bf{L}}}{{\bf{v}}_t}$\end{small},
respectively. Using the inequality
\begin{small}$\left\|{\bf{Q}}\right\|_2^2 \left\| {{\bf{Q}}^T}\left({{{\bf{d}}_t} - {{\bf{v}}_t}}\right) \right\|_2^2 \ge
\left\|{{{\bf{d}}_t} - {{\bf{v}}_t}}\right\|_2^2$\end{small},
we further simplify 
the optimization problem (\ref{4}) as
\begin{eqnarray}\label{41}
\resizebox{0.45\hsize}{!}{$
\begin{aligned}
\mathop {\min }\limits_{{{\bf{v}}_t}} \left\| {{{\bf{d}}_t} - {{\bf{v}}_t}} \right\|_2^2 + \alpha {{\bf{v}}_t}^T{\tilde {\bf{L}}}{{\bf{v}}_t}.
\end{aligned} $}
\end{eqnarray}
Therefore, the proof is complete.
\end{myproof}

\section{Conclusion}
We study the problem of graph learning for spatiotemporal signals
with long- and short-term correlations.
By leveraging the spatiotemporal smoothness that reflects the temporal and graph structural information, as well as the low-rank property of the spatiotemporal signal,
we
formulate graph learning as a joint problem of estimating
low-rank components and the graph topology.
A new graph learning method, GL-LRSS, is proposed by
applying alternating minimization and the ADMM 
to solve the formulated problem. 
These two optimization strategies improve each other,
fostering better graph learning.
Experiments on synthetic datasets
verify a substantial performance improvement of the proposed GL-LRSS over state-of-the-art graph learning and low-rank component estimation methods. In addition,
experiments on three real-world datasets demonstrate
that the proposed GL-LRSS outperforms the baseline methods in practice.
In our future work, we plan to study a more general signal model with 
an arbitrary
transition matrix and explore effective graph learning approaches.

\appendices
\section{Derivation of the closed-form solution in (\ref{Uniquesolution})} \label{AppendixA}
Being prepared for the following analysis, we
first introduce the property of the vec-operator as follows:
\begin{eqnarray} \label{vecopertor}
\resizebox{0.45\hsize}{!}{$
\begin{aligned}
{\rm{tr}}\left( {{{\bf{A}}^T}{\bf{B}}} \right) = {\rm{vec}}{\left( {\bf{A}} \right)^T}{\rm{vec}}\left( {\bf{B}} \right).
\end{aligned} $}
\end{eqnarray}
Then the second term in (\ref{admm21rewitten}) can be transformed as
\begin{eqnarray} \label{devariation}
\resizebox{0.8\hsize}{!}{$
\begin{aligned}
&{\rm{tr }}\left( {\mathcal{D}{{\left( {\bf{X}} \right)}^T}{\bf{L}}\mathcal{D}\left( {\bf{X}}  \right)} \right)
= {\rm{vec}}\left( {{\bf{X}}  - {\bf{R}}{\bf{X}} {\bf{B}}} \right)^T{\rm{vec}}\left[ {{\bf{L}}\left( {{\bf{X}}  - {\bf{R}}{\bf{X}} {\bf{B}}} \right)} \right]\\
&= \left[ {{\rm{vec}}\left( {\bf{X}}  \right)^T - {\rm{vec}}\left( {\bf{X}}  \right)^T\left( {{\bf{B}} \otimes {\bf{R}}} \right)} \right]\cdot \\
&\quad\quad\quad\quad\quad\quad\quad\quad\left[ {\left( {{{\bf{I}}_M} \otimes {\bf{L}}} \right){\rm{vec}}\left( {\bf{X}}  \right)- \left( {{{\bf{B}}^T} \otimes {\bf{L}}{\bf{R}}} \right){\rm{vec}}\left( {\bf{X}}  \right)} \right]\\
& = {\rm{vec}}\left( {\bf{X}}  \right)^T{{\bf{T}}_d}\left( {{{\bf{I}}_M} \otimes {\bf{L}}} \right){{\bf{T}}_d}^T{\rm{vec}}\left( {\bf{X}}  \right).
\end{aligned} $}\nonumber
\end{eqnarray}
Similarly, the first term in (\ref{admm21rewitten})
is given by
\begin{eqnarray} \label{devariation1}
\resizebox{1\hsize}{!}{$
\begin{aligned}
\left\| {\mathcal{D}\left( {{\bf{X}}  - {\bf{Y}}} \right)} \right\|_F^2
&= {\rm{tr}}\left( {\mathcal{D}{{\left( {{\bf{X}}  - {\bf{Y}}} \right)}^T}\mathcal{D}\left( {{\bf{X}}  - {\bf{Y}}} \right)} \right)= {\rm{vec}}\left( {{\bf{X}}  - {\bf{Y}}} \right)^T{{\bf{T}}_d}{{\bf{T}}_d}^T{\rm{vec}}\left( {{\bf{X}}  - {\bf{Y}}} \right),
\end{aligned} $}\nonumber
\end{eqnarray}
and thus the objective function in problem (\ref{admm21rewitten}) can be equivalently written as
\begin{eqnarray} \label{devariation2}
\resizebox{0.9\hsize}{!}{$
\begin{aligned}
{{\tilde f}_X}\left( {\boldsymbol{\upsilon }} \right)
&=
\left( {{{\boldsymbol{\upsilon }}^T} - {\rm{vec}}{{\left( {\bf{Y}} \right)}^T}} \right){{\bf{T}}_d}{{\bf{T}}_d}^T\left( {{\boldsymbol{\upsilon }} - {\rm{vec}}\left( {\bf{Y}} \right)} \right) + \alpha {{\boldsymbol{\upsilon }}^T}{\bf{G}}{\boldsymbol{\upsilon }}\\
&+\frac{\rho }{2}\left[ {{{\boldsymbol{\upsilon }}^T} - {\rm{vec}}{{\left( {\bf{P}} \right)}^T} + {{{\rm{vec}}{{\left( {\bf{Q}} \right)}^T}} \mathord{\left/
 {\vphantom {{{\rm{vec}}{{\left( {\bf{Q}} \right)}^T}} \rho }} \right.
 \kern-\nulldelimiterspace} \rho }} \right]\left[ {{\boldsymbol{\upsilon }} - {\rm{vec}}\left( {\bf{P}} \right) + {{{\rm{vec}}\left( {\bf{Q}}\right)} \mathord{\left/
 {\vphantom {{{\rm{vec}}\left( {\bf{Q}}\right)} \rho }} \right.
 \kern-\nulldelimiterspace} \rho }} \right],
\end{aligned} $}\nonumber
\end{eqnarray}
where
\begin{small}${\bf{G}}={{\bf{T}}_d}\left( {{{\bf{I}}_M} \otimes {\bf{L}}} \right){{\bf{T}}_d}^T\in {{\mathbb {R}}^{NM \times NM}}$\end{small}
and
\begin{small}${\boldsymbol{\upsilon }}={\rm{vec}} \left({\bf{X}}\right)$\end{small}. The gradient of \begin{small}${{\tilde f}_X}\left( {\boldsymbol{\upsilon }} \right)$\end{small} can
be deduced as
\begin{eqnarray} \label{pfvector}
\resizebox{0.7\hsize}{!}{$
\begin{aligned}
\nabla {{\tilde f}_X}\left( {\boldsymbol{\upsilon }} \right) &=
2{\kern 1pt}{{\bf{T}}_d}{{\bf{T}}_d}^T{\boldsymbol{\upsilon }} - 2{\kern 1pt}{{\bf{T}}_d}{{\bf{T}}_d}^T{\rm{vec}}\left( {\bf{Y}} \right) + 2{\kern 1pt}\alpha {\bf{G}}{\boldsymbol{\upsilon }}\\
&+{\rm{vec}}\left( {\bf{Q}} \right)+\rho{\kern 1pt}{\boldsymbol{\upsilon }}-\rho{\kern 1pt}
{\rm{vec}}\left( {\bf{P}} \right).
\end{aligned} $}
\end{eqnarray}
The proof is accomplished by setting (\ref{pfvector})
to zero.


\begin{thebibliography}{1}

\bibitem{ref1}
A. R. McIntosh, W. K. Chau, A. B. Protzner, ``Spatiotemporal analysis of event-related fMRI data using partial least squares,''\emph{ Neuroimage}, vol. 23, no. 2, pp. 764-775, 2004.

\bibitem{ref2}
I. Kompatsiaris, and M. Strintzis, ``Spatiotemporal Segmentation and Tracking of Objects for Visualization of Videoconference Image Sequences," \emph{IEEE Transactions on Circuits and Systems for Video Technology,} vol. 10, no. 8, pp. 1388-1402, 2000.

\bibitem{ref3}
H. Pham, C. Shahabi, Y. Liu, ``EBM: an entropy-based model to infer social strength from spatiotemporal data,'' in \emph{Proceedings of the 2013 ACM SIGMOD International Conference on Management of Data}, pp. 265-276, 2013.

\bibitem{ref4}
N. Eckert, E. Parent, R. Kies, H. Baya, ``A spatio-temporal modelling framework for assessing the fluctuations of avalanche occurrence resulting from climate change: application to 60 years of data in the northern French Alps,'' \emph{Climatic Change},
vol. 101, no. 3, pp. 515-553, 2010.

\bibitem{ref5}
A. Sandryhalia, and J. M. F. Moura, ``Big data analysis with signal processing on graphs: Representation and processing of massive data sets with irregular structure,'' \emph{IEEE Signal Process. Mag.}, vol. 31, no. 5, pp. 80-90, 2014.

\bibitem{ref37}
D. I. Shuman, S. K. Narang, P. Frossard, A. Ortega, and P. Vandergheynst, ``The emerging field of signal processing on graphs: Extending high-dimensional data analysis to networks and other irregular domains,'' \emph{IEEE Signal Process. Mag.}, vol. 30, no. 3, pp. 83-98, May. 2013.


\bibitem{ref8}
X. Zhu and M. Rabbat, ``Approximating signals supported on graphs,'' in \emph{Proc. 37th IEEE Int. Conf. Acoust., Speech, Signal Process.}, 2012, pp. 3921-3924.


\bibitem{ref10}
D. Romero, M. Ma, and G. B. Giannakis, ``Kernel-based reconstruction of graph signals,'' \emph{IEEE Trans. Signal Process.}, vol. 65, no. 3, pp. 2547-2560, May. 2017.

\bibitem{ref6}
A. Sandryhalia, and J. M. F. Moura, ``Discrete signal processing on graphs: Graph filters,'' in \emph{Proc. 38th IEEE Int. Conf. Acoust., Speech, Signal Process.}, IEEE, 2013, pp. 6163-6166.

\bibitem{ref112}
J. Cheng, Q. Ye, H. Jiang, D. Wang, and C. Wang, ``STCDG:An efficient data gathering algorithm based on matrix completion for wireless sensor networks,'' \emph{IEEE Transactions on Wireless Communications}, vol. 12, no. 2, pp. 850-861, 2013.

\bibitem{ref13}
X. Mao, K. Qiu, T. Li, and Y. Gu, ``Spatio-Temporal Signal Recovery Based on Low Rank and Differential Smoothness,'' \emph{IEEE Trans. Signal Process.}, vol. 66, no. 23, pp. 6281 - 6296, Dec. 2018.

\bibitem{ref35}
L. Rui, H. Nejati, S. H. Safavi, and N. M. Cheung, ``Simultaneous low rank component and graph estimation for high-dimensional graph signals: Application to brain imaging,'' in \emph{Proc. 42th IEEE Int. Conf. Acoust., Speech, Signal Process.}, IEEE, 2017, pp. 4134-4138.

\bibitem{ref24}
Y. Liu, L. Yang, K. You, W. Guo and W. Wang, ``Graph learning based on spatiotemporal smoothness for time-varying graph signal,'' \emph{IEEE ACCESS}, vol. 7, pp. 62372 - 62386, 2019.

\bibitem{ref20}
X. Dong, D. Thanou, P. Frossard, and P. Vandergheynst, ``Learning Laplacian matrix in smooth graph signal representations,'' \emph{IEEE Trans. Signal Process.}, vol. 64, no. 23, pp. 6160-6173, Dec. 2016.

\bibitem{ref21}
V. Kalofolias, ``How to learn a graph from smooth signals,'' in \emph{Proc. 19th Int. Conf. Artif. Intell. Statist.}, pp. 920-929, May. 2016.

\bibitem{ref16}
N. Shahid, V. Kalofolias, X. Bresson, M. Bronstein, and P. Vandergheynst, ``Fast robust PCA on graphs,'' \emph{IEEE J. Sel. Topics Signal Process.}, vol. 10, no. 4, pp. 740-756, Jun. 2016.

\bibitem{ref9}
X. Dong, D. Thanou, M. Rabbat, and P. Frossard, ``Learning
graphs from data: A signal representation perspective,'' \emph{IEEE
Signal Process. Mag.,} vol. 36, no. 3, pp. 44-63, May. 2019.

\bibitem{ref54}
G. Mateos, S. Segarra, A. G. Marques, and A. Ribeiro, ``Connecting the dots: Identifying network
structure via graph signal processing,'' \emph{IEEE Signal Process.
Mag.}, vol. 36, no. 3, pp. 16-43, May. 2019.

\bibitem{ref11}
M. T. Bahadori, Q. R. Yu, and Y. Liu, ``Fast multivariate spatio-temporal analysis
via low rank tensor learning,'' in \emph{Advances in neural information processing systems}, 2014,
pp. 3491-3499.

\bibitem{ref12}
X. Piao, Y. Hu, Y. Sun, B. Yin, and J. Gao, ``Correlated spatio-temporal data collection in wireless sensor networks based on low rank matrix approximation and optimized node sampling,'' \emph{Sensors}, vol. 14, no. 12, pp. 23137-23158, 2014.



\bibitem{ref14}
B. Jiang, C. Ding, and J. Tang, ``Graph-laplacian PCA: Closed-form solution and robustness,'' in \emph{Proc. IEEE Conf. Comput. Vis. Pattern Recog. (CVPR)}, 2013, pp. 3492-3498.

\bibitem{ref15}
N. Shahid, V. Kalofolias, X. Bresson, M. Bronstein, and P. Vandergheynst, ``Robust principal component analysis on graphs,'' in \emph{Proceedings of International Conference on Computer Vision}, Santiago, Chile, 2015, pp. 2812-2820.


\bibitem{ref53}

A. Jung, ``Learning the Conditional Independence Structure of Stationary Time Series: A Multitask Learning Approach,'' in \emph{IEEE Transactions on Signal Processing}, vol. 63, no. 21, pp. 5677-5690, Nov. 1, 2015.

\bibitem{ref17}
J. Friedman, T. Hastie, and R. Tibshirani, ``Sparse inverse covariance estimation with the graphical lasso,'' \emph{Biostatistics.}, vol. 9, no. 3, pp. 432-441, 2008.

\bibitem{ref18}
R. Mazumder and T. Hastie, ``Exact covariance thresholding into connected components for large-scale graphical lasso,'' \emph{J. Mach. Learn. Res.}, vol. 13, no. 1, pp. 781-794, 2012.

\bibitem{ref19}
B. M. Lake and J. B. Tenenbaum, ``Discovering structure by learning sparse graph,'' in \emph{Proc. 33rd Annual Cognitive Science conf.}, 2010, pp. 778-783.


\bibitem{ref22}
H. E. Egilmez, E. Pavez, and A. Ortega, ``Graph learning from data under structural and laplacian constraints,'' \emph{IEEE J. Sel. Topics Signal Process.}, vol. 11, no. 6, pp. 825-841, Sept. 2017.

\bibitem{ref23}
S. P. Chepuri, S. Liu, G. Leus, and A. O. Hero, ``Learning sparse graphs under smoothness prior,'' in \emph{Proc. 42th IEEE Int. Conf. Acoust., Speech, Signal Process.}, IEEE, 2017, pp. 6508-6512.

\bibitem{ref25}
M. G. Rabbat, ``Inferring sparse graphs from smooth signals with theoretical guarantees,'' in \emph{Proc. 42th IEEE Int. Conf. Acoust., Speech, Signal Process.}, IEEE, 2017, pp. 6533-6537.

\bibitem{ref26}
V. Kalofolias, A. Loukas, D. Thanou, and P. Frossard, ``Learning time varying graphs,'' in \emph{Proc. 42th IEEE Int. Conf. Acoust., Speech, Signal Process.}, IEEE, 2017, pp. 2826-2830.

\bibitem{ref27}
Yamada, Koki, Yuichi Tanaka, and Antonio Ortega. ``Time-varying Graph Learning Based on Sparseness of Temporal Variation.'' in \emph{Proc. 44th IEEE Int. Conf. Acoust., Speech, Signal Process.}, IEEE, 2019,
pp. 5411-5415.

\bibitem{ref28}
S. Segarra, A. G. Marques, G. Mateos, and P. Vandergheynst, ``Network topology inference from spectral templates,'' \emph{IEEE Trans. Signal Inf. Process. Netw.}, vol. 3, no. 3, pp. 467-483, Sept. 2017.

\bibitem{ref29}
B. Pasdeloup, V. Gripon, G. Mercier, D. Pastor, and M. G. Rabbat, ``Characterization and inference of graph diffusion processes from observations of stationary signals,'' \emph{IEEE Trans. Signal Inf. Process. Netw.}, vol. PP, no. 99, pp. 1-16, 2017.

\bibitem{ref30}
R. Shafipour, S. Segarra, A. G. Marques, and G. Mateos, ``Identifying undirected network structure via semidefinete relaxation,'' in \emph{Proc. 43th IEEE Int. Conf. Acoust., Speech, Signal Process.}, IEEE, 2018, pp. 4049-4053.

\bibitem{ref31}
D. Thanou, X. Dong, D. Kressner, and P. Frossard, ``Learning heat diffusion graphs,'' \emph{IEEE Trans. Signal Inf. Process. Netw.}, vol. 3, no. 3, pp. 484-499, Sept, 2017.

\bibitem{ref32}
J. Mei and J. M. F. Moura, ``Signal processing on graphs: Causal modeling of unstructured data,'' \emph{IEEE Trans. Signal Process.}, vol. 65, no. 8, pp. 2077-2092, Apr. 2017.

\bibitem{ref33}
B. Baingana and G. B. Giannakis, ``Tracking switched dynamic network topologies from information cascades,'' \emph{IEEE Trans. Signal Process.}, vol. 65, no. 4, pp. 985-997, Feb. 2017.

\bibitem{ref34}
Y. Shen, B. Baingana, and G. B. Giannakis, ``Topology inference of directed graphs using nonlinear structural vector autoregressive models,'' in \emph{Proc. 42th IEEE Int. Conf. Acoust., Speech, Signal Process.}, IEEE, 2017, pp. 6513-6517.


\bibitem{ref36}
E. J. Cand{\`e}s, X. Li, Y. Ma, and J. Wright, ``Robust principal component analysis,'' \emph{Jouranl of the ACM}, vol. 58, no. 3, pp. 1-37, May, 2011.


\bibitem{ref38}
N. Cressie and C. K. Wikle, \emph{Statistics for spatio-temporal data}. John Wiley \& Sons, 2011.

\bibitem{ref51}
S. Li, K. Li, and Y. Fu, ``Temporal subspace clustering for human motion segmentation,'' \emph{in Proceedings of the IEEE
International Conference on Computer Vision (ICCV)}, pp. 4453-4461, 2015.

\bibitem{ref52}
F. Grassi, A. Loukas, N. Perraudin, and B. Ricaud, ``Scalable processing and meaningful representations
for time-series on graphs,'' \emph{IEEE Trans. Signal Process.}, vol. 66, no. 3, pp. 817-829, 2018.


\bibitem{ref60}
A. Bolstad, B. D. Van Veen, and R. Nowak, ``Causal network inference via group sparse regularization,'' \emph{IEEE Trans. Signal Process.}, vol. 59, no. 6, pp. 2628-2641, 2011.


\bibitem{ref39}
S. Boyd, N. Parikh, E. Chu, B. Peleato, and J. Eckstein, ``Distributed optimization and statistical learning via the alternating direction method of multipliers,'' \emph{Foundations and Trends in Machine Learning}, vol. 3, no. 1, pp. 1-122, 2011.

\bibitem{ref40}
J. R. Shewchuk, ``An introduction to the conjugate gradient method
without the agonizing pain,'' Pittsburgh, PA, USA, Tech. Rep., 1994.

\bibitem{ref41}
S. Boyd and L. Vandenberghe, \emph{Convex optimization}. New York, NY, USA: Cambridge University Press, 2004.

\bibitem{ref42}
J. Cai, E. J. Cand{\`e}s, and Z. Shen, ``A singular value thresholding algorithm
for matrix completion,'' \emph{SIAM J. on Optimization}, vol. 20, no. 4, pp. 1956-1982, 2010.

\bibitem{ref43}
G. H. Golub and C. F. Van Loan, \emph{Matrix Computations}. The Johns Hopkins University Press,
Baltimore, MD, USA, third edition, 1996.

\bibitem{ref44}
Dynamic mesh for a dancing man, [Online]. Available: https://epfl-lts2.github.io/rrp-html/start.html.

\bibitem{ref45}
National Centers for Environmental Information, 2018, [Online]. Available: https://www.ncdc.noaa.gov/cdo-web/datasets.

\bibitem{ref46}
California Irrigation Management Infromation System, 2018, [Online]. Available: https://cimis.water.ca.gov.

\bibitem{ref47}
C. D. Manning, P. Raghavan, and H. Schutze, \emph{Introduction to infromation retrieval.} Cambridge University Press, 2008.

\bibitem{ref48}
W. M. Rand, ``Objective criteria for the evaluation of clustering methods,'' \emph{J. Amer. Statist. Assoc.}, vol. 66, no. 336, pp. 846-850, 1971.

\bibitem{ref49}
U. Von Luxburg, ``A tutorial on spectral clustering,'' \emph{Statist. Comput.}, vol. 17, no. 4, pp. 395-416, 2007.

\bibitem{ref50}
Evapotranspiration Zones in California, 2018, [Online]. Available: https://www.cimis.water.ca.gov/App-Themes/images/etozonemap.jpg.

\end{thebibliography}
\end{document}